\documentclass[useAMS,usenatbib]{mn2e} 
\usepackage[fleqn]{amsmath}
\usepackage{amssymb} 
\usepackage{graphicx} 
\usepackage{subfigure}
\usepackage{float} 
\usepackage{relsize}
\usepackage[mathscr]{euscript}

\title[Dust and gas mixtures in SPH]{Two fluid dust and gas mixtures in SPH: A
semi-implicit approach} \author[Pablo Lor\'en-Aguilar and Matthew R. Bate]{Pablo
Loren-Aguilar$^{1}$ and Matthew R. Bate$^{1}$ \thanks{E-mail:pablo@astro.ex.ac.uk
(PLA); mbate@astro.ex.ac.uk (MRB)} \\ $^{1}$ School of Physics and Astronomy, 
University of Exeter, Stocker Road, Exeter EX4 4QL, United Kingdom}

\begin{document}

\date{12 June 2014.}

\pagerange{\pageref{firstpage}--\pageref{lastpage}} \pubyear{2014} \maketitle
\label{firstpage} 

\begin{abstract} A method to avoid the explicit time
integration of small dust grains in the two fluid gas /dust smoothed particle
hydrodynamics (SPH) approach is proposed. By assuming a very simple exponential
decay model for the relative velocity between the gas and dust components, all
the effective characteristics of the drag force can be reproduced. A series of
tests has been performed to compare the accuracy of the method with
analytical and explicit integration results. We find that the method performs
well on a wide range of tests, and can provide large speed ups over explicit
integration when the dust stopping time is small. We have also found that the 
method is much less dissipative than conventional explicit or implicit two-fluid
SPH approaches when modelling dusty shocks. \end{abstract}


\section{Introduction}

Gas and dust mixtures are ubiquitously present in nature, so a correct numerical
prescription of its evolution turns out to be of the uttermost importance. In
many astrophysical applications, dust can be described as a set of particles
immersed in a fluid
phase (gas). Mathematically, such a system can be described using the \cite{S}
notation, by the following set of equations
%
% MISSING in the submitted version. There is an error in the equations!!!
%
\begin{equation} \begin{aligned} \hat{m}_{\rm D}\mathscr{D}_{\rm t,D}\textbf{v}_{\rm D}(t,\textbf{r}) &= 
\hat{m}_{\rm D} \left( \frac{\partial \textbf{v}_{\rm D}}{\partial t}(t,\textbf{r})+ (\textbf{v}_{\rm D}\cdot \nabla)\textbf{v}_{\rm D}
(t,\textbf{r}) \right)\\
&= \textbf{f}_{\rm ext} - K_{\rm s}(\textbf{v}_{\rm
D}-\textbf{v}_{\rm G}), \end{aligned} \label{Eu1} \end{equation}
\begin{equation} \begin{aligned} \rho_{\rm G}\mathscr{D}_{\rm t,G}\textbf{v}_{\rm G}(t,\textbf{r}) &=  \rho_{\rm G} \left( \frac{\partial \textbf{v}_{\rm G}}{\partial t}(t,\textbf{r})+ (\textbf{v}_{\rm G}\cdot \nabla)\textbf{v}_{\rm G}(t,\textbf{r}) \right) \\
&=- \mathbf{\nabla} P + \textbf{f}^{\rm V}_{\rm ext} +
n_{\rm D}K_{\rm s} (\textbf{v}_{\rm D}-\textbf{v}_{\rm G}), \end{aligned}
\label{Eu2} \end{equation}
\begin{equation} \begin{aligned} \rho_{\rm G}\mathscr{D}_{\rm t,G}u_{\rm G}(t,\textbf{r}) &= \rho_{\rm G} \left( \frac{\partial u_{\rm G}}{\partial t}(t,\textbf{r})+ (\textbf{v}_{\rm G}\cdot \nabla)u_{\rm G}(t,\textbf{r}) \right) \\
&=-P(\nabla \cdot \textbf{v}_{\rm G}) + n_{\rm D}K_{\rm s}(\textbf{v}_{\rm
D}-\textbf{v}_{\rm G})^2, \end{aligned} \label{Eu3} \end{equation}
\begin{equation} \frac{\partial n_{\rm D}}{\partial t} + \nabla(n_{\rm
D}\textbf{v}_{\rm D}) = 0, \label{Cont1} \end{equation}
\begin{equation} \frac{\partial \rho_{\rm G}}{\partial t} + \nabla(\rho_{\rm
G}\textbf{v}_{\rm G}) = 0, \label{Cont2} \end{equation}
%
% MISSING in the submitted version. There is an error in the equations!!!
%
where $n_{\rm D}$ and $\hat{m}_{\rm D}$ are the dust particles'
number density and mass respectively, $\rho_{\rm G}$ is the gas density,
$\textbf{v}_{\rm D}$ and $\textbf{v}_{\rm G}$ are the dust and gas 
velocities, $u_{\rm G}$ is the gas thermal energy, $K_{\rm s}$ is the 
drag coefficient for a single particle, $P$ represents the gas pressure, and 
$\textbf{f}_{\rm ext}$ stands for any external forces, like gravity 
or radiation pressure. Note than in equation \ref{Eu2}, the external 
force per unit volume $\textbf{f}^{\rm V}_{\rm ext}$ is required for the gas. 
$D/Dt$ is the Lagrangian derivative, and its specific form will be 
discussed in section 2. The effects of forces related to the intrinsic 
volume of the dust particles have been ignored, since in normal 
astrophysical applications they become negligible. 

In the present work, we will concentrate on
the study of drag forces. The form of the 
drag force of gas on a single dust grain may vary considerably as a 
function of the grain and gas properties \citep{Wei}. If the mean 
free path of the gas molecules is bigger than the dust particle radius
s (assuming spherical grains), the expression of the drag coefficient 
on a single dust grain becomes
\begin{figure*} \centering \includegraphics[width=80mm]{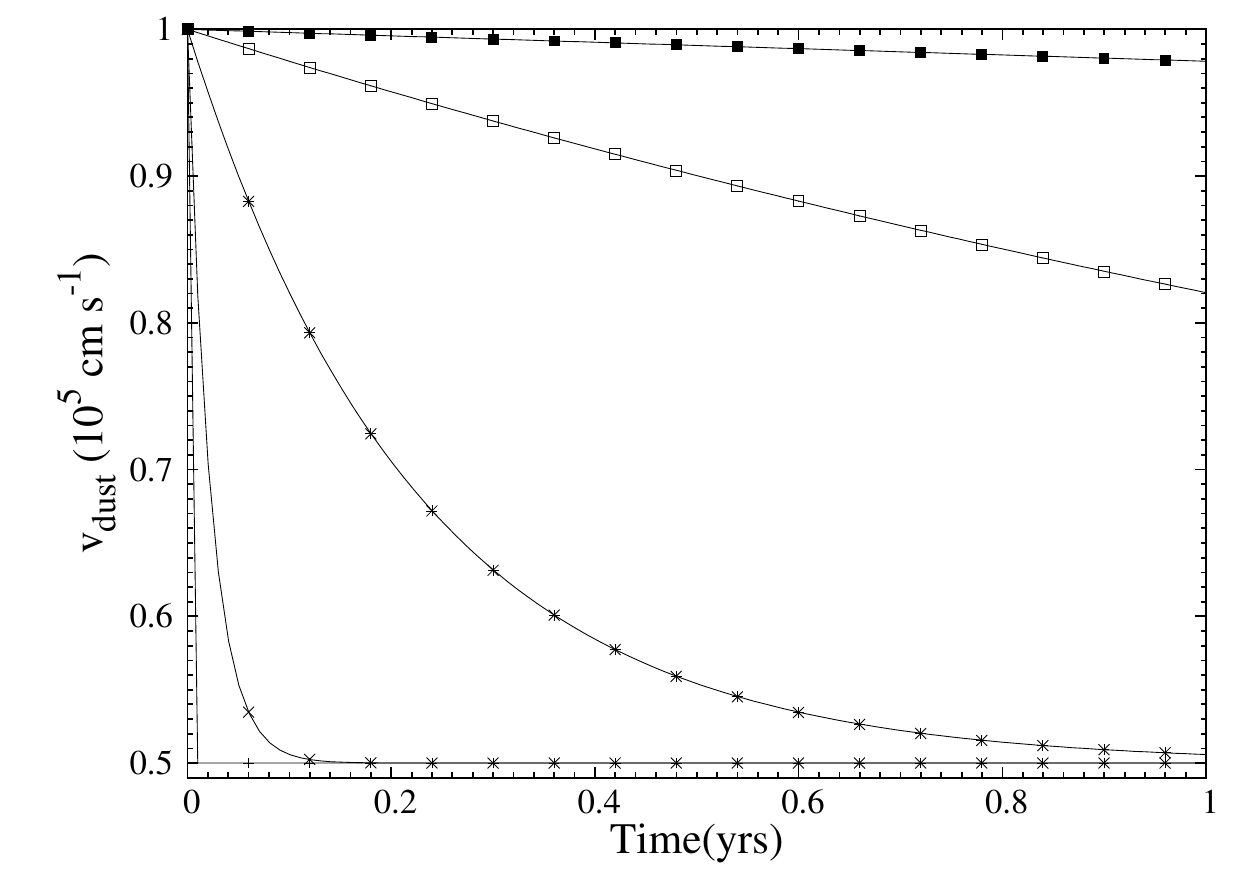}
\hspace{1cm} \includegraphics[width=80mm]{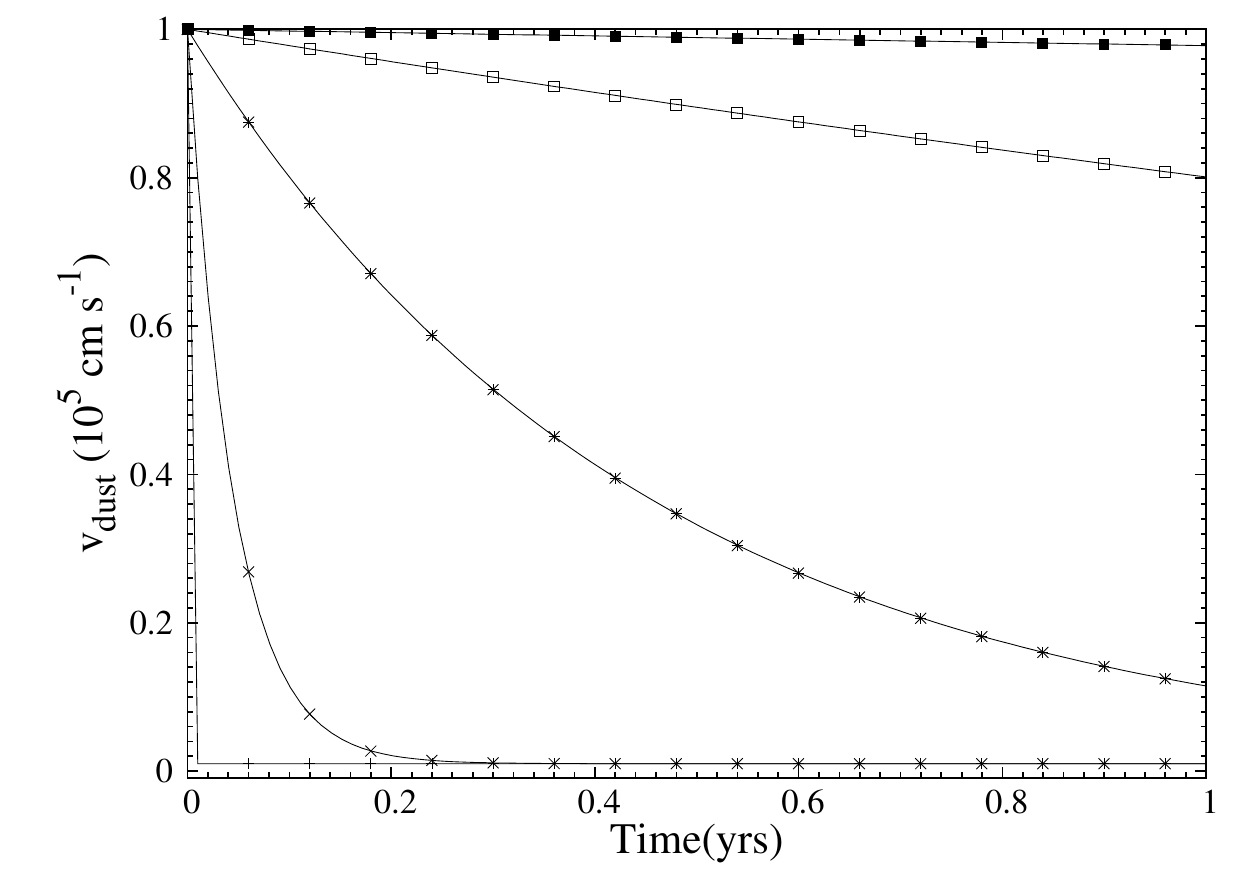}
\caption{Time evolution of a single SPH dust particle velocity in the 
\textsc{dustybox} test, for several different
dust grain sizes: $s=1~$mm, $1~$m, $10~$m, $100~$m, and $1~$km from
bottom to top. The adopted physical conditions are those appropriate
for a dust particle at the mid-plane of a protoplanetary disk at 1AU: 
$\rho_G=10^{-9}~$g~cm$^{-3}$, $v_{\rm th}\approx 10^{5}$~cm~s$^{-1}$,
and $\hat{\rho}_{\rm D}=3$~g~cm$^{-3}$. The computational domain 
comprises a total volume of 1 cubic AU. The method has been tested 
with two different dust-to-gas ratios, $\rho_{\rm D}/\rho_{\rm G}=1$
(left figure), and $\rho_{\rm D}/\rho_{\rm G}=0.01$ (right figure). 
A total of $20^3$ gas and $20^3$ dust particles have been used for the
test. Dotted lines represent the analytical solutions for the problem 
for each dust grain size.}
\label{fig:test1a} \end{figure*}
\begin{equation} K_{\rm s} \simeq \frac{4\pi}{3}\rho_{\rm G}s^2v_{\rm th},
~~~\text{(Epstein drag)}, \label{KEp} \end{equation}
where
\begin{equation} v_{\rm th} = \sqrt{\frac{8k_{\rm B}T}{\pi\mu m_{H}}},
\label{vth} \end{equation}
is the velocity of the gas molecules due to thermal motion, $T$ 
the gas temperature, $\mu$ is the mean molecular weight and $m_{\rm H}$ 
is the atomic mass of hydrogen. If on the contrary, the mean free path 
of the gas molecules is smaller than the dust particle radius, the 
expression of the drag force on a single dust particle becomes
\begin{equation} K_{\rm s} \simeq \frac{1}{2}C_{\rm D}\pi s^2 \rho_{\rm
G}|\textbf{v}_{\rm D}-\textbf{v}_{\rm G}|, \label{KSt}
\end{equation}
where the dimensionless coefficient $C_{\rm D}$ will be given by
\citep{Whi}
\begin{equation} C_{\rm D} \simeq 24R_e^{-1}, ~\text{for}~ R_e <
1,~~~\text{(Stokes drag)}, \label{CD1} \end{equation}
\begin{equation} C_{\rm D} \simeq
24R_e^{-0.6}, ~\text{for}~ 1< R_e < 800, \label{CD2} \end{equation}
\begin{equation} C_{\rm D}
\simeq 0.44, ~\text{for}~ R_e > 800, \label{CD3} \end{equation}
where $R_e=2s\rho_{\rm G}|\textbf{v}_{\rm D}-\textbf{v}_{\rm
G}|/\nu$ is the Reynolds number and $\nu$ is the molecular viscosity of
the gas. Under certain circumstances (typically for small dust grain sizes), the
acceleration experienced by the dust can become very large, leading to very short stopping
times. The occurrence of such short stopping times may become, under certain
circumstances, a very severe problem in the numerical simulation of dust and gas
mixtures. In protoplanetary disks, for example, the typical range of body sizes
spreads from micron-sized dust grains, up to kilometre-sized planetesimals.
Consequently, the ranges of dust-gas coupling intensities and stopping times
will be large, leading to a large range of dynamical time scales.

The first attempt to study gas and dust mixtures in the framework of the SPH
method was developed by \cite{MK}, and was subsequently improved by \cite{MNc}
by the inclusion of an implicit time-integration scheme. The main problem with
the method was its incapacity to guarantee a convergent solution under certain
circumstances. \cite{LPa,LPb} proposed a variation of \cite{MK} method. Despite
being capable of providing stable and convergent solutions, their method still
%
% MISSING in the submitted version. There are changes in the following paragraph!!!
%
suffers three main difficulties, intrinsic to any two fluid approach: (i) an inclination 
to produce artificial dust clumps whenever the dust is concentrated below the gas 
resolution, due to the pressureless nature of the dust component, 
(ii) the necessity of a very high spatial resolution, in the high drag regime in 
order to avoid overdissipation, and (iii) the necessity of a very high number of 
iterations in the implicit time integration scheme, or a very high number of 
time-steps in the explicit scheme, for the high drag regime. More recently, 
a new one fluid approach has been proposed by the same authors 
\citep{LaiPri2014a,LaiPri2014b}. In this new approach, both fluids are 
evolved as a single fluid by using the barycentric velocity as a common
reference frame. Through this approach, most of the aforementioned 
problems are avoided. However, in its present state, the one fluid 
method struggles with the low drag regime when dust and gas 
are not well described as a mixture and the velocity field should be 
multi-valued \citep{LaiPri2014b}, whereas a two-fluid method handles this
situation with ease.
%
% MISSING in the submitted version. There are changes in the previous paragraph!!!
%
%
% MISSING in the submitted version. The following paragraph is missing!!!
%
In this paper, a new two-fluid SPH method will be investigated in order to solve the third 
of the aforementioned problems. A simple semi-analytical model is proposed, 
in order to approximate the time evolution of the dust component, and thus 
avoid the need for a numerical integration of its time evolution. Special
attention will also be paid to the impact of overdissipation in the method. In
particular, it will be shown that the method is much better at resolving dusty 
shocks in the limit of short stopping times than other explicit or
implicit two-fluid SPH methods.  Whenever possible, an
estimation of the resolution requirements of the method will be provided.
%
% MISSING in the submitted version. The previous paragraph is missing!!!
%

This paper is organized as follows. In section 2, the possibility of imposing an
analytical decay model as an approximate solution for the small dusty grains
evolution will be discussed. In section 3, a series of numerical tests will be
presented in order to compare the accuracy of the present method with more
traditional approaches. Finally, in section 4, we will draw our conclusions.

\begin{figure*} \centering \includegraphics[width=80mm]{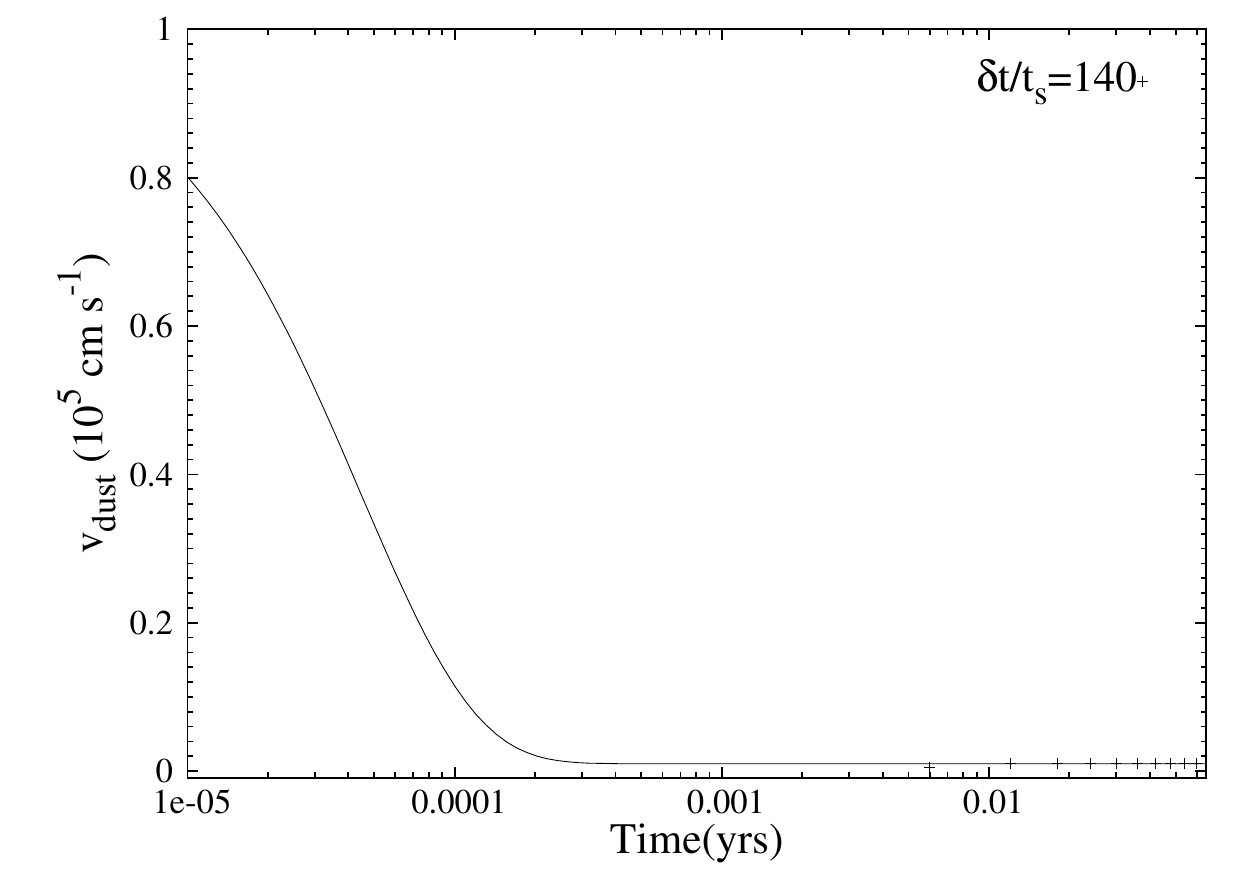} \hspace{12mm}
\includegraphics[width=80mm]{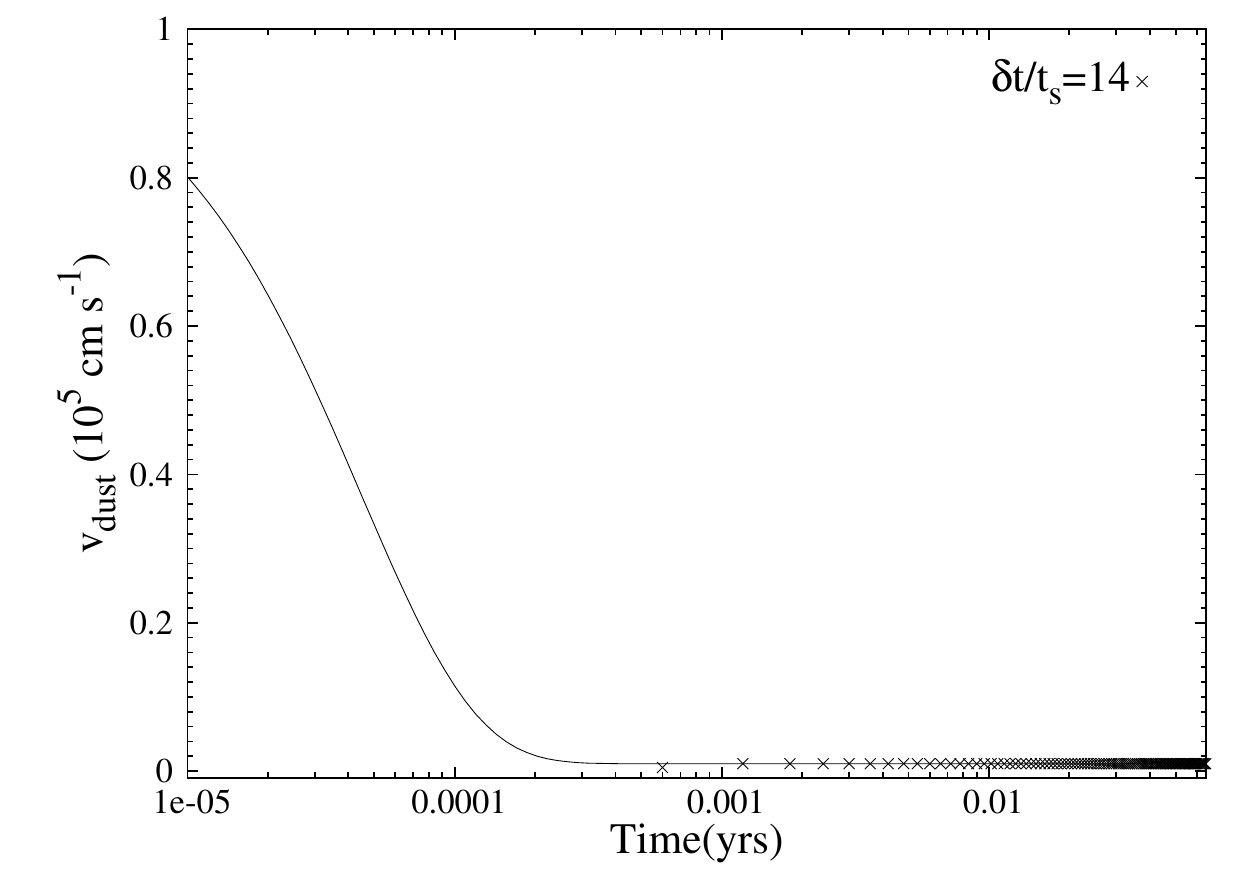}\\
\includegraphics[width=80mm]{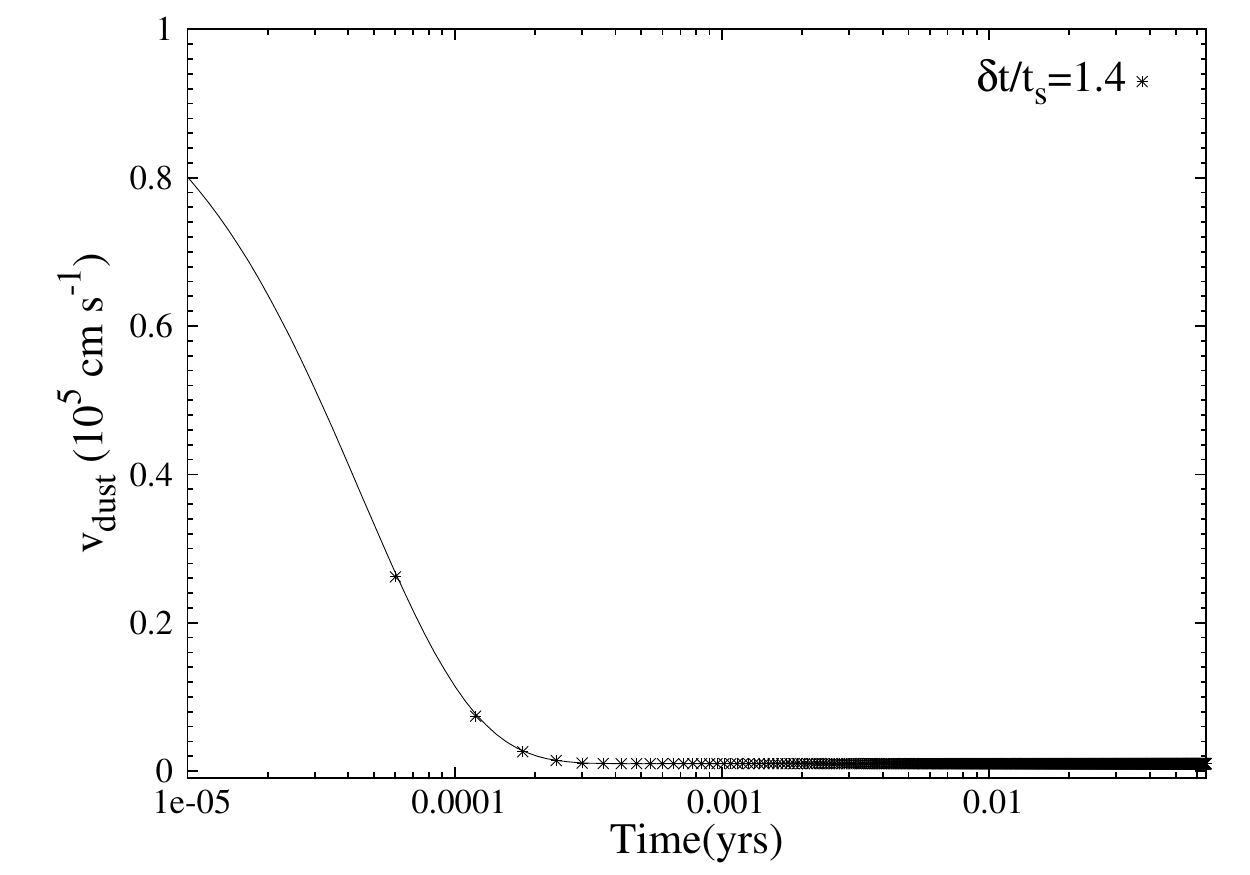} \hspace{12mm}
\includegraphics[width=80mm]{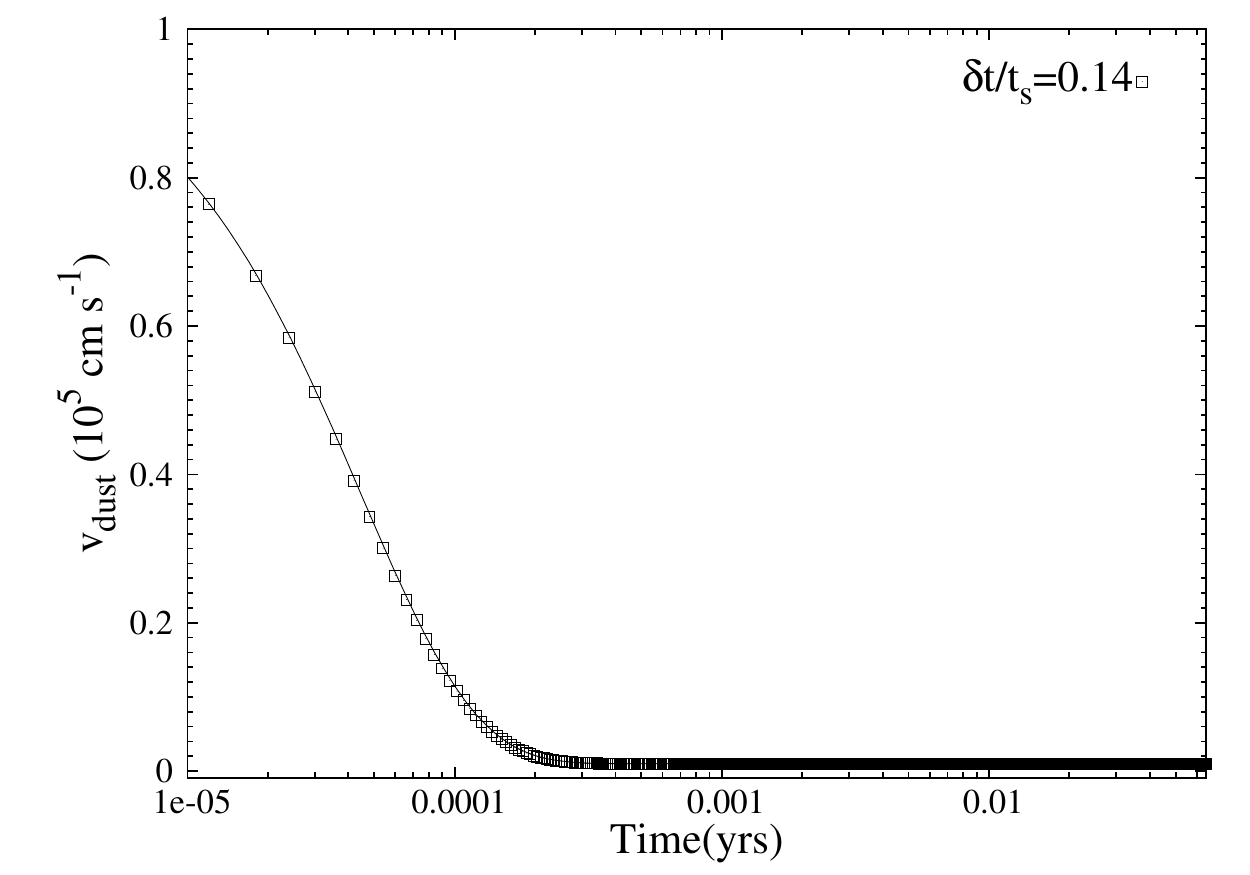} \caption{Time evolution of a single SPH
dust particle velocity in the \textsc{dustybox} test for a 
dust grain size $s=1$~mm. The adopted physical conditions are those appropriate
for a dust particle at the mid-plane of a protoplanetary disk at 1 AU: 
$\rho_G=10^{-9}$~g~cm$^{-3}$, $v_{\rm th}\approx 10^{5}$~cm~s$^{-1}$, and
$\hat{\rho}_{\rm D}=3$~g~cm$^{-3}$. A dust-to-gas ratio 
$\rho_{\rm D}/\rho_{\rm G}=0.01$ has been used in this case. In each
figure a different integration time-step $\delta t$ has been used, in order
to illustrate the behaviour of the method when $\delta t/t_{\rm s} >1$.}
\label{fig:test1b} \end{figure*}

\section{numerical method}

\subsection{Dust evolution in the Epstein regime}

As mentioned in the introduction, the objective of the present work is to avoid
the need for a full numerical integration of the velocity evolution of small
dust grains, whenever the stopping time becomes prohibitively short. In order to
do so, one could try to estimate the total change in velocity of a dust
particle, after having interacted through drag with the gas, for a certain time
$\delta t$. As seen in the introduction, if we concentrate exclusively on the
drag interaction, the equations of motion for the time-evolution of an arbitrary pair of dust and gas fluid elements (represented in a two-fluid 
SPH method by a pair of particles located at positions $\textbf{r}_{\rm D}$ 
and $\textbf{r}_{\rm G}$) are
\begin{equation}\begin{aligned} \mathscr{D}_{\rm t,D}\textbf{v}_{\rm D}(t,\textbf{r}_{\rm D}) &= \frac{\partial \textbf{v}_{\rm D}}{\partial t}(t,\textbf{r}_{\rm D})+ (\textbf{v}_{\rm D}\cdot \nabla)\textbf{v}_{\rm D}
(t,\textbf{r}_{\rm D})\\
&=-\frac{K^{\rm E}_{\rm s}} {\hat{m}_{\rm D}}\rho_{\rm G}\textbf{v}_{\rm DG}(t,\textbf{r}_{\rm D}), \label{EuEp1} \end{aligned} \end{equation}
\begin{equation} \begin{aligned} \mathscr{D}_{\rm t,G}\textbf{v}_{\rm G}(t,\textbf{r}_{\rm G}) &= \frac{\partial \textbf{v}_{\rm G}}{\partial t}(t,\textbf{r}_{\rm G})+ (\textbf{v}_{\rm G}\cdot \nabla)\textbf{v}_{\rm G}
(t,\textbf{r}_{\rm G})\\
&= \frac{K^{\rm E}_{\rm s}}{\hat{m}_{\rm D}}\rho_{\rm D}\textbf{v}_{\rm DG}(t,\textbf{r}_{\rm G}), \label{EuEp2} \end{aligned} \end{equation}
\begin{equation} \begin{aligned} \mathscr{D}_{\rm t,G}u_{\rm G}(t,\textbf{r}_{\rm G}) &= \frac{\partial u_{\rm G}}{\partial t}(t,\textbf{r}_{\rm G})+ (\textbf{v}_{\rm G}\cdot \nabla)u_{\rm G}(t,\textbf{r}_{\rm G})\\
&= \frac{K^{\rm E}_{\rm s}}{\hat{m}_{\rm D}}\rho_{\rm D}\textbf{v}^2_{\rm DG}(t,\textbf{r}_{\rm G}), \end{aligned} \label{EuEp3} \end{equation}
where $\rho_{\rm D}=\hat{m}_{\rm D}n_{\rm D}$ is the volume density of the dust component, $\textbf{v}_{\rm DG}(t,\textbf{r}) \equiv \textbf{v}_{\rm
D}(t,\textbf{r})-\textbf{v}_{\rm G} (t,\textbf{r})$ and we consider the Epstein 
regime where we have defined $K_{\rm s}^{\rm E} \equiv K_{\rm s}/\rho_{\rm G} = 
 4\pi s^2 v_{\rm th}/3$. In the present work,  the adopted 
evolutionary equations for the dust and gas components are
\begin{equation} \begin{aligned} \textbf{v}_{\rm D}(t+\delta t,\textbf{r}_{\rm D}) &= \textbf{v}_{\rm D}(t,\textbf{r}_{\rm D}) - \left(\frac{1-e^{-\delta t/t_{\rm s}}}
{1+\rho_{\rm D}/\rho_{\rm G}}\right)\textbf{v}_{\rm DG}(t,\textbf{r}_{\rm D}),
\end{aligned} \label{vD_sol} \end{equation}
\begin{equation} \begin{aligned} \textbf{v}_{\rm G}(t+\delta t,\textbf{r}_{\rm G}) &=\textbf{v} _{\rm G}(t,\textbf{r}_{\rm G}) \\ & +\frac{\rho_{\rm D}}{\rho_{\rm G}}\left(\frac{1-e^{-\delta t/t_{\rm s}}}{1+\rho_{\rm D}/\rho_{\rm
G}}\right)\textbf{v} _{\rm DG}(t,\textbf{r}_{\rm G}), \end{aligned} 
\label{vG_sol} \end{equation}
\begin{equation} \begin{aligned} u_{\rm G}(t+\delta t,\textbf{r}_{\rm G}) &=
u_{\rm G}(t,\textbf{r}_{\rm G}) \\ & + \frac{\rho_{\rm
D}}{2\rho_{\rm G}}\left(\frac{1-e^{-2\delta t/t_{\rm s}}}
{1+\rho_{\rm D}/\rho_{\rm G}}\right)\textbf{v}^2_{\rm DG}(t,\textbf{r}_{\rm G}),
\end{aligned} \label{uG_sol} \end{equation}
where
\begin{equation} t_{\rm s} \equiv \frac{\hat{m}_{\rm D}}{K^{\rm E}_{\rm s}
\rho_{\rm G}(1+\rho_{\rm D}/\rho_{\rm G})}. \label{ts} \end{equation}
Equations \ref{vD_sol}, \ref{vG_sol} and \ref{uG_sol} will constitute an 
approximate solution for the equations of motion,  as long as dust and gas 
densities can be considered as approximately constant along the integration 
time-step $\delta t$, since
\begin{equation} \begin{aligned}
 \mathscr{D}_{\rm t,D}\textbf{v}_{\rm D}(t,\textbf{r}_{\rm D}) &= 
 \lim_{\delta t \rightarrow 0} \frac{\textbf{v}_{\rm D}(t+\delta t,\textbf{r}_{\rm D})-\textbf{v}_{\rm D}(t,\textbf{r}_{\rm D})}{\delta t}\\
 &= -\lim_{\delta t \rightarrow 0} \left(\frac{1-e^{-\delta t/t_{\rm s}}}
 {1+\rho_{\rm D}/\rho_{\rm G}}\right)\frac{\textbf{v}_{\rm DG}(t,\textbf{r}_{\rm D})}{\delta t}\\
 &= -\frac{K^{\rm E}_{\rm s}} {\hat{m}_{\rm D}}\rho_{\rm G}\textbf{v}_{\rm DG}(t,\textbf{r}_{\rm D}),
\end{aligned} \end{equation}
\begin{equation} \begin{aligned}
 \mathscr{D}_{\rm t,G}\textbf{v}_{\rm G}(t,\textbf{r}_{\rm G}) &= 
 \lim_{\delta t \rightarrow 0} \frac{\textbf{v}_{\rm G}(t+\delta t,\textbf{r}_{\rm G})-\textbf{v}_{\rm G}(t,\textbf{r}_{\rm G})}{\delta t}\\
 &= \lim_{\delta t \rightarrow 0} \frac{\rho_{\rm D}}{\rho_{\rm G}}\left(\frac{1-e^{-\delta t/t_{\rm s}}} {1+\rho_{\rm D}/\rho_{\rm G}}\right)\frac{\textbf{v}_{\rm DG}(t,\textbf{r}_{\rm G})}{\delta t}\\
 &= \frac{K^{\rm E}_{\rm s}}{\hat{m}_{\rm D}}\rho_{\rm D}\textbf{v}_{\rm DG}(t,\textbf{r}_{\rm G}),
\end{aligned} \end{equation}
\begin{equation} \begin{aligned}
 \mathscr{D}_{\rm t,G}u_{\rm G}(t,\textbf{r}_{\rm G}) &= 
 \lim_{\delta t \rightarrow 0} \frac{u_{\rm G}(t+\delta t,\textbf{r}_{\rm G})-u_{\rm G}(t,\textbf{r}_{\rm G})}{\delta t}\\
 &= \lim_{\delta t \rightarrow 0} \frac{\rho_{\rm D}}{2\rho_{\rm G}}\left(\frac{1-e^{-2\delta t/t_{\rm s}}} {1+\rho_{\rm D}/\rho_{\rm G}}\right)\frac{\textbf{v}^2_{\rm DG}(t,\textbf{r}_{\rm G})}{\delta t}\\
 &= \frac{K^{\rm E}_{\rm s}}{\hat{m}_{\rm D}}\rho_{\rm D}\textbf{v}^2_{\rm DG}(t,\textbf{r}_{\rm G}). \end{aligned} \end{equation}
The main attractive of equations \ref{vD_sol}, \ref{vG_sol} and \ref{uG_sol} is that they can be used to approximately describe both strong and weak drag regimes. If $\delta t/t_{\rm s} \ll 1$, equations \ref{vD_sol} to \ref{uG_sol} become
\begin{equation} \begin{aligned}
\textbf{v}_{\rm D}(t+\delta t,\textbf{r}_{\rm D}) &\approx
 \textbf{v}_{\rm D}(t,\textbf{r}_{\rm D})-\frac{K^{\rm E}_{\rm s}} {\hat{m}_{\rm D}}\rho_{\rm G}\textbf{v}_{\rm DG}(t,\textbf{r}_{\rm D})\delta t,
\end{aligned} \end{equation}
\begin{equation} \begin{aligned}
\textbf{v}_{\rm G}(t+\delta t,\textbf{r}_{\rm G}) &\approx \textbf{v}_{\rm G}(t,\textbf{r}_{\rm G})
+ \frac{K^{\rm E}_{\rm s}}{\hat{m}_{\rm D}}\rho_{\rm D}\textbf{v}_{\rm DG}(t,\textbf{r}_{\rm G})\delta t, \end{aligned} \end{equation}
\begin{equation} \begin{aligned}
u_{\rm G}(t+\delta t,\textbf{r}_{\rm G}) &\approx u_{\rm G}(t,\textbf{r}_{\rm G})
+ \frac{K^{\rm E}_{\rm s}}{\hat{m}_{\rm D}}\rho_{\rm D}\textbf{v}^2_{\rm DG}(t,\textbf{r}_{\rm G})\delta t, \end{aligned} \end{equation}
whereas if $\delta t/t_{\rm s} \gg 1$, equations \ref{vD_sol}, \ref{vG_sol} and \ref{uG_sol} simply read
\begin{equation} \begin{aligned}
 \textbf{v}_{\rm D}(t+\delta t,\textbf{r}_{\rm D}) & = \frac{\rho_{\rm D}\textbf{v}_{\rm D}(t,\textbf{r}_{\rm D})+\rho_{\rm G}\textbf{v}_{\rm G}(t,\textbf{r}_{\rm D})}{\rho_{\rm D}+\rho_{\rm G}}, \end{aligned} \end{equation}
\begin{equation} \begin{aligned}
 \textbf{v}_{\rm G}(t+\delta t,\textbf{r}_{\rm G}) & = \frac{\rho_{\rm D}\textbf{v}_{\rm D}(t,\textbf{r}_{\rm G})+\rho_{\rm G}\textbf{v}_{\rm G}(t,\textbf{r}_{\rm G})}{\rho_{\rm D}+\rho_{\rm G}}, \end{aligned} \end{equation}
\begin{equation} \begin{aligned} u_{\rm G}(t+\delta t,\textbf{r}_{\rm G}) &=
u_{\rm G}(t,\textbf{r}_{\rm G}) + \frac{1}{2}\left(\frac{\rho_{\rm D}}
{\rho_{\rm D}+\rho_{\rm G}}\right)\textbf{v}^2_{\rm DG}(t,\textbf{r}_{\rm G}),
\end{aligned} \end{equation}
which is the expected solution for the equations of motion of a
strongly coupled dust and gas mixture. 
%
% MISSING in the submitted version. Should be "in the dust frame" and not "at the dust frame"!!!
%
Another attractive feature of equations \ref{vD_sol} to \ref{uG_sol} is that they naturally incorporate, due to its fully Lagrangian nature, perfect advection into the numerical scheme. If one calculates the time evolution of the relative velocity between dust and gas, in the dust frame, one gets
%
% MISSING in the submitted version. Should be "in the dust frame" and not "at the dust frame"!!!
%
%
% MISSING in the submitted version. There is some extra formatting of the equations!!!
%
\begin{equation} \begin{aligned}
\mathscr{D}_{\rm t,D} \textbf{v}_{\rm DG}(t,\textbf{r}_{\rm D}) &= \frac{\partial \textbf{v}_{\rm DG}}{\partial t}(t,\textbf{r}_{\rm D})+ (\textbf{v}_{\rm D}\cdot \nabla)\textbf{v}_{\rm DG}(t,\textbf{r}_{\rm D})\\
%&= \frac{\partial \textbf{v}_{\rm D}}{\partial t}(t,\textbf{r}_{\rm D})+ %(\textbf{v}_{\rm D}\cdot \nabla)\textbf{v}_{\rm D}(t,\textbf{r}_{\rm D})\\
%&-\frac{\partial \textbf{v}_{\rm G}}{\partial t}(t,\textbf{r}_{\rm D})- %(\textbf{v}_{\rm D}\cdot \nabla)\textbf{v}_{\rm G}(t,\textbf{r}_{\rm D})\\
&= \frac{\partial \textbf{v}_{\rm D}}{\partial t}(t,\textbf{r}_{\rm D})+ (\textbf{v}_{\rm D}\cdot \nabla)\textbf{v}_{\rm D}(t,\textbf{r}_{\rm D})\\
&~~~~ -\frac{\partial \textbf{v}_{\rm G}}{\partial t}(t,\textbf{r}_{\rm D})- (\textbf{v}_{\rm G}\cdot \nabla)\textbf{v}_{\rm G}(t,\textbf{r}_{\rm D})\\
&~~~~ -(\textbf{v}_{\rm DG}\cdot\nabla)\textbf{v}_{\rm G}(t,\textbf{r}_{\rm D})\\
&= \mathscr{D}_{\rm t,D} \textbf{v}_{\rm D}(t,\textbf{r}_{\rm D}) -
\mathscr{D}_{\rm t,G} \textbf{v}_{\rm G}(t,\textbf{r}_{\rm D})\\
&~~~~ -(\textbf{v}_{\rm DG}\cdot\nabla)\textbf{v}_{\rm G}(t,\textbf{r}_{\rm D})\\
&= -\frac{K^{\rm E}_{\rm s}}{\hat{m}_{\rm D}}\textbf{v}_{\rm DG}(t,\textbf{r}_{\rm D}) - \frac{K^{\rm E}_{\rm s}}{\hat{m}_{\rm D}}\rho_{\rm D}\textbf{v}_{\rm DG}(t,\textbf{r}_{\rm D}) \\
&~~~~ -(\textbf{v}_{\rm DG}\cdot\nabla)\textbf{v}_{\rm G}(t,\textbf{r}_{\rm D})\\
&=-\frac{\textbf{v}_{\rm DG}(t,\textbf{r}_{\rm D})}{t_{\rm s}}
-(\textbf{v}_{\rm DG}\cdot\nabla)\textbf{v}_{\rm G}(t,\textbf{r}_{\rm D}).
\label{LagDG} \end{aligned} \end{equation}
%
% MISSING in the submitted version. There is some extra formatting of the equations!!!
%
So, as long as the velocity evolution for each phase is calculated
by using the local acceleration on each frame, the extra term related to the 
differential velocity of the frames, will be naturally included into the scheme.
This property, although not completely intuitive, can clearly be seen if one 
considers the case of ballistic particles and a gas that do not interact at 
all (something that SPH can treat very easily).  The Lagrangian equations 
that describe the evolution of such a system, which are the equations that 
would be solved by an SPH implementation, are
\begin{equation} \mathscr{D}_{\rm t,D}\textbf{v}_{\rm D}(t,\textbf{r}_{\rm D}) = 0, 
\label{Trivial1} \end{equation}
\begin{equation} \mathscr{D}_{\rm t,G}\textbf{v}_{\rm G}(t,\textbf{r}_{\rm G}) = 
- \frac{\nabla P_{\rm G}}{\rho_{\rm G}}. \label{Trivial2} \end{equation}
If one now calculates the time variation of the relative velocity between the
phases as in equation \ref{LagDG}, we obtain
\begin{equation}
\mathscr{D}_{\rm t,D}\textbf{v}_{\rm DG}(t,\textbf{r}_{\rm D}) = \frac{\nabla P_{\rm G}}{\rho_{\rm G}} - 
(\textbf{v}_{\rm DG}\cdot\nabla)\textbf{v}_{\rm G}(t,\textbf{r}_{\rm G}),
\label{vDG_nodrag}
\end{equation}
where the second term on the right hand side just reflects that we have had to choose between the dust and the gas when defining our Lagrangian derivative.  It is not an extra term that needs to be implemented. Note that in some recent SPH one-fluid prescriptions, the extra advection terms do need to be explicitly calculated (e.g. equation 14 of \cite{LaiPri2014a}). This characteristic should
be clearly considered as an advantage of our method. 

The key to our two fluid method for modelling a dusty gas is that we now operator split the differential equations that describe the evolution of gas and dust, so that we solve everything except the drag term using standard explicit
integration methods, and subsequently modify the resulting velocities by applying the drag term separately.  For example, to include gas pressure and drag forces between the dust and the gas, we first use standard explicit SPH integration to apply
\begin{equation} \mathscr{D}_{\rm t,D}\textbf{v}_{\rm D}(t,\textbf{r}_{\rm D}) = 0, 
\label{Trivial3} \end{equation}
\begin{equation} \mathscr{D}_{\rm t,G}\textbf{v}_{\rm G}(t,\textbf{r}_{\rm G}) = 
- \frac{\nabla P_{\rm G}}{\rho_{\rm G}}, \label{Trivial4} \end{equation}
\begin{equation} \mathscr{D}_{\rm t,G}u_{\rm G}(t,\textbf{r}_{\rm G}) =
-\frac{P_{\rm G}(\nabla \cdot \textbf{v}_{\rm G})}{\rho_{\rm G}}, \end{equation}
and then, we apply equations \ref{vD_sol} to \ref{uG_sol} to the obtained intermediate velocities and thermal energy $\tilde{\textbf{v}}_{\rm D}(t+\delta t,\textbf{r}_{\rm D})$, $\tilde{\textbf{v}}_{\rm G}(t+\delta t,\textbf{r}_{\rm G})$ and $\tilde{u}_{\rm G}(t+\delta t,\textbf{r}_{\rm G})$
%
% MISSING in the submitted version. There is some extra formatting in the equations!!!
%
\begin{equation} \begin{aligned}
\textbf{v}_{\rm D}(t+\delta t,\textbf{r}_{\rm D}) &= \tilde{\textbf{v}}_{\rm D}(t+\delta t,\textbf{r}_{\rm D}) \\
&~~~~ -\left(\frac{1-e^{-\delta t/t_{\rm s}}}{1+\rho_{\rm D}/\rho_{\rm G}}\right)\tilde{\textbf{v}}_{\rm DG}(t+\delta t,\textbf{r}_{\rm D}), \label{vD_sol2} \end{aligned} \end{equation}
\begin{equation} \begin{aligned}
\textbf{v}_{\rm G}(t+\delta t,\textbf{r}_{\rm G}) &= \tilde{\textbf{v}}_{\rm G}(t+\delta t,\textbf{r}_{\rm G}) \\
&~~~~ + \frac{\rho_{\rm D}}{\rho_{\rm G}}\left(\frac{1-e^{-\delta t/t_{\rm s}}}{1+\rho_{\rm D}/\rho_{\rm G}}\right)\tilde{\textbf{v}}_{\rm DG}(t+\delta t,\textbf{r}_{\rm G}), \label{vG_sol2} 
\end{aligned} \end{equation}
\begin{equation} \begin{aligned} u_{\rm G}(t+\delta t,\textbf{r}_{\rm G}) &=
\tilde{u}_{\rm G}(t,\textbf{r}_{\rm G}) \\ 
&~~~~  + \frac{\rho_{\rm
D}}{2\rho_{\rm G}}\left(\frac{1-e^{-2\delta t/t_{\rm s}}}
{1+\rho_{\rm D}/\rho_{\rm G}}\right)\tilde{\textbf{v}}^2_{\rm DG}(t+\delta t,
\textbf{r}_{\rm G}). \label{uG_sol2}  \end{aligned} \end{equation}
%
% MISSING in the submitted version. There is some extra formatting in the equations!!!
%
\begin{figure*} \centering \includegraphics[width=80mm]{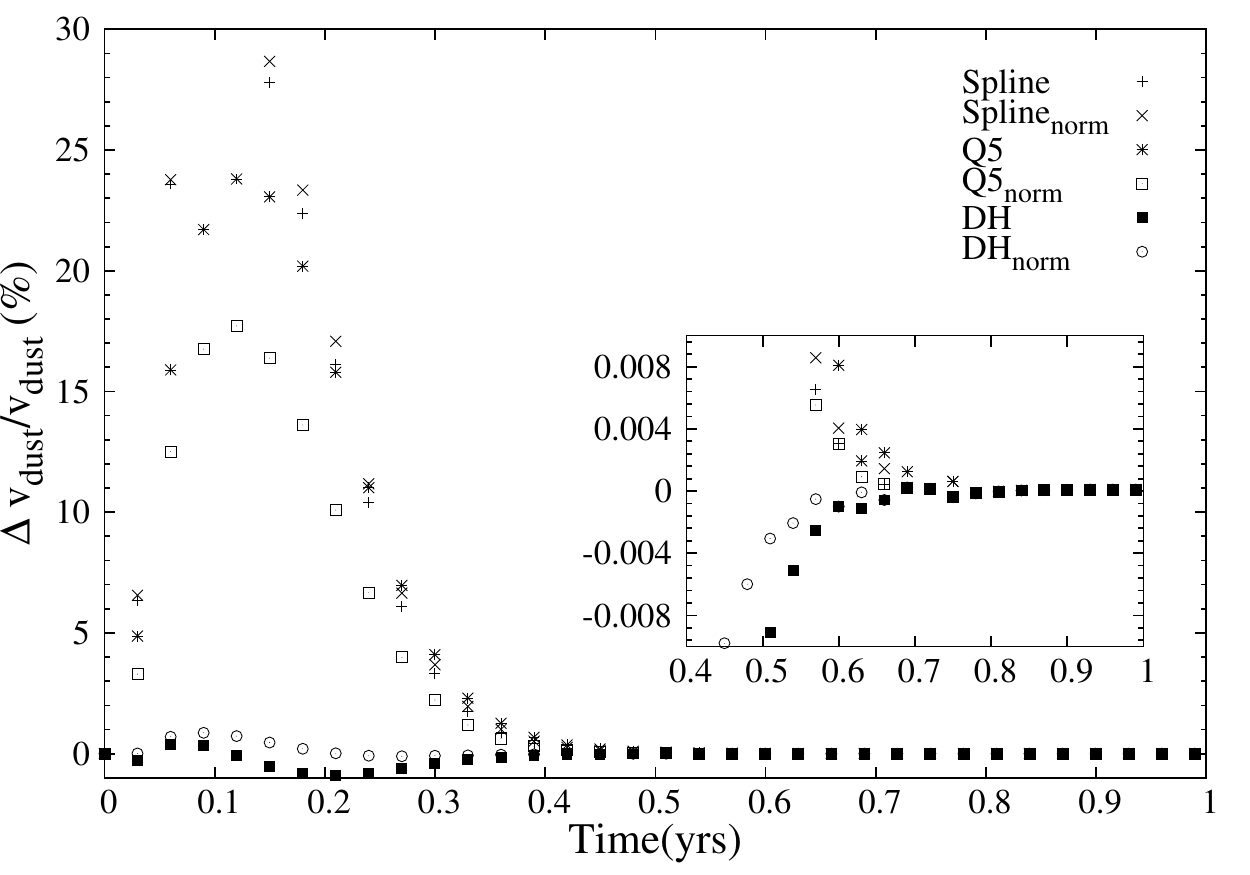}
\hspace{12mm} \includegraphics[width=80mm]{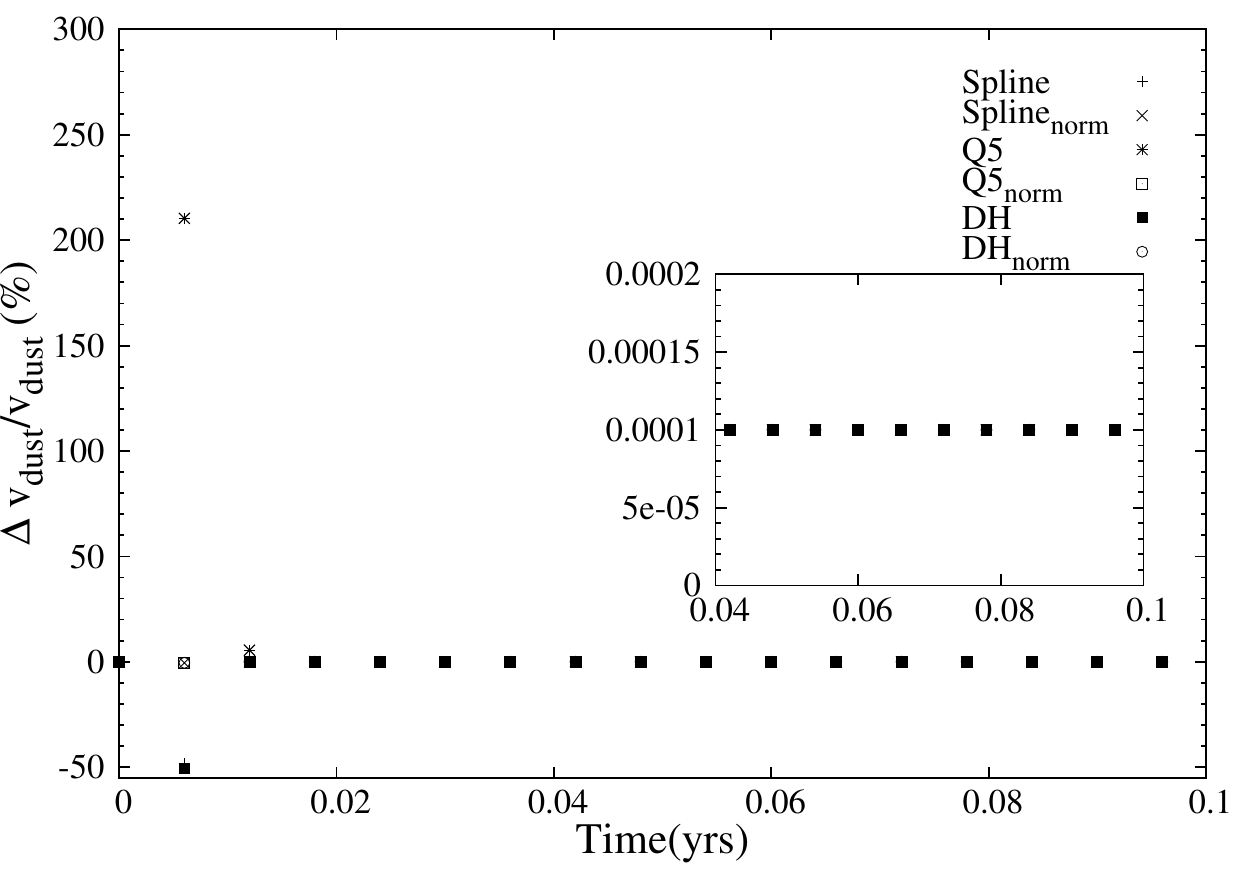}\\
\caption{Time evolution of the relative error in the \textsc{dustybox} test with a
dust-to-gas ratio 0.01. Left figure corresponds to the $s=1$~m case 
and right figure corresponds to the $s=1$~mm case. As can be seen, the
limit velocity is correctly predicted, irrespectively of the kernel used, with
an extremely high precision. The maximum errors are obtained during the velocity
decay phase. If the double hump kernel with the normalization condition is
used, the maximum relative errors in the decay phase are $\lesssim 1 \%$}
\label{fig:test1c} \end{figure*}

%
% MISSING in the submitted version. It should read "..the gas and dust
% elements ARE discretized".!!!
%
In order to apply equations \ref{vD_sol2}, \ref{vG_sol2}, and \ref{uG_sol2}
in the SPH two fluid approach, the gas and dust elements are discretized
into a set of mass elements, often called particles. Any continuous
quantity will be thus reconstructed by means of an interpolation method
%
% MISSING in the submitted version. It should read "..the gas and dust
% elements ARE discretized".!!!
%
\begin{equation} A(\textbf{r}) = \sum_{\rm k} \frac{m_{\rm k}}{\rho_{\rm
k}}A_{\rm k}W(\mid\textbf{r}-\textbf{r}_{\rm k}\mid ,h_{\rm k}),
\label{Interp1} \end{equation}
\begin{equation} \nabla A(\textbf{r}) = \sum_{\rm k} \frac{m_{\rm
k}}{\rho_{\rm k}}A_{\rm k}\nabla W(\mid\textbf{r}-\textbf{r}_{\rm k}\mid
,h_{\rm k}), \label{Interp2} \end{equation}
where $m_{\rm k}$ is the mass of each SPH particle, $h_{\rm k}$ 
is the smoothing length of each SPH particle, and $W$ is
the interpolating function, called the kernel \citep[see for example][]{MN}. 
In general, in the two fluid scheme, the value of the gas velocity at a
dust location (and vice versa) will be unknown, so in equations \ref{vD_sol2},
\ref{vG_sol2}, and \ref{uG_sol2} the use of equations 
\ref{Interp1} and \ref{Interp2} will be necessary.  In particular, using the 
$i$ index to refer to dust particles, $j$ to gas particles, and
$k$ to the neighbours of opposite type, we can evaluate the difference 
between the dust and gas velocities as
\begin{equation} \textbf{v}_{\rm DG}(t,\textbf{r}_{\rm i}) = \sum^{\rm
gas}_{k}\frac{m_{\rm k}}{\rho_{\rm k}}\textbf{v}_{\rm ik}
W(|\textbf{r}_{\rm ik}|,h_{\rm k}), \label{vDG_sum1} \end{equation}
\begin{equation} \textbf{v}_{\rm DG}(t,\textbf{r}_{\rm j}) = \sum^{\rm
dust}_{k}\frac{m_{\rm k}}{\rho_{\rm k}}\textbf{v}_{\rm kj}
W(|\textbf{r}_{\rm kj}|,h_{\rm j}), \label{vDG_sum2} \end{equation}
where $\textbf{r}_{\rm ik}\equiv \textbf{r}_{\rm i}-
\textbf{r}_{\rm k}$, $\textbf{r}_{\rm kj}\equiv \textbf{r}_{\rm k}-
\textbf{r}_{\rm j}$, $\textbf{v}_{\rm ik}\equiv \textbf{v}_{\rm i}-
\textbf{v}_{\rm k}$, and $\textbf{v}_{\rm kj}\equiv \textbf{v}_{\rm k}-
\textbf{v}_{\rm j}$. By using SPH interpolation, equations \ref{vD_sol2}, 
\ref{vG_sol2} and \ref{uG_sol2} can be discretized
\begin{equation} \begin{aligned} \textbf{v}^{\rm i}_{\rm D}(t&+\delta t,
\textbf{r}_{\rm i}) = \tilde{\textbf{v}}^{\rm i}_{\rm D}(t+\delta t,\textbf{r}_{\rm i}) \\ 
&-\frac{\nu}{N_{\rm i}}\sum_{\rm k}^{\rm gas}\frac{m_{\rm k}} {\rho_{\rm k}}
\frac{1-e^{-\delta t/t^{\rm i}_{\rm s}}}{1+\rho_{\rm i}/\rho_{\rm k}} 
(\tilde{\textbf{v}}_{\rm ik}\cdot\hat{\textbf{r}}_{\rm ik})\hat{\textbf{r}}_{\rm ik}
W(|\textbf{r}_{\rm ik}|,h_{\rm k}), \label{vD_SPH1} \end{aligned} \end{equation}
\begin{equation} \begin{aligned} \textbf{v}^{\rm j}_{\rm G}(t&+\delta t,
\textbf{r}_{\rm j}) = \tilde{\textbf{v}}^{\rm j}_{\rm G}(t+\delta t,\textbf{r}_{\rm j}) \\
&+\nu\sum_{\rm k}^{\rm dust}\frac{m_k}{N_{\rm k}\rho_{\rm j}}
\frac{1-e^{-\delta t/t^{\rm k}_{\rm s}}}{1+\rho_{\rm k}/\rho_{\rm
j}}(\tilde{\textbf{v}}_{\rm kj}\cdot\hat{\textbf{r}}_{\rm kj})\hat{\textbf{r}}_{\rm kj}
W(|\textbf{r}_{\rm kj}|,h_{\rm j}), \end{aligned} \label{vG_SPH1} \end{equation}
\begin{equation} \begin{aligned} u^{\rm j}_{\rm G}(t&+\delta t,\textbf{r}_{\rm j})
= \tilde{u}^{\rm j}_{\rm G}(t+\delta t,\textbf{r}_{\rm j}) \\ &+ \frac{\nu}{2}\sum_{\rm k}^{\rm dust}
\frac{m_{\rm k}}{N_{\rm k}\rho_{\rm j}}\frac{1-e^{-2\delta t/t^{\rm k}_{\rm s}}}{1+\rho_{\rm k}/\rho_{\rm j}}(\textbf{v}_{\rm kj}\cdot\hat{\textbf{r}}_{\rm kj})(\tilde{\textbf{v}}_{\rm kj}\cdot\hat{\textbf{r}}_{\rm kj})W(|\textbf{r}_{\rm
kj}|,h_{\rm j}).  \label{uG_SPH1} \end{aligned} \end{equation}
%
% MISSING in the submitted version. There are changes in the following paragraph!!!
%
where $N_{\rm i}$ and $N_{\rm k}$ are normalisation factors (see below).
Physically, SPH particles should be understood as finite mass elements  
of each one of the components. In particular, SPH 
dust particles should be interpreted as homogeneous ensembles of dust
particles of radius $s$, intrinsic mass $\hat{m}_{\rm D}$, and number
density $n_{\rm D}$. Therefore, for each SPH dust particle one can assign
a volume density $\rho_{\rm D}$ which will represent the total dust mass contained within the volume of the SPH dust particle (determined by its kernel support
radius). Smoothing lengths and volume densities for both components can be calculated 
by a standard iterative SPH manner, solving
%
% MISSING in the submitted version. There are changes in the previous paragraph!!!
%
\begin{equation}
h = \sigma \left(\frac{m}{\rho}\right)^{1/3},
\end{equation} 
through a Newton-Raphson method \citep{PriMon2004}, where $\sigma=1.2$ for the standard cubic spline kernel, and the dust and gas densities are given by
\begin{equation}
\rho_D(\textbf{r}_{\rm i}) = \sum_{\rm k}^{\rm dust}m_{\rm k}W(|\textbf{r}_{\rm
i}-\textbf{r}_{\rm k}|,h_{\rm i}), \label{summation1} \end{equation}
\begin{equation}
\rho_G(\textbf{r}_{\rm j}) = \sum_{\rm k}^{\rm gas}m_{\rm k}W(|\textbf{r}_{\rm
j}-\textbf{r}_{\rm k}|,h_{\rm j}). \label{summation2} \end{equation}
This procedure is equivalent to solving the continuity equations \ref{Cont1} 
and \ref{Cont2}. SPH particle masses will be assigned by dividing the total 
mass of each component present in the simulation, by the number of particles 
of the component.

%
% MISSING in the submitted version. Should read "..the gas and dust mass fraction
% contained within the interpolation sphere..."!!!
%
In order to calculate the dust-to-gas ratio at a given dust particle location, 
we estimate the gas and dust mass fraction contained within the interpolation sphere
of the SPH dust particle. That is, we take
%
% MISSING in the submitted version. Should read "..the gas and dust mass fraction
% contained within the interpolation sphere..."!!!
%
\begin{equation}
\frac{\rho_{\rm D}}{\rho_{\rm G}} = \frac{m_{\rm D}}{m_{\rm G}} = 
\frac{m_{\rm D}}{\rho_{\rm G}}\left(\frac{\sigma}{h_{\rm D}}\right)^3.
\label{dust-to-gas}\end{equation}
This prescription is chosen due to its greater stability, in
comparison with the simpler dust and gas densities quotient. We have found
that, whenever discontinuities are present in the computational domain (for 
example in the shock tube test), the fluctuations in the dust density can lead 
to high stopping time fluctuations if the ratio $\rho_{\rm D}/\rho_{\rm G}$ is used directly in equation \ref{ts}. If equation \ref{dust-to-gas} is used, 
because the mass of the SPH dust particle is constant, the fluctuations are 
avoided. Furthermore, this approach allows us to calculate dust evolution 
even with a very low number of SPH dust particles, since it does not rely 
on the validity of the fluid approximation for the dust component.

Also, and in order to minimize fluctuations if a low number of neighbours is present, a normalization factor $N_{\rm i}$ has also been included in the SPH dust summation \citep{RL}, equal to
\begin{equation} N_{\rm i} = \sum_{\rm k}^{\rm gas}\frac{m_{\rm
k}}{\rho_{\rm k}}W(|\textbf{r}_{\rm ik}|,h_{\rm k}). \label{norm} \end{equation}
Due to the symmetric structure of equations \ref{vD_SPH1} and \ref{vG_SPH1} linear momentum is preserved during the interaction,and a projection of the relative velocity along the line of sight of the particlesis introduced in order to guarantee angular momentum conservation \citep{MK}. A normalization factor $\nu$, equal to the number of the spatial dimensions of the system, is necessary to guarantee the equivalence of the projection method with equations \ref{vD_sol}, \ref{vG_sol} and \ref{uG_sol} up to a second order approximation (see \cite{LPa} for an excellent discussion). For the same reason, energy can also be shown to be conserved. The kinetic energy of the mixture, at $t+\delta t$, will be expressible as
%
% MISSING in the submitted version. There is some extra formatting in the equations!!!
%
\begin{equation} \begin{aligned}
&E_{\rm k} (t+\delta t) = \frac{1}{2}\sum_{\rm i}^{Dust}m_{\rm i}(\textbf{v}_{\rm i}+\delta \textbf{v}_{\rm i})^2 + \frac{1}{2}\sum_{\rm j}^{Gas}m_{\rm j}(\textbf{v}_{\rm j}+\delta \textbf{v}_{\rm j})^2\\
&= \frac{1}{2}\sum_{\rm i}^{Dust}m_{\rm i}(\textbf{v}_{\rm i})^2 +
\sum_{\rm i}^{Dust}m_{\rm i}(\textbf{v}_{\rm i}\cdot\delta \textbf{v}_{\rm i})+\frac{1}{2}\sum_{\rm i}^{Dust}m_{\rm i}(\delta \textbf{v}_{\rm i})^2\\
&~~~~ +\frac{1}{2}\sum_{\rm j}^{Gas}m_{\rm j}(\textbf{v}_{\rm j})^2 +
\sum_{\rm j}^{Gas}m_{\rm j}(\textbf{v}_{\rm j}\cdot\delta \textbf{v}_{\rm j})+\frac{1}{2}\sum_{\rm j}^{Gas}m_{\rm j}(\delta \textbf{v}_{\rm j})^2\\
&=E_{\rm K}(t)+\sum_{\rm i}^{Dust}m_{\rm i}(\textbf{v}_{\rm i}\cdot\delta \textbf{v}_{\rm i})+\frac{1}{2}\sum_{\rm i}^{Dust}m_{\rm i}(\delta \textbf{v}_{\rm i})^2\\
&~~~~ +\sum_{\rm j}^{Gas}m_{\rm j}(\textbf{v}_{\rm j}\cdot\delta \textbf{v}_{\rm j})+\frac{1}{2}\sum_{\rm j}^{Gas}m_{\rm j}(\delta \textbf{v}_{\rm j})^2.\end{aligned} \end{equation}
So the change in kinetic energy will be
\begin{equation} \begin{aligned}
\Delta E_{\rm K} &= \sum_{i}^{Dust}m_{\rm i}\textbf{v}_{\rm i}\cdot\delta\textbf{v}_{\rm D}^{\rm i} + \frac{1}{2}\sum_{i}^{Dust}m_{\rm i}(\delta \textbf{v}_{\rm D}^{\rm i})^2\\
&~~~~ +\sum_{j}^{Gas}m_{\rm j}\textbf{v}_{\rm j}\cdot\delta\textbf{v}_{\rm G}^{\rm j} +
\frac{1}{2}\sum_{j}^{Gas}m_{\rm j}(\delta \textbf{v}_{\rm G}^{\rm j})^2.
\end{aligned} \end{equation}
Then, by assuming that
\begin{equation} \begin{aligned}
\delta \textbf{v}_{\rm D}^{\rm i} &= -\frac{1-e^{\delta t/t_{\rm s}}}{1+\rho_{\rm D}/\rho_{\rm G}}\textbf{v}_{\rm DG}(t,\textbf{r}_{\rm i}) \equiv -\xi\textbf{v}_{\rm DG}(t,\textbf{r}_{\rm i}),\\
\delta \textbf{v}_{\rm G}^{\rm j} &= \frac{\rho_{\rm D}}{\rho_{\rm G}}\frac{1-e^{\delta t/t_{\rm s}}}{1+\rho_{\rm D}/\rho_{\rm G}}\textbf{v}_{\rm DG}(t,\textbf{r}_{\rm j}) \equiv \frac{\rho_{\rm D}}{\rho_{\rm G}}\xi\textbf{v}_{\rm DG}(t,\textbf{r}_{\rm j}),\end{aligned} \end{equation}
one finds, after introducing the SPH summations
\begin{equation} \begin{aligned}
\Delta E_{\rm K} & = -\nu\sum_{ik}\frac{m_{\rm i}m_{\rm k}} {N_{\rm i}\rho_{\rm k}}\xi\textbf{v}_{\rm i}(\textbf{v}_{\rm ik}\cdot\hat{\textbf{r}}_{\rm ik})\hat{\textbf{r}}_{\rm ik}W(|\textbf{r}_{\rm ik}|,h_{\rm k})\\
&~~~~ +\frac{1}{2}\nu\sum_{ik}\frac{m_{\rm i}m_{\rm k}} {N_{\rm i}\rho_{\rm k}}\xi^2(\textbf{v}_{\rm ik}\cdot\hat{\textbf{r}}_{\rm ik})^2 W(|\textbf{r}_{\rm ik}|,h_{\rm k})\\
&~~~~ +\nu\sum_{kj}\frac{m_{\rm k}m_{\rm j}} {N_{\rm k}\rho_{\rm j}}
\xi\textbf{v}_{\rm j}(\textbf{v}_{\rm kj}\cdot\hat{\textbf{r}}_{\rm kj})\hat{\textbf{r}}_{\rm kj}W(|\textbf{r}_{\rm kj}|,h_{\rm j})\\
&~~~~ +\frac{1}{2}\nu\sum_{kj}\frac{m_{\rm k}m_{\rm j}} {N_{\rm k}\rho_{\rm j}}\frac{\rho_{\rm k}}{\rho_{\rm j}}\xi^2(\textbf{v}_{\rm kj}\cdot\hat{\textbf{r}}_{\rm kj})^2
W(|\textbf{r}_{\rm kj}|,h_{\rm j})\\
&=-\nu\sum_{ij}\frac{m_{\rm i}m_{\rm j}} {N_{\rm i}\rho_{\rm j}}
\xi(\textbf{v}_{\rm ij}\cdot\hat{\textbf{r}}_{\rm ij})^2W(|\textbf{r}_{\rm ij}|,h_{\rm j})\\
&~~~~ +\frac{1}{2}\nu\sum_{ij}\frac{m_{\rm i}m_{\rm j}} {N_{\rm i}\rho_{\rm j}}\left(1+\frac{\rho_{\rm i}}{\rho_{\rm j}}\right)\xi^2(\textbf{v}_{\rm ij}\cdot\hat{\textbf{r}}_{\rm ij})^2 W(|\textbf{r}_{\rm ij}|,h_{\rm j})\\
&=-\nu\sum_{\rm ij}\frac{m_{\rm i}m_{\rm j}} {N_{\rm i}\rho_{\rm j}}
\xi\left(1-\frac{1}{2}\xi\left(1+\frac{\rho_{\rm i}}{\rho_{\rm j}}\right)\right)(\textbf{v}_{\rm ij}\cdot\hat{\textbf{r}}_{\rm ij})^2W(|\textbf{r}_{\rm ij}|,h_{\rm j})\\
&=-\sum_{\rm j}^{Gas}m_{\rm j}\nu\sum_{i}^{Dust} \frac{m_{\rm i}} {2N_{\rm i}\rho_{\rm j}}
\frac{1-e^{-2\delta t/t_{\rm s}}}{1+\rho_{\rm i}/\rho_{\rm j}}(\textbf{v}_{\rm ij}\cdot\hat{\textbf{r}}_{\rm ij})^2W(|\textbf{r}_{\rm ij}|,h_{\rm j})\\
&=-\sum_{\rm j}^{Gas}m_{\rm j} \delta u_{\rm G}^{\rm j} = -\Delta U_{\rm G}.\end{aligned}\end{equation}
%
% MISSING in the submitted version. There is some extra formatting in the previous equations!!!
%
%
The method has been tested with two different integrators, a second order
Runge-Kutta Fehlberg \citep{RK,Wetz}, and a second order predictor-corrector
\citep{Sea}. The obtained results with the two integrators have been equivalent
in all cases, except in the sound wave test (section 3.2) where the Runge-Kutta
scheme leads to a poorer energy and momentum conservation. Please, see
appendices A and B for a detailed explanation of both integration methods.

\subsection{Stability and convergence of the method}

To investigate the stability of the numerical scheme, equations
\ref{vD_sol} and \ref{vG_sol} may be written in the following form
\begin{equation} \frac{\textbf{v}^{\rm n+1}_{\rm D}- \textbf{v}^{\rm n}_{\rm
D}}{\xi} = -\textbf{v}^{\rm n}_{\rm DG}, \label{vD_stab} \end{equation}
\begin{equation} \frac{\textbf{v}^{n+1}_{\rm G} - \textbf{v}^{n}_{\rm G}}{\xi}
=\frac{\rho_{\rm D}}{\rho_{\rm G}}\textbf{v}^{\rm n}_{\rm DG},
\label{vG_stab} \end{equation}
where
\begin{equation} \xi \equiv \frac{1-e^{-\delta t/t_{\rm s}}}{1+\rho_{\rm
D}/\rho_{\rm G}}. \label{chi} \end{equation}
As can be seen, equations \ref{vD_stab} and \ref{vG_stab} can be 
interpreted as a forward Euler method, where velocity is evolved with 
respect to $\xi$ instead of time. Following \cite{LPa}, a von Newmann
analysis can be done. If the dust and gas components are perturbed
with a monochromatic plane wave
\begin{equation} \textbf{v}^{\rm n}_{\rm D} = \textbf{V}^{\rm n}_{\rm
D}e^{i\textbf{k}\cdot\textbf{x}}, \end{equation} \label{vD_vN}
\begin{equation} \textbf{v}^{\rm n}_{\rm G} = \textbf{V}^{\rm n}_{\rm
G}e^{i\textbf{k}\cdot\textbf{x}}, \label{vG_vN} \end{equation}
equations \ref{vD_stab} and \ref{vG_stab}  may be written 
as the following linear system
\begin{equation} \left( \begin{array}{c} \textbf{V}_{\rm D} \\ \textbf{V}_{\rm
G} \end{array} \right)^{n+1} = \left( \begin{array}{cc} 1- \xi & \xi\\
\xi\frac{\rho_{\rm D}}{\rho_{\rm G}} & 1- \xi\frac{\rho_{\rm D}}{\rho_{\rm
G}}\end{array} \right)\left( \begin{array}{c} \textbf{V}_{\rm D} \\
\textbf{V}_{\rm G} \end{array} \right)^{n}. \label{vDG_stab} \end{equation}
The corresponding two eigenvalues of the system are
\begin{equation} \lambda_{\pm} = 1 - \frac{\xi}{2}\left(1+\frac{\rho_{\rm
D}}{\rho_{\rm G}}\right) \pm \frac{\xi}{2} \left(1+\frac{\rho_{\rm D}}{\rho_{\rm
G}}\right), \label{lambda} \end{equation}
and the system will remain numerically stable ($\lambda_{-}<1$)
whenever
\begin{equation} \xi <\frac{1}{1+\rho_{\rm D}/\rho_{\rm G}},
\label{Cour1} \end{equation}
which will always occur, given the definition of $\xi$, except in
the limit $t_{\rm s} \rightarrow 0$. In this case, equation \ref{Cour1} will
act as a Courant-like condition. In order to keep stability, it will be
enough to decrease the integration $\xi$-step by a factor of 2, and 
evolve the system of equations \ref{vD_stab} and \ref{vG_stab} in two steps.
Because of the existing linear relation between the velocity and $\xi$,
the accuracy of the solution will not be affected by the number of steps 
performed, like in an ordinary explicit integration scheme.

\begin{figure*} \centering \includegraphics[width=80mm]{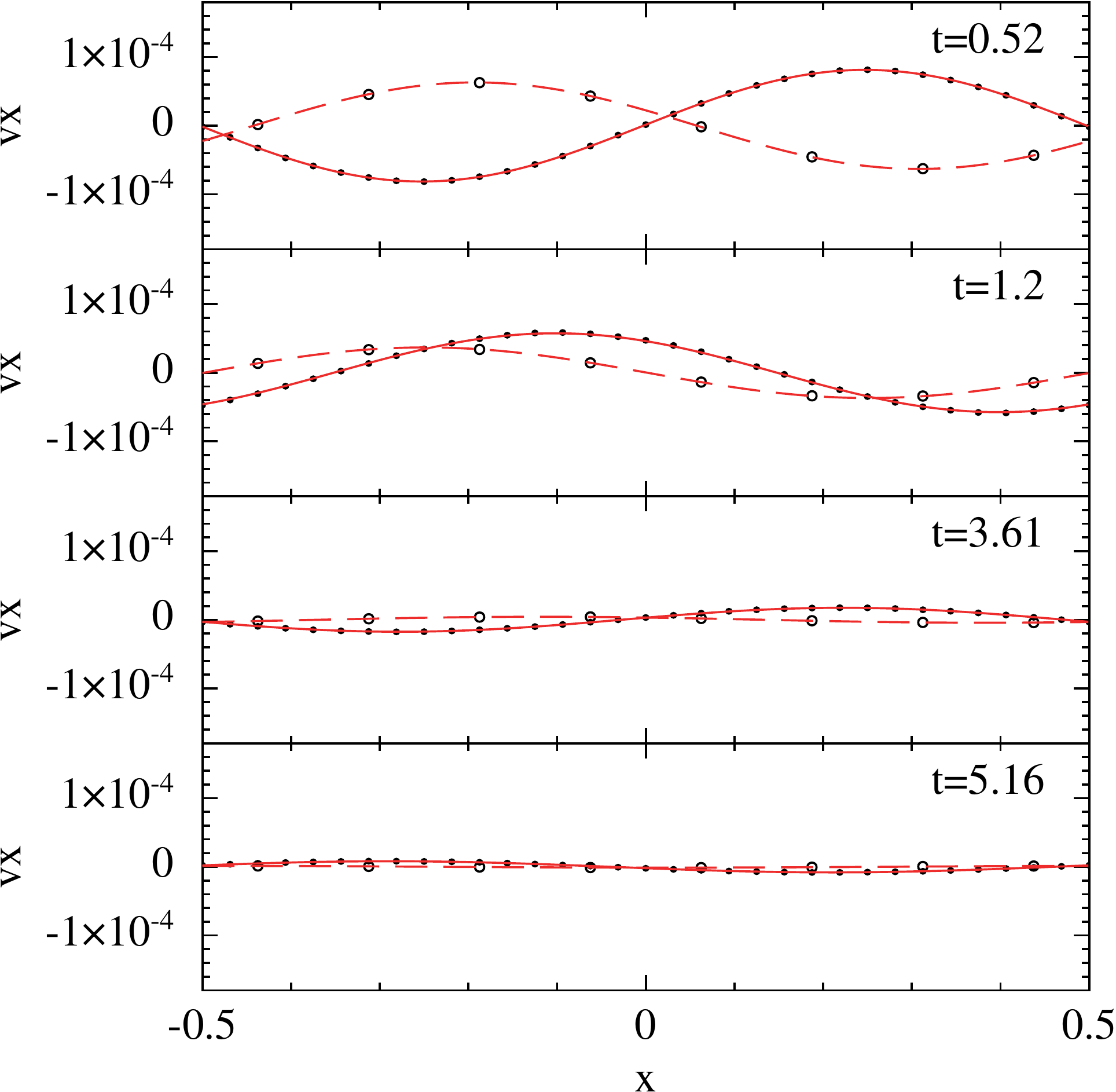}
\hspace{14mm}\includegraphics[width=80mm]{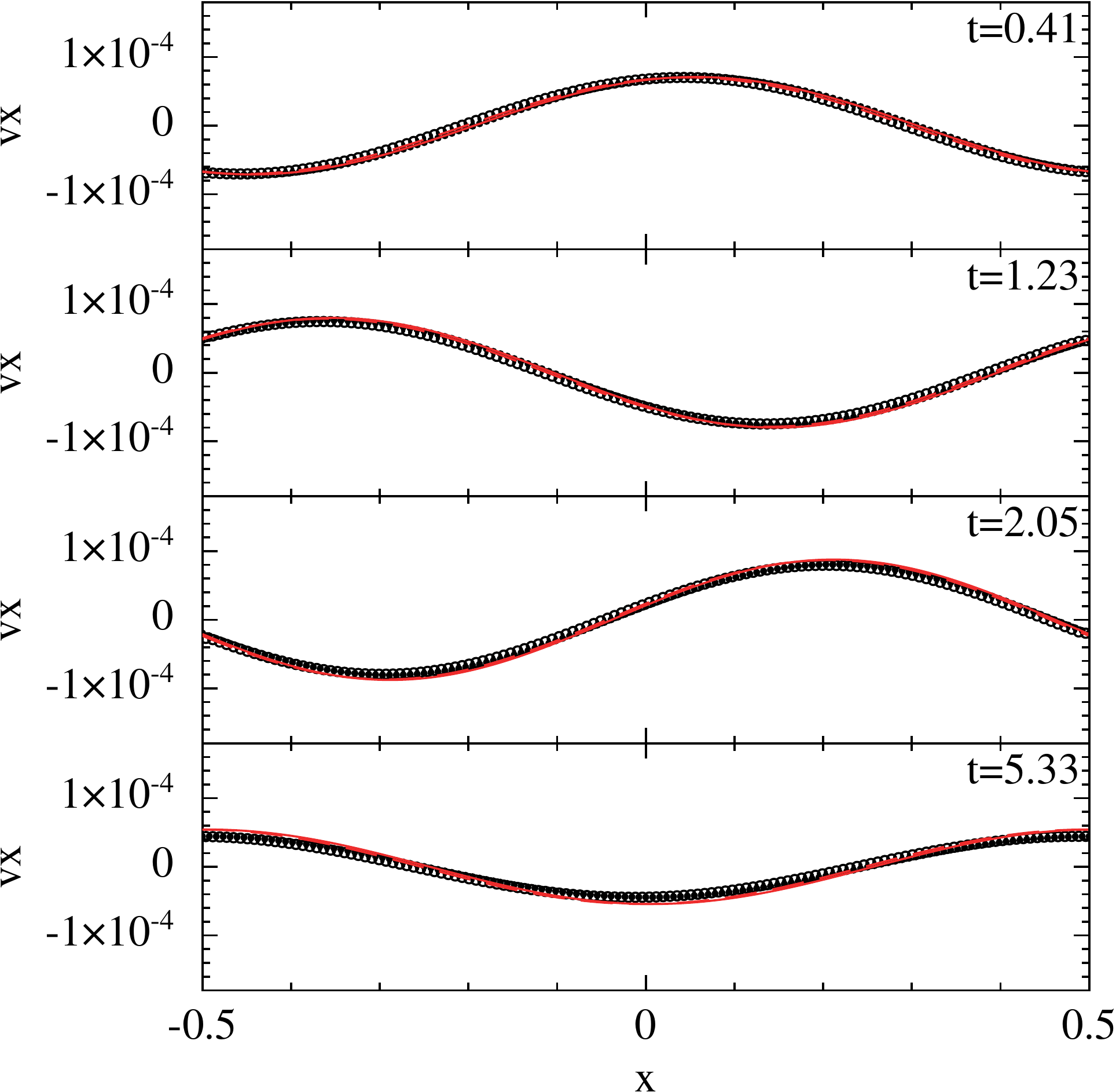} \caption{Time evolution of
the gas (stars) and dust (open circles) components in the \textsc{dustywave} test
with $\rho_{\rm D}/\rho_{\rm G}=1$ case. Dashed (dust) and solid (gas) lines
represent the analytical solutions for the gas and dust components respectively.
Left panels correspond to a low drag regime with $\delta t/t_s \approx 10^{-3}$
($K_{\rm const}=1$), where 32 and 8 particles have been respectively
used for the gas and dust components. The right panels correspond to a strong
drag regime with $\delta t/t_s \approx 0.1$ ($K_{\rm const}=100$). A
total of 128 dust and gas particles have been necessary in this case in order to
reproduce the solution. In order to quantify the deviations of the numerical
solutions with respect to the analytical solutions, the error norms can be
calculated for both cases. At $t=5.33$, $L1=1.8\times 10^{-2}$,
$L2=2\times10^{-2}$, and $L3=3.4\times10^{-2}$ for the $K_{\rm const}=1$
case, while $L1=1.2\times 10^{-1}$,$L2=1.3\times10^{-1}$, and
$L\infty=1.9\times10^{-1}$ for the $K_{\rm const}=100$ case. In the
latter case, a higher deviation from the analytical solution can be observed due
to the presence of overdissipation.} \label{fig:test2a} \end{figure*}

The present method possesses two very different regimes, depending on the ratio
between the gas integration time-scale $\delta t$, and the dust stopping time
$t_{\rm s}$. If $\delta t/t_s \ll 1$ (when an explicit integration could be
used)
\begin{equation} \xi \approx \frac{K^{\rm E}_{\rm s}}{\hat{m}_{\rm
D}}\rho_{\rm G}\delta t, \label{chi_low} \end{equation}
and equations \ref{vD_SPH1}, \ref{vG_SPH1} and \ref{uG_SPH1} thus become
%
% MISSING in the submitted version. Extra formatting in the equations
%
\begin{equation} \begin{aligned} \textbf{v}_{\rm D}^i(&t+\delta t,\textbf{r}_{\rm i}) \approx 
\textbf{v}_{\rm D}^i(t,\textbf{r}_{\rm i})  \\ 
& ~~~~ - \frac{\nu}{N_{\rm i}}
\sum_{\rm k}^{\rm gas}m_{\rm k}
\frac{K^{\rm E}_{\rm s}}{\hat{m}_{\rm D}}(\textbf{v}_{\rm ik}\cdot
\hat{\textbf{r}}_{\rm ik})\hat{\textbf{r}}_{\rm ik}W(|\textbf{r}_{\rm ik}|,h_{\rm k})
\delta t \\ 
&= \textbf{v}_{\rm D}^{\rm i}(t,\textbf{r}_{\rm i}) - 
\frac{K^{\rm E}_{\rm s}\rho_{\rm G}}{\hat{m}_{\rm D}} \textbf{v}_{\rm DG}(t,\textbf{r}_{\rm i}) \delta t,
\end{aligned} \label{vD_SPH3} \end{equation} 
\begin{equation} \begin{aligned} \textbf{v}_{\rm G}^j(&t+\delta t,\textbf{r}_{\rm j}) \approx
\textbf{v}_{\rm G}^j(t,\textbf{r}_{\rm j}) -\left.\frac{\nabla P_{\rm G}}{\rho_{\rm G}}
\right\vert_{\textbf{r}_{\rm j}}\delta t \\
& ~~~~ + \nu\sum_{\rm k}^{\rm dust}\frac{m_{\rm k}}{N_{\rm k}}\frac{K^{\rm E}_{\rm s,k}}{\hat{m}_{\rm D}}(\textbf{v}_{\rm kj} \cdot\hat{\textbf{r}}_{\rm kj})
\hat{\textbf{r}}_{\rm kj}W(|\textbf{r}_{\rm jk}|,h_{\rm j})
\delta t  \\ &= \textbf{v}_{\rm G}^{\rm j}(t,\textbf{r}_{\rm j})
-\left.\frac{\nabla P_{\rm G}}{\rho_{\rm G}}\right\vert_{\textbf{r}_{\rm j}}\delta t + \frac{K^{\rm E}_{\rm s}\rho_{\rm D}}{\hat{m}_{\rm D}}\textbf{v}_{\rm DG}(t,\textbf{r}_{\rm j})\delta t,\label{vG_SPH3} \end{aligned} \end{equation}
\begin{equation} \begin{aligned} u_{\rm G}^j(&t+\delta t,\textbf{r}_{\rm j}) \approx u_{\rm G}^j(t,\textbf{r}_{\rm j}) -\left.\frac{P_{\rm G}(\nabla\cdot\textbf{v}_{\rm G})}{\rho_{\rm G}}\right\vert_{\textbf{r}_{\rm j}}\delta t\\ 
& ~~~~ + \nu\sum_{\rm k}^{\rm dust}\frac{m_{\rm
k}}{N_{\rm k}}\frac{K^{\rm E}_{\rm s,k}}{\hat{m}_{\rm D}}
(\textbf{v}_{\rm kj}\cdot\hat{\textbf{r}}_{\rm kj})^2
W(|\textbf{r}_{\rm jk}|,h_{\rm j})\delta t \\
&= u_{\rm G}^{\rm j}(t,\textbf{r}_{\rm j}) -\left.\frac{P_{\rm G}(\nabla\cdot\textbf{v}_{\rm G})}{\rho_{\rm G}}\right\vert_{\textbf{r}_{\rm j}}\delta t
+\frac{K^{\rm E}_{\rm s}\rho_{\rm D}}{\hat{m}_{\rm D}}
\textbf{v}^2_{\rm DG}(t,\textbf{r}_{\rm j}) \delta t.\end{aligned}
\label{uG_SPH3} \end{equation}
%
% MISSING in the submitted version. Extra formatting in the equations
%
In this limit, the stability condition \ref{Cour1} becomes
\begin{equation} \delta t < \frac{\hat{m}_{\rm D}}{K^{\rm E}_{\rm s}
\rho_{\rm G}(1+\rho_{\rm D}/\rho_{\rm G})}=t_{\rm s}, \end{equation}
coinciding with the Courant condition of an explicit integration as
shown by \cite{LPa}. Therefore, equations \ref{vD_SPH3} to \ref{uG_SPH3} will
be equivalent to an explicit SPH two fluid method \citep{LPa} as long as very
sharp density gradients are absent from the gas component. To quantify the
errors produced by this approximation, the behaviour of the
algorithm in the presence of strong density gradients (shocks) will be tested
in section 3.3.

If, on the contrary, $\delta t/t_{\rm s} \gg 1$ (i.e. in the strong drag regime)
\begin{equation} \xi \approx \frac{1}{1+\rho_{\rm D}/\rho_{\rm G}},
\label{chi_high} \end{equation}
then equations \ref{vD_SPH1}, \ref{vG_SPH1}, and \ref{uG_SPH1} become
%
% MISSING in the submitted version. Extra formatting in the equations
%
\begin{equation} \begin{aligned} \textbf{v}_{\rm D}^i(t&+\delta t,\textbf{r}_{\rm i}) =
\nu\sum_{\rm k}^{\rm gas}\frac{m_{\rm k}}{\rho_{\rm k}}\left[\frac{\rho_{\rm i}
\textbf{v}_{\rm i}+\rho_{\rm k}\textbf{v}_{\rm k}} {\rho_{\rm i}+\rho_{\rm
k}}\cdot\hat{\textbf{r}}_{\rm ik}\right]\hat{\textbf{r}}_{\rm ik}W(|\textbf{r}_{\rm
ik}|,h_{\rm k})  \\ 
&~~~~ -\nu\sum_{\rm k}^{\rm gas}\frac{m_{\rm k}}{\rho_{\rm k}}\left(\frac{\nabla P_{\rm k}}{\rho_{\rm i}+\rho_{\rm k}}\cdot\hat{\textbf{r}}_{\rm ik}\right)\hat{\textbf{r}}_{\rm ik}W(|\textbf{r}_{\rm ik}|,h_{\rm k})\delta t \\
& = \frac{\rho_{\rm D}\textbf{v}^{\rm i}_{\rm D}
(t,\textbf{r}_{\rm i}) + \rho_{\rm G}\textbf{v}_{\rm G}(t,\textbf{r}_{\rm i})}{\rho_{\rm D} + \rho_{\rm G}}-\left.\frac{\nabla P_{\rm G}}{\rho_{\rm D}+\rm \rho_{\rm G}}\right\vert_{\textbf{r}_{\rm i}}\delta t,
\end{aligned} \label{vD_SPH4} \end{equation} 
\begin{equation} \begin{aligned} \textbf{v}_{\rm G}^j(t&+\delta t,\textbf{r}_{\rm j}) = 
\nu\sum_{\rm k}^{\rm dust}\frac{m_{\rm k}}{\rho_{\rm j}}
\left[\frac{\rho_{\rm j}\textbf{v}_{\rm j}+\rho_{\rm k}
\textbf{v}_{\rm k}} {\rho_{\rm j}+\rho_{\rm k}}\cdot
\hat{\textbf{r}}_{\rm kj}\right]\hat{\textbf{r}}_{\rm kj}
W(|\textbf{r}_{\rm kj}|,h_{\rm j})  \\
&~~~~ -\nu\sum_{\rm k}^{\rm dust}\frac{m_{\rm k}}{\rho_{\rm k}}\left(\frac{\nabla P_{\rm j}}{\rho_{\rm k}+\rho_{\rm j}}\cdot\hat{\textbf{r}}_{\rm kj}\right)\hat{\textbf{r}}_{\rm kj}W(|\textbf{r}_{\rm kj}|,h_{\rm j})\delta t \\
&=\frac{\rho_{\rm D} \textbf{v}_{\rm D}(t,\textbf{r}_{\rm j}) + \rho_{\rm G}
\textbf{v}^{\rm j}_{\rm G}(t,\textbf{r}_{\rm j})}{\rho_{\rm D} + \rho_{\rm G}}
-\left.\frac{\nabla P_{\rm G}}{\rho_{\rm D}+\rm \rho_{\rm G}}\right\vert_{\textbf{r}_{\rm j}}\delta t,
\end{aligned} \label{vG_SPH4} \end{equation}
\begin{equation} \begin{aligned} &u_{\rm G}^j(t+\delta t,\textbf{r}_{\rm j})
= u_{\rm G}^j(t,\textbf{r}_{\rm j}) -\left.\frac{P_{\rm G}(\nabla\cdot\textbf{v}_{\rm G})}{\rho_{\rm G}}\right\vert_{\textbf{r}_{\rm j}}\delta t  \\
& + \frac{\nu}{2}\sum_{\rm k}^{\rm dust}\frac{m_{\rm
k}}{\rho_{\rm j}+\rho_{\rm k}}(\textbf{v}_{\rm
kj}\cdot\hat{\textbf{r}}_{\rm kj}+\frac{\nabla P_{\rm j}}{\rho_{\rm j}}\cdot\hat{\textbf{r}}_{\rm kj}\delta t)^2W(|\textbf{r}_{\rm jk}|,h_{\rm j}),
\end{aligned} \label{uG_SPH4} \end{equation}
%
% MISSING in the submitted version. Extra formatting in the equations
%
%
% MISSING in the submitted version. Should be "incorporate" not "incorporates!!!
%
which means that both components will be travelling, after the drag
interaction, at the barycentric velocity of the fluid. Note that the last term
of equation  \ref{uG_SPH4}, just incorporate all the relative kinetic energy between the phases into thermal energy. In this regime, the algorithm removes all relative dust and gas motion by setting them in the barycentric velocity, and then applies an equal amount of
pressure to both phases (equations \ref{vD_SPH4} and \ref{vG_SPH4}). So, effectively, dust and gas phases behave as a single fluid with a modified sound speed \citep[see for example][]{MRB}
%
% MISSING in the submitted version. Should be "incorporate" not "incorporates!!!
%
\begin{equation} \hat{c}_{\rm s}=\frac{c_{\rm s}}{\sqrt{1+ \rho_{\rm D}/
\rho_{\rm G}}}. \label{cs} \end{equation}
It is interesting to note that in this limit, the evolution of the system
is analogous to the one-fluid \textit{zeroth order approximation} of \cite{LaiPri2014a}. In this limit, if the gas resolution is set too low, the first term in
the right hand side of equations \ref{vD_SPH4} and \ref{vG_SPH4} will lead to an unphysical energy dissipation. One can easily visualize this phenomena by 
setting up a wave where gas particles are located in the wave antinodes and 
dust particles in the nodes. In this fiducial case, if equations \ref{vD_SPH4} 
and \ref{vG_SPH4} are applied, the resulting barycentric velocity will be zero, 
thus destroying all wave features. It is thus important to have a minimum 
gas resolution in order to guarantee a correct behaviour of the barycentric
term. Additionally, for high dust-to-gas ratios, it will also be important
to have equal gas and dust resolutions. If dust resolution is set too low,
and dust and gas particles possess very different masses, the dust velocity
will dominate in the barycentric term, and the one fluid limit will not be
recovered. For low dust-to-gas ratios, overdissipative effects are 
reduced, since the fraction of momentum transferred between the
phases (and thus the dissipated energy) will be diminished. Thus, it is
possible to obtain the correct strong drag limit with an arbitrarily low number
of dust particles. Overdissipation in the strong coupling limit, and the
behaviour of the method as a function of the dust and gas resolutions will be
tested in section 3.2.

\begin{figure*} \centering \includegraphics[width=0.45\textwidth]{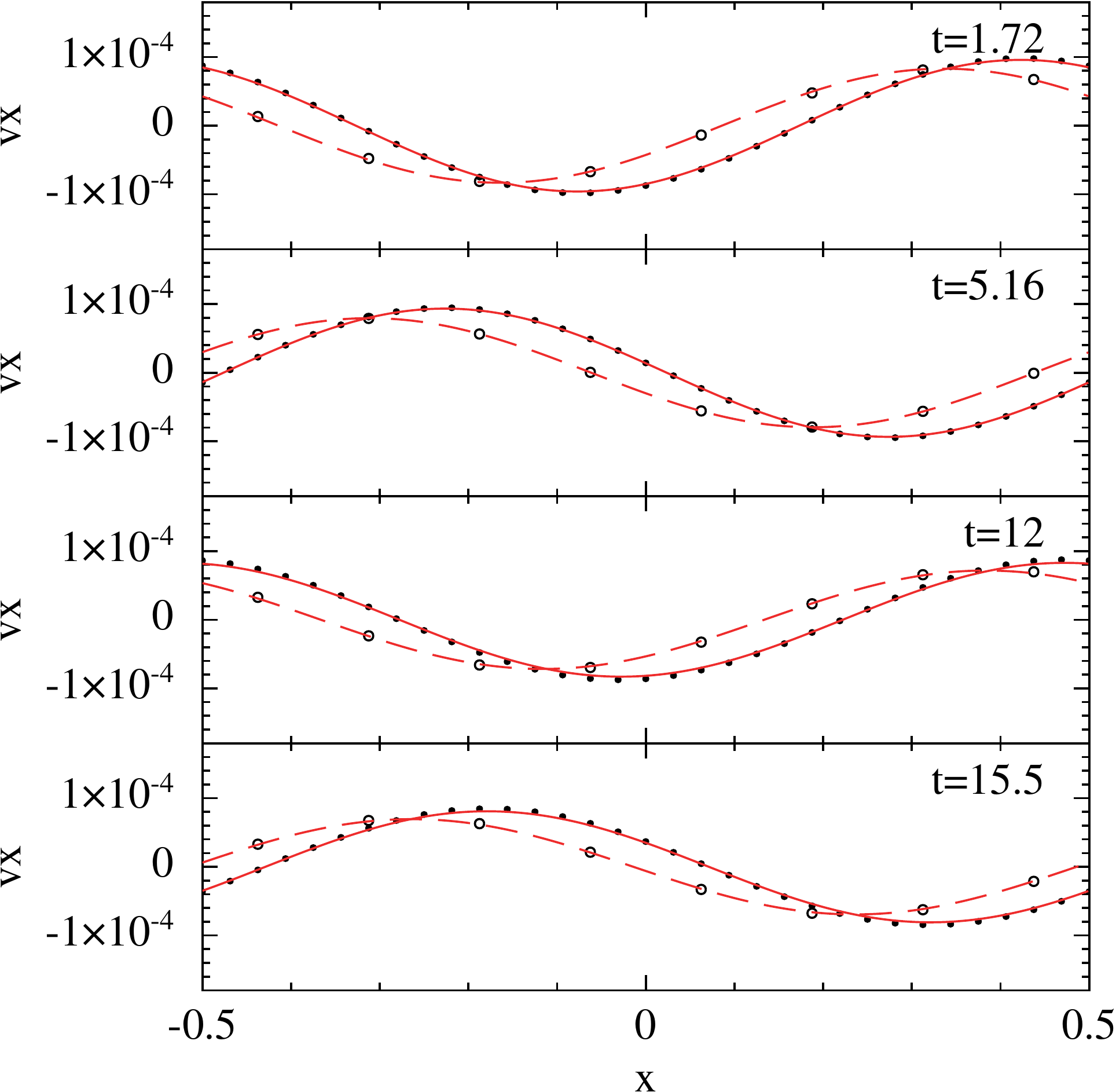}
\hspace{12mm} \includegraphics[width=0.45\textwidth]{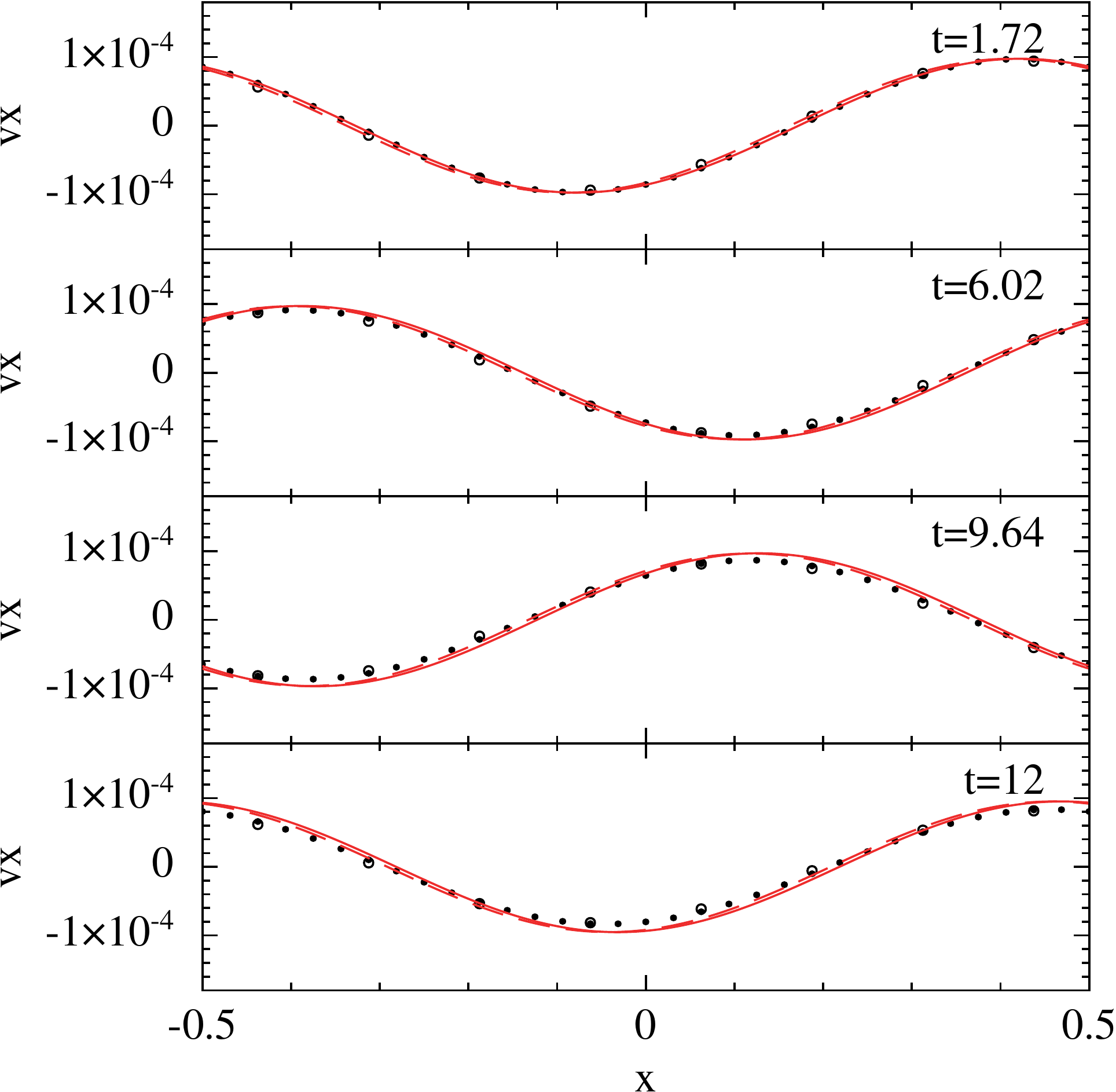} 
\caption{Time evolution of the gas (stars) and dust (open circles) 
components in the \textsc{dustywave} test with $\rho_{\rm D}/\rho_{\rm G}=0.01$. 
Dashed (dust) and solid (gas) lines represent the analytical solutions for
the gas and dust components respectively. Left panels correspond to a low
drag regime with $\delta t/t_s \approx 10^{-2}$ ($K_{\rm const}=0.1$),
while right panels correspond to a high drag regime with $\delta t/t_s 
\approx 10$ ($K_{\rm const}=100$). A total of 32 gas particles and 8 dust
particles have been used in both cases. Because of the relatively low
fraction of momentum being transferred between the dust and gas phases,
overdissipation has a negligible impact on the simulation, even if an arbitrarily
small number of dust particles are used. The error norms for the $K_{\rm
const}=0.1$ case at $t=15.5$ are $L1=2.9\times10^{-2}$, $L2=3.2\times10^{-2}$,
and $L\infty=4.7\times10^{-2}$, while the error norms for the $K_{\rm
const}=100$ case at $t=12$ are  $L1=7.3\times10^{-2}$, $L2=8.4\times10^{-2}$, and
$L\infty=1.25\times10^{-1}$.} \label{fig:test2b} \end{figure*}

It is also interesting to check whether the method can reproduce the properties
of the dust and gas mixture in the so-called \textit{terminal velocity approximation} 
(see for example \cite{LaiPri2014a} and references therein). When dust and gas
are strongly coupled, the dust reaches a constant relative velocity with respect
to the gas, which is small but still finite. Such a relative velocity is proportional
to the pressure gradient and the stopping time $t_{\rm s}$. One can see this by using
\begin{equation} \begin{aligned}
&\mathcal{D}_{\rm t,G} \textbf{v}_{\rm G}(t,\textbf{r}_{\rm G}) = 
\lim_{\delta t \rightarrow 0}
\frac{{\bf v}_{\rm G}(t+\delta t,{\bf r}_{\rm G})-{\bf v}_{\rm G}(t,{\bf r}_{\rm G})}{\delta t} \\
&= \lim_{\delta t \rightarrow 0} \sum_{\rm k}^{Dust}\frac{m_{\rm k}}{\delta t\rho_{\rm k}}
\left[-(1-\xi)\frac{\nabla P_{\rm j}}{\rho_{\rm j}}\delta t+\xi(\textbf{v}_{\rm kj}\cdot\hat{\textbf{r}}_{\rm kj})\hat{\textbf{r}}_{\rm kj}\right]W(r_{\rm kj})
\end{aligned} \end{equation}
where we have introduced
\begin{equation}
\xi \equiv \frac{1-e^{-\delta t/t_{\rm s}}}{1+\rho_{\rm D}{\rho_{\rm G}}}
\label{vlim1}
\end{equation}
to simplify the notation. Now, in order to reach the terminal velocity,
pressure gradient and drag forces must balance each other, leading to
\begin{equation} \begin{aligned}
0&=\sum_{\rm k}^{Dust}\frac{m_{\rm k}}{\delta t\rho_{\rm k}}
\lim_{\delta t \rightarrow 0} \left[-(1-\xi)\frac{\nabla P_{\rm j}}{\rho_{\rm j}}\delta t+\xi(\textbf{v}_{\rm kj}\cdot\hat{\textbf{r}}_{\rm kj})\hat{\textbf{r}}_{\rm kj}
\right]W(r_{\rm kj})\\
&= \sum_{\rm k}^{Dust}\frac{m_{\rm k}}{\rho_{\rm k}}\left[-\frac{\nabla P_{\rm j}}{\rho_{\rm j}}+\frac{1}{t_{\rm s}}(\textbf{v}_{\rm kj}\cdot\hat{\textbf{r}}_{\rm kj})\hat{\textbf{r}}_{\rm kj}
\right]W(r_{\rm kj})\end{aligned} \end{equation}
which is simply the SPH equivalent of

\begin{equation}
\textbf{v}_{\rm DG}(t,\textbf{r}_{\rm G}) = t_{\rm s}\frac{\nabla P_{\rm G}}{\rho_{\rm G}}
\end{equation}

\subsection{Dust evolution in the non-linear regime}

The procedure followed in section 2.1 can be extended to the non-linear drag
regimes as long as an approximate analytic solution can be found for the time
evolution of the dust grains. For example, in a full non-linear regime
(equations \ref{KSt} and \ref{CD3}), a procedure analogous to the one in
section 2.1 can be followed. In such a regime,  the equations of motion 
of the dust and gas components can always be expressed as

\begin{equation} \mathscr{D}_{\rm t,D}\textbf{v}_{\rm D}(t,\textbf{r}_{\rm D}) =
-\frac{K^{\rm St}_{\rm s}}{\hat{m}_{\rm D}}\rho_{\rm G}
|\textbf{v}_{\rm DG}(t,\textbf{r}_{\rm D})|\textbf{v}_{\rm DG}
(t,\textbf{r}_{\rm D}),\label{Eu_St1} \end{equation} 
\begin{equation} \mathcal{D}_{\rm D}\textbf{v}_{\rm G}(t,\textbf{r}_{\rm G}) = 
\frac{K^{\rm St}_{\rm s}}{\hat{m}_{\rm D}}\rho_{\rm G}
|\textbf{v}_{\rm DG}(t,\textbf{r}_{\rm G})|\textbf{v}_{\rm DG}
(t,\textbf{r}_{\rm G}), \label{Eu_St2} \end{equation}
where $K_{\rm s}^{\rm St} \equiv \frac{1}{2}C_{\rm D}\pi
s^2$. In this case, the chosen equations for a pair of arbitrary dust and gas fluid elements located at points $\textbf{r}_{\rm D}$ and $\textbf{r}_{\rm G}$ 
are
\begin{equation} \begin{aligned} \textbf{v}_{\rm D}(t+\delta t,\textbf{r}_{\rm D}) &= \textbf{v}_{\rm D}(t,\textbf{r}_{\rm D}) \\ & - \left(\frac{1-\frac{1}{1+\delta t/t_{\rm s}}}
{1+\rho_{\rm D}/\rho_{\rm G}}\right)\textbf{v}_{\rm DG}
(t,\textbf{r}_{\rm D}), \end{aligned} \label{vD_sol3}\end{equation} 
\begin{equation} \begin{aligned} \textbf{v}_{\rm G}(t+\delta
t,\textbf{r}_{\rm G}) &= \textbf{v}_{\rm G}(t,\textbf{r}_{\rm G})\\
& + \frac{\rho_{\rm D}}{\rho_{\rm G}}
\left(\frac{1-\frac{1}{1+\delta t/t_{\rm s}}}{1+\rho_{\rm D}
/\rho_{\rm G}}\right) \textbf{v}_{\rm DG}(t,\textbf{r}_{\rm G}),
\end{aligned} \label{vG_sol3} \end{equation}
\begin{equation} \begin{aligned} u_{\rm G}(t+\delta
t,\textbf{r}_{\rm G}) &= u_{\rm G}(t,\textbf{r}_{\rm G})\\
& + \frac{1}{2}\frac{\rho_{\rm D}}{\rho_{\rm G}}
\left(\frac{1-\frac{1}{1+2\delta t/t_{\rm s}}}{1+\rho_{\rm D}
/\rho_{\rm G}}\right) \textbf{v}^2_{\rm DG}(t,\textbf{r}_{\rm G}), 
\end{aligned} \label{uG_sol3} \end{equation}
where
\begin{equation} t_{\rm s} \equiv \frac{\hat{m}_{\rm D}}{K_{\rm s}^{\rm St}
\rho_{\rm G}(1+\rho_{\rm D}/\rho_{\rm G})|\textbf{v}_{\rm DG}|}.
\label{ts2}\end{equation}

\begin{figure*} \centering \includegraphics[width=0.45\textwidth]{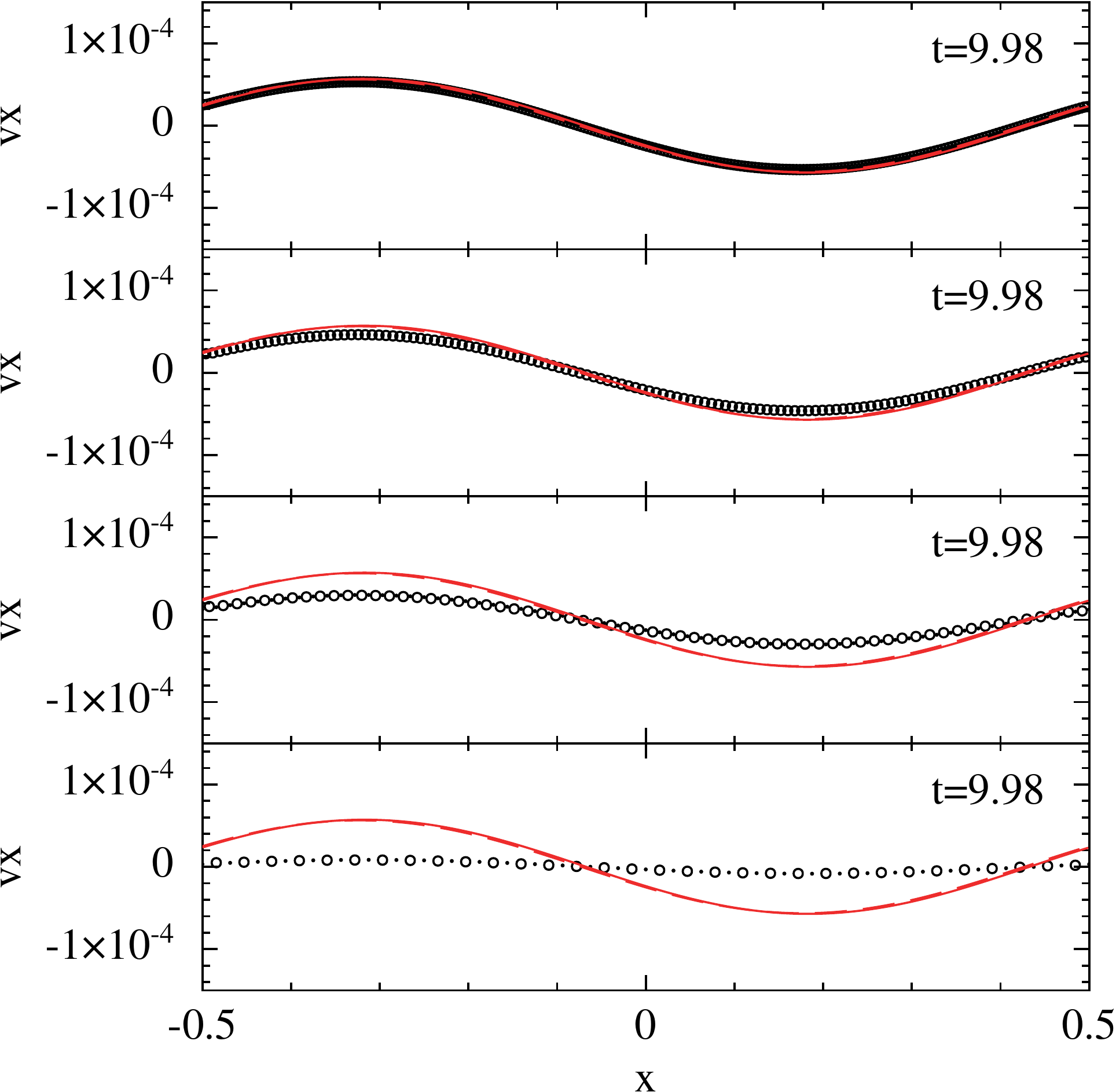}
\hspace{12mm} \includegraphics[width=0.45\textwidth]{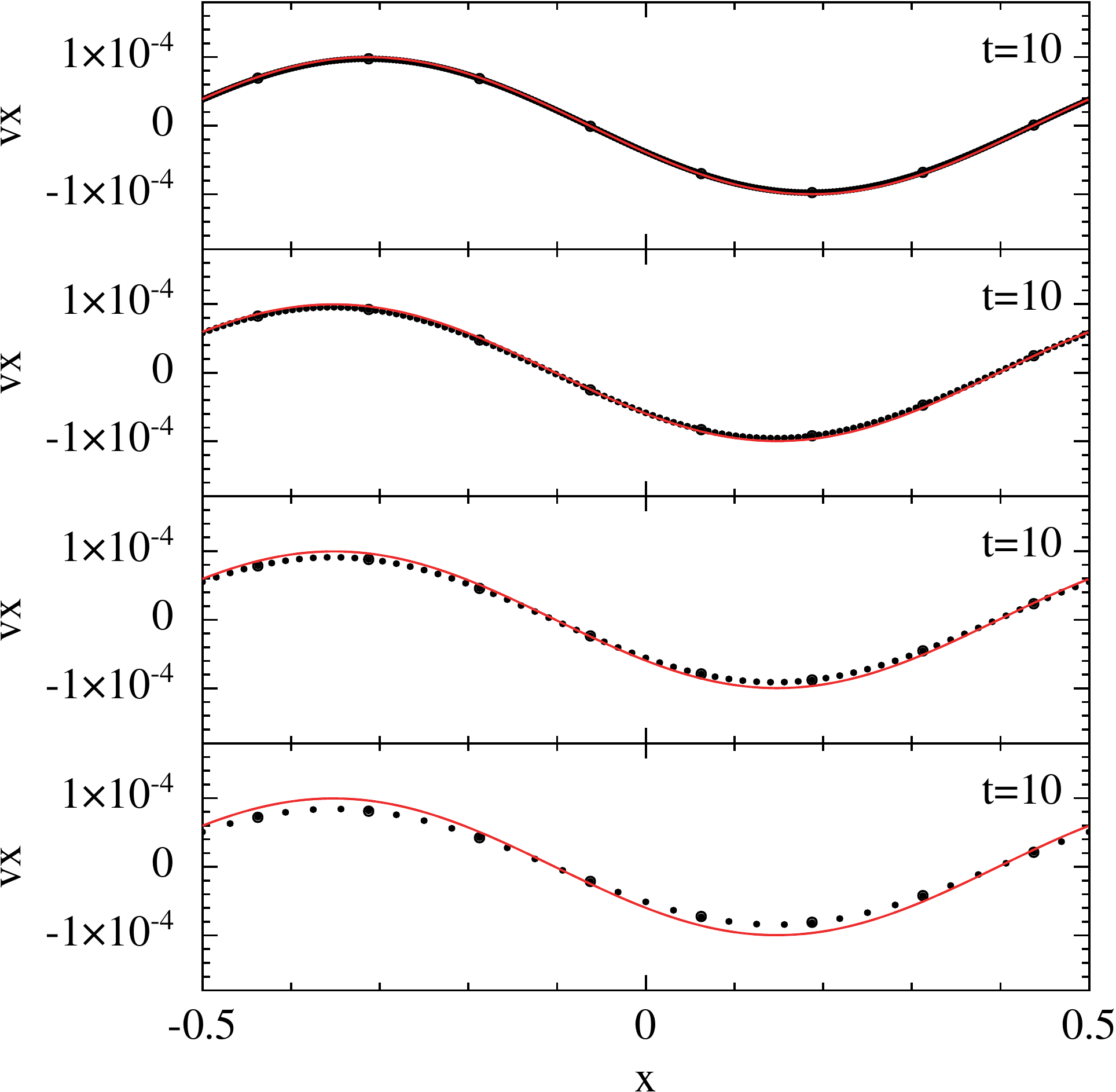}
\caption{Comparison of the gas and dust velocities for several different
resolutions in the \textsc{dustywave} test for a high drag regime ($K_{\rm const}=100$). 
From top to bottom a total of 256, 128, 64, and 32 particles have been used for 
the gas component. Left figure corresponds to a $\rho_{\rm D}/\rho_{\rm G}=1$ 
case with equal numbers of gas and dust particles, while right figure 
corresponds to a $\rho_{\rm D}/\rho_{\rm G}=0.01$, case with only 8 dust 
particles. In complete agreement with Laibe \& Price (2012a,b) an excess of
dissipation is found for low resolutions. However, for low dust-to-gas ratios, 
overdissipation effects becomes much less important even if a very low 
resolution is used.} \label{fig:test2c} \end{figure*}

By using  SPH discretization, equations \ref{vD_sol3}, \ref{vG_sol3} and 
\ref{uG_sol3} become
\begin{equation} \begin{aligned} \textbf{v}^{\rm i}_{\rm D}(t+\delta t,\textbf{r}_{\rm i})
&= \textbf{v}^{\rm i}_{\rm D}(t,\textbf{r}_{\rm i}) \\ & - \nu\sum_{\rm k}^{\rm
gas}\frac{m_{\rm k}}{\rho_{\rm k}}\frac{1-\frac{1}{1+\delta
t/t^{\rm i}_{\rm s}}}{1+\rho_{\rm i}/\rho_{\rm k}}
(\textbf{v}_{\rm ik}\cdot\hat{\textbf{r}}_{\rm
ik})\hat{\textbf{r}}_{\rm ik}W(|\textbf{r}_{\rm ik}|,h_{\rm k}),
\end{aligned} \label{vD_SPH5} \end{equation}
\begin{equation} \begin{aligned} \textbf{v}^{\rm j}_{\rm G}(t+\delta t,\textbf{r}_{\rm j})
&= \textbf{v}^{\rm j}_{\rm G}(t,\textbf{r}_{\rm j}) \\ & + \nu\sum_{\rm k}^{\rm
dust}\frac{m_{\rm k}}{\rho_{\rm j}}\frac{1-\frac{1}{1+\delta
t/t^{\rm k}_{\rm s}}}{1+\rho_{\rm k}/\rho_{\rm
j}}(\textbf{v}_{\rm kj}\cdot\hat{\textbf{r}}_{\rm
kj})\hat{\textbf{r}}_{\rm kj}W(|\textbf{r}_{\rm kj}|,h_{\rm j}),
\end{aligned} \label{vG_SPH5} \end{equation}
\begin{equation} \begin{aligned} u^{\rm j}_{\rm G}(t+\delta t,\textbf{r}_{\rm j}) &=
u^{\rm j}_{\rm G}(t,\textbf{r}_{\rm j}) \\ & + \frac{\nu}{2}\sum_{\rm k}^{\rm dust}\frac{m_{\rm
k}}{\rho_{\rm j}}\frac{1-\frac{1}{1+2\delta t/t^{\rm k}_{\rm
s}}}{1+\rho_{\rm k}/\rho_{\rm j}}(\textbf{v}_{\rm
kj}\cdot\hat{\textbf{r}}_{\rm kj})^2W(|\textbf{r}_{\rm kj}|,h_{\rm j}),
\end{aligned} \label{uG_SPH5} \end{equation}
where in this case, an additional SPH summation is necessary to
calculate $t^{\rm i}_{\rm s}$, since it depends on the relative velocity of the
components at the dust particle location.

\section{Numerical tests}

To perform most of the numerical tests, the dragging algorithm was
implemented in a purpose-built SPH code. The code included
self-consistent $\rho$ and $h$ calculation, grad-h terms \citep{SH, MNb}, and
Riemann solver-like artificial viscosity with thermal conductivity whenever
neeeded \citep{MNa}. To perform the Sedov test, the dragging algorithm was
implemented into a well tested three-dimensional SPH code. For the sake 
of conciseness, the exact details of the SPH code will not be presented
here, but the interested reader is referred to \cite{Aea}.

\begin{figure*} \centering \includegraphics[width=80mm]{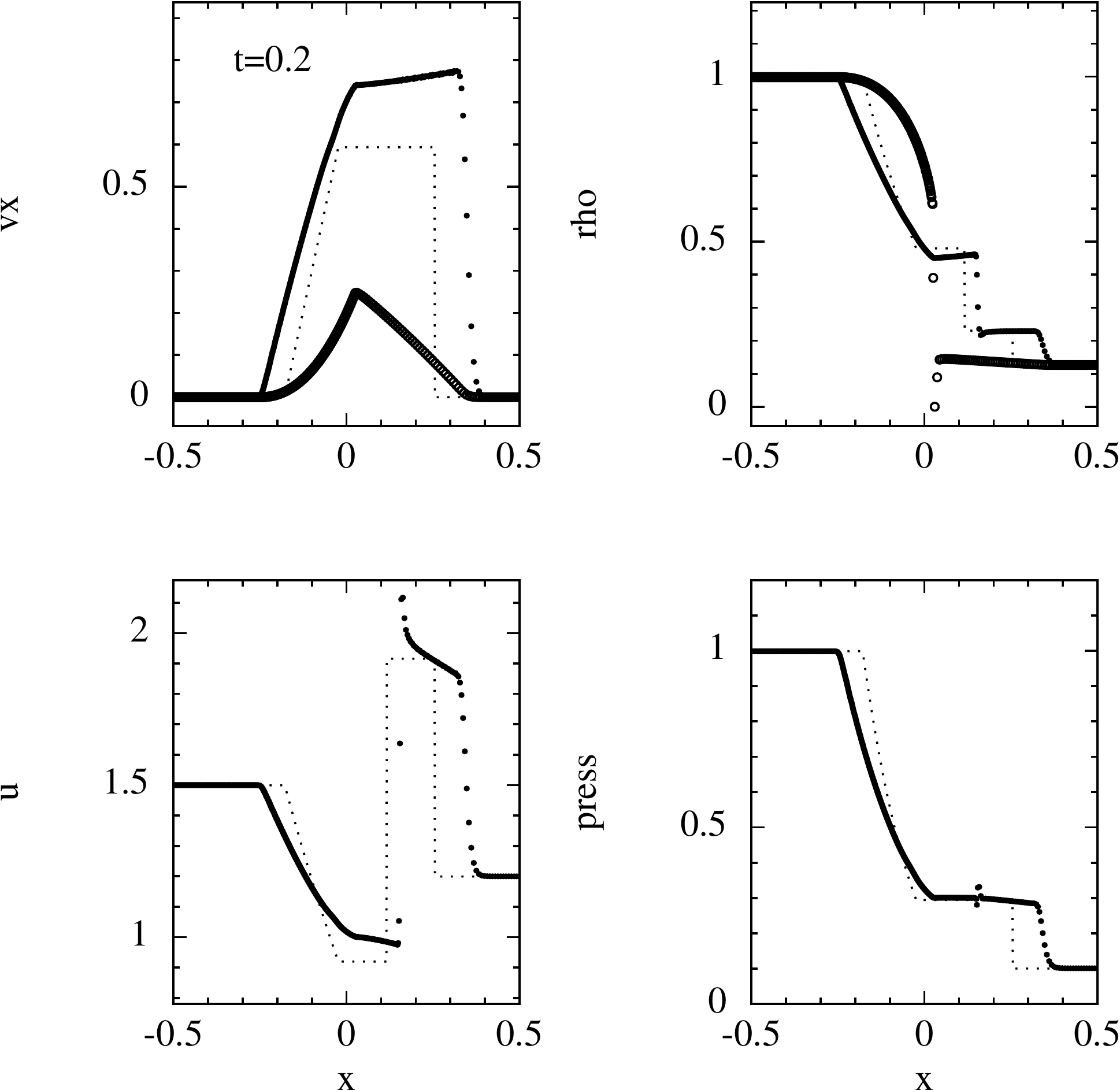} \hspace{12mm}
\includegraphics[width=80mm]{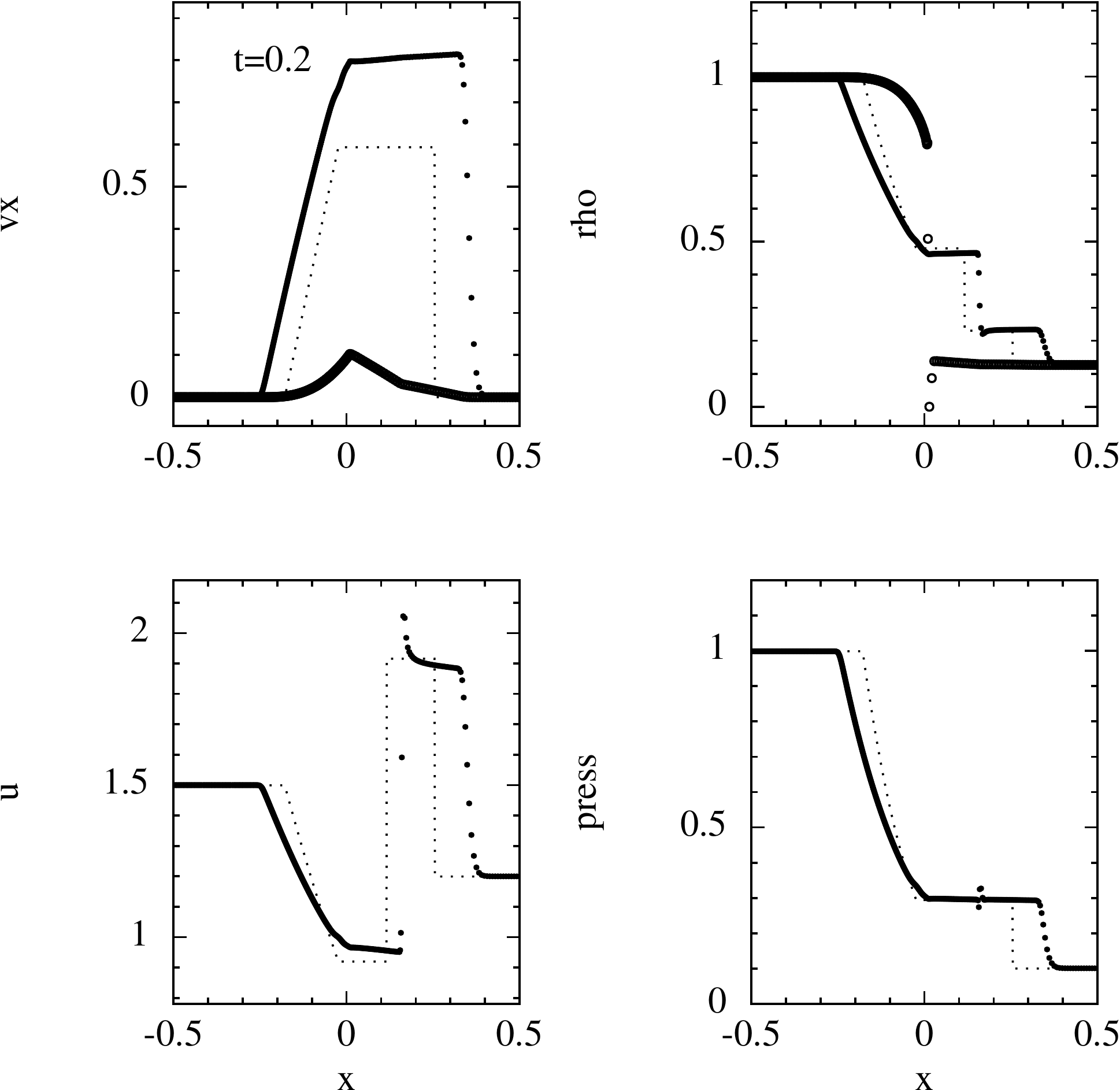} \caption{Results of the gas (stars) and
dust (open circles) components of a shock-tube test with $\rho_{\rm G}/\rho_{\rm D}=1$ 
and 569 particles per phase. The left panels correspond to a
constant drag regime with $K_{\rm s,const}/\hat{m}_{\rm D}=2$, whilst the
right panels correspond to a non-linear regime (equations \ref{KSt} and 
\ref{CD3}) with $K^{\rm St}_{\rm s}/\hat{m}_{\rm D}=2$. Dotted lines correspond 
to the long-term stationary solution of the problem, and have been 
added only as a guide. It has to be stressed out
that no analytical solution exists for the transient case in this problem.}
\label{fig:test3} \end{figure*}

\subsection{DUSTYBOX test in the Epstein regime}

The \textsc{dustybox} test \citep{LPc} was performed in order to prove the capacity of
the method to reproduce the expected asymptotic behaviour of the drag force. 
A set of $20^3$ dust and $20^3$ gas particles with homogeneous densities 
$\rho_{\rm G}$ and $\rho_{\rm D}$ are placed in a periodic box with an 
initial velocity $\textbf{v}_{\rm D}=1$ and $\textbf{v}_{\rm G}=0$. In 
order to construct the initial model, particles are evenly distributed along
a cubic lattice with $-0.5 \le x,y,z \le 0.5$. The dust lattice is shifted, 
with respect to the gas, by half of the gas particles separation in each 
direction. The mass of each SPH particle is equal to
\begin{equation} \centering m = \frac{V\rho}{N}, \end{equation}
where $V$ is the computational domain volume, and $N$ the number of 
particles in each phase. An isothermal equation of state is adopted 
($P=c_{\rm s}^2\rho_{\rm G}$), and in this case no artificial viscosity is
used. The physical units of the problem are chosen such that 
$\rho_{\rm G}=10^{-9}$~g~cm$^{-3}$, $\hat{\rho}_{\rm D}=3$ ~g~cm$^{-3}$,
$v_{\rm therm}=c_{\rm s}=10^5$~cm~s$^{-1}$. These are the appropriate
conditions for a dust particle at the mid-plane of a protoplanetary
disk at 1 AU from the central star (see for example \cite{Ar}).
The computational domain comprises a total volume of 1 cubic AU, 
and the total mass of gas inside the domain is $3.4\times10^{30}$~g.
The integration time-step $\delta t$ is calculated by finding the
minimum value, for all gas particles, of
\begin{equation}
\delta t = \left(\frac{h}{c_{\rm s}}\right),
\label{courant0a} \end{equation}
and
\begin{equation}
\delta t = 0.1\left(\frac{h}{|\textbf{a}|}\right)^{1/2},
\label{courant0b} \end{equation}
where $h$ is the SPH particle smoothing length and $\textbf{a}$
is the gas particle acceleration. Since the pressure gradient is zero,
the exact solution for equation (4) is easy to find in this case
\begin{equation} \begin{aligned} \textbf{v}_{\rm DG}(t) &= \textbf{v}_{\rm
DG}(0)e^{-(t/t_{\rm s})}, \end{aligned} \end{equation}
allowing a direct comparison of the results obtained. In
Fig.~\ref{fig:test1a} the time evolution of the velocity of a single
SPH dust particle is presented for several different values of the dust grain
size $s$. Left figure corresponds to a case with $\rho_{\rm D}/\rho_{\rm G}=1$ 
while right figure corresponds to a case with $\rho_{\rm D}/\rho_{\rm G}=0.01$.
As can be seen, irrespectively of the dust grain size, the correct terminal
velocity between gas and dust components is reached in all cases. Whenever
the gas integration time step becomes smaller than the dust stopping time,
the algorithm is capable of following the velocity decay of the dust 
component towards its limiting velocity. If the dust stopping time becomes
much smaller than the gas integration time step, the algorithm simply
tries to put both components on their barycentric velocity, right from the
start. If under any circumstance, resolving the velocity decay becomes essential,
one can always artificially decrease the gas integration time-step by reducing the
gas Courant time condition (equation \ref{courant0a}) by an arbitrary factor.
In Fig.~\ref{fig:test1b} time evolution of the dust component velocity in
the three-dimensional $\rho_{\rm D}/\rho_{\rm G}=0.01$, $s=1$~mm case is
shown, for different values of the gas integration time step. As can be seen,
since the stopping time is much shorter than the gas integration time-step
($t_{\rm s} \approx 4.5\times10^{-5}$~yrs), an artificially reduced gas integration
time-step is needed to start resolving the dust component velocity decay.

In order to better appreciate the precision of the adopted approximation,
in Fig.~\ref{fig:test1c}, the relative errors in the 
$\rho_{\rm D}/\rho_{\rm G}=0.01$ test with $s=1$~m (left plot) and 
$s=1$~mm (right plot) are presented, for several different kernels. We test the
standard $M_4$ cubic spline kernel \citep{MN}, the $M_6$ quintic spline kernel, 
and the double hump cubic kernel \citep{FQ,LPa}.  As can be
seen, the correct terminal velocity is obtained, irrespectively of the used 
kernel, with very high precision (the relative error between the numerical 
and analytical results is $\lesssim 10^{-4}\%$). The greatest departures from the analytical solution are obtained during the velocity decay phase. In this phase, 
only the double hump kernel keeps errors under acceptable limits. A 
similar result was also found by \cite{LPa} in their study. 
One can also see from the right plot of Fig.~\ref{fig:test1c}, that the normalization
condition (equation \ref{norm}), helps to reduce the errors further.
%If the double hump kernel is used without the normalization condition, 
%an error of $~50 \%$ is obtained in the first integration time step. On 
%the contrary, if the normalization condition is used, the integration 
%error is reduced to a $0.4\%$.
%
% MISSING in the submitted version. Should be ACCURATELY and not PRECISELY!!!
%
If the double hump kernel is used
in conjunction with the normalization condition, the maximum relative
error is $\lesssim 1\%$ in all the tested cases. This is a very important results,
since it shows that the trajectories of dust particles of arbitrary size
can be accurately calculated, in a protoplanetary-like environment, without the
need for excessive resolution.
%
% MISSING in the submitted version. Should be ACCURATELY and not PRECISELY!!!
%

\begin{figure*} \centering \includegraphics[width=80mm]{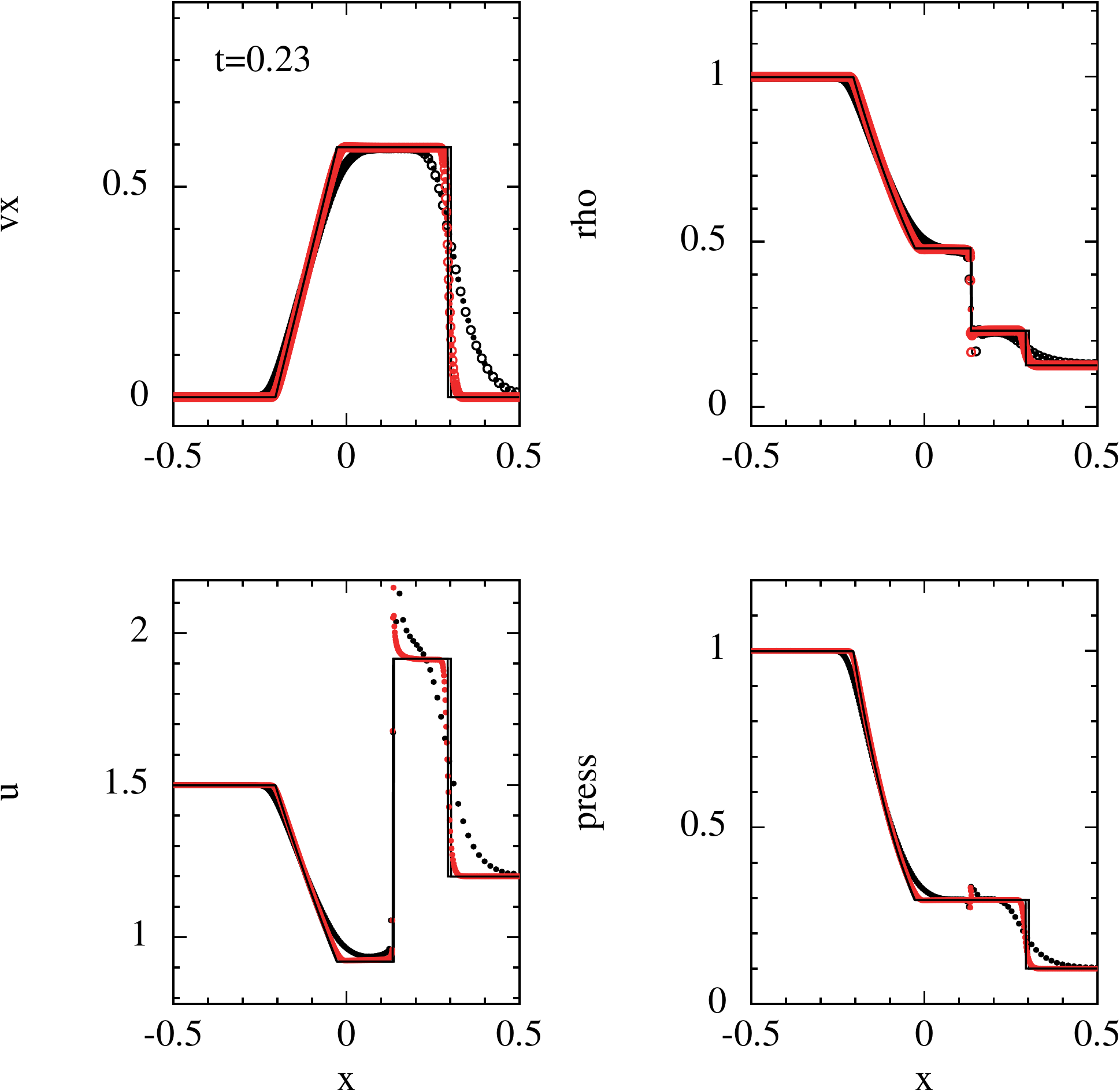} \hspace{1cm}
\includegraphics[width=80mm]{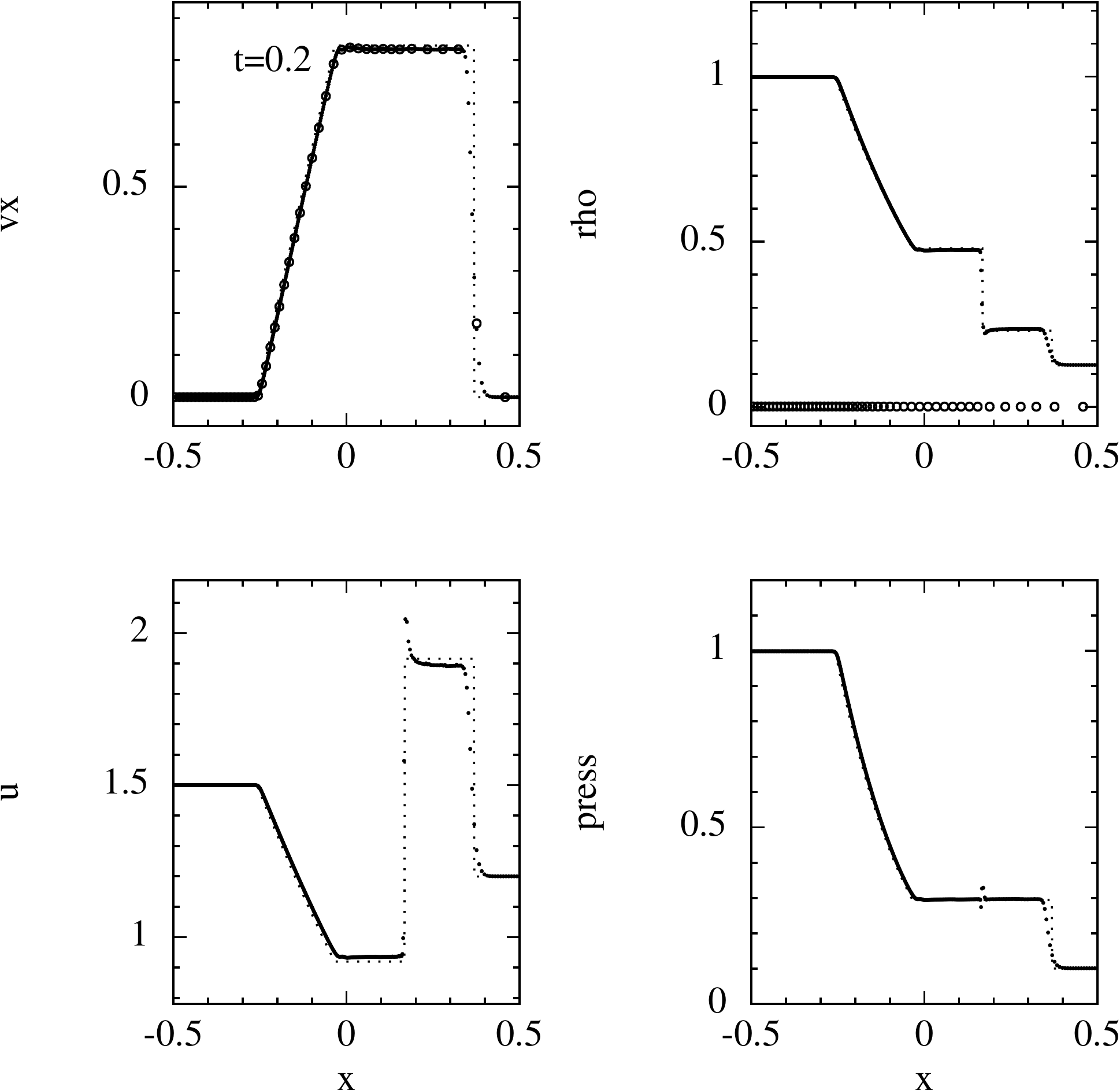} \caption{In the left plot, 
the $\rho_{\rm D}/\rho_{\rm G}=1$ shock-tube results in a high non-linear (equations \ref{KSt} and \ref{CD3})
drag case with $K^{\rm St}_{\rm s}/\hat{m}_{\rm D} = 100$ for two different
resolutions; 256 particles per phase (red) and 2048 particles per phase (black).
Results clearly converge towards the theoretical solution of the problem with the
increase in particle resolution. Error norms for the velocity solution in the high
resolution case are $L1=5.8\times10^{-3}$, $L2=2.5\times10^{-2}$ and 
$L\infty=5.1\times10^{-1}$. In the right plot, a highly dragged case with 
$K^{\rm St}_{\rm s}/\hat{m}_{\rm D}=100$ with $\rho_{\rm D}/\rho_{\rm G}=0.01$ 
is  presented for the non-linear regime. 569 gas and 50 dust particles have been used.
As can be seen, despite the low number of dust particles used, no evidence of 
overdissipation is found.} \label{fig:test3b} \end{figure*}

\subsection{DUSTYWAVE test}

The second test performed was the study of the propagation of a sound wave in a
dust-gas mixture in a constant drag regime, also known as the \textsc{dustywave} 
test \citep{LPc}. This can be done by setting the drag
coefficients on a single grain, in the Epstein regime, to be equal to
\begin{equation} K^{\rm E}_{\rm s} = \frac{K_{\rm s,const}}{\rho_{\rm G}}.
\end{equation}
As a consequence, the equations of motion for the dust and gas
components become
\begin{equation} \mathcal{D}_{\rm t,D}\textbf{v}_{\rm D}(t,\textbf{r}) =
-\frac{K_{\rm s,const}}{\hat{m}_{\rm D}}\textbf{v}_{\rm DG}
(t,\textbf{r}) = -\frac{K_{\rm const}}{\rho_{\rm D}}
\textbf{v}_{\rm DG}(t,\textbf{r}), \end{equation}
\begin{equation} \begin{aligned} \mathcal{D}_{\rm t,G}\textbf{v}_{\rm G}(t,\textbf{r})
& = \frac{K_{\rm s, const}}{\hat{m}_{\rm D}}\frac{\rho_{\rm D}}{\rho_{\rm G}}
\textbf{v}_{\rm DG}(t,\textbf{r}) \\ &=-\frac{K_{\rm const}}{\rho_{\rm
G}}\textbf{v}_{\rm DG}(t,\textbf{r}),\end{aligned} \end{equation}
where we have introduced the drag coefficient per unit volume 
$K_{\rm const} \equiv K_{\rm s,const}\rho_{\rm D}/\hat{m}_{\rm D}$. This is a 
particularly interesting problem, because as in the previous case, it 
possesses an analytic solution \citep{LPc}. To set up the test, an ensemble
of dust and gas particles with homogeneous densities $\rho_{\rm D}$ and
$\rho_{\rm G}$ are evenly distributed over a periodic one-dimensional
domain $-0.5 \le x \le 0.5$. Particle masses are assigned in the same 
way as in the previous section. No artificial viscosity is used in this
case in order to avoid introducing non-physical energy dissipation in the
test. The integration time-step $\delta t$ is again calculated by finding the
minimum value given by equations \ref{courant0a} and \ref{courant0b}. 
An isothermal equation of state 
$P=c^2_{\rm s}\rho_{\rm G}$ with $c_{\rm s}=1$ is used in this case.
In order to create the waves, a sinusoidal perturbation is introduced 
for each particle, both in position and velocity
\begin{equation} x_{\rm p} = x - \delta_x \cos (2\pi x), \end{equation}
\begin{equation} v_{\rm x,p} = -\delta_{v} \sin (2\pi x), \end{equation}
where $x$ is the original position of each particle, and
$\delta_{v}=10^{-4}$, so that the velocity perturbation of the wave
is $\delta v/c_{\rm s}=10^{-4}$. The spatial perturbation $\delta_x$
will be different for every resolution, and is selected in each case
so that the density perturbation of the wave is always
$\delta \rho /\rho = 10^{-4}$. After introducing the perturbation,
the propagation of the resulting sound wave within the domain is 
followed. As previously mentioned, we used the predictor-corrector 
integrator for this test, since it gives better long-term energy and
momentum conservation.

\begin{figure*} \centering \includegraphics[width=80mm]{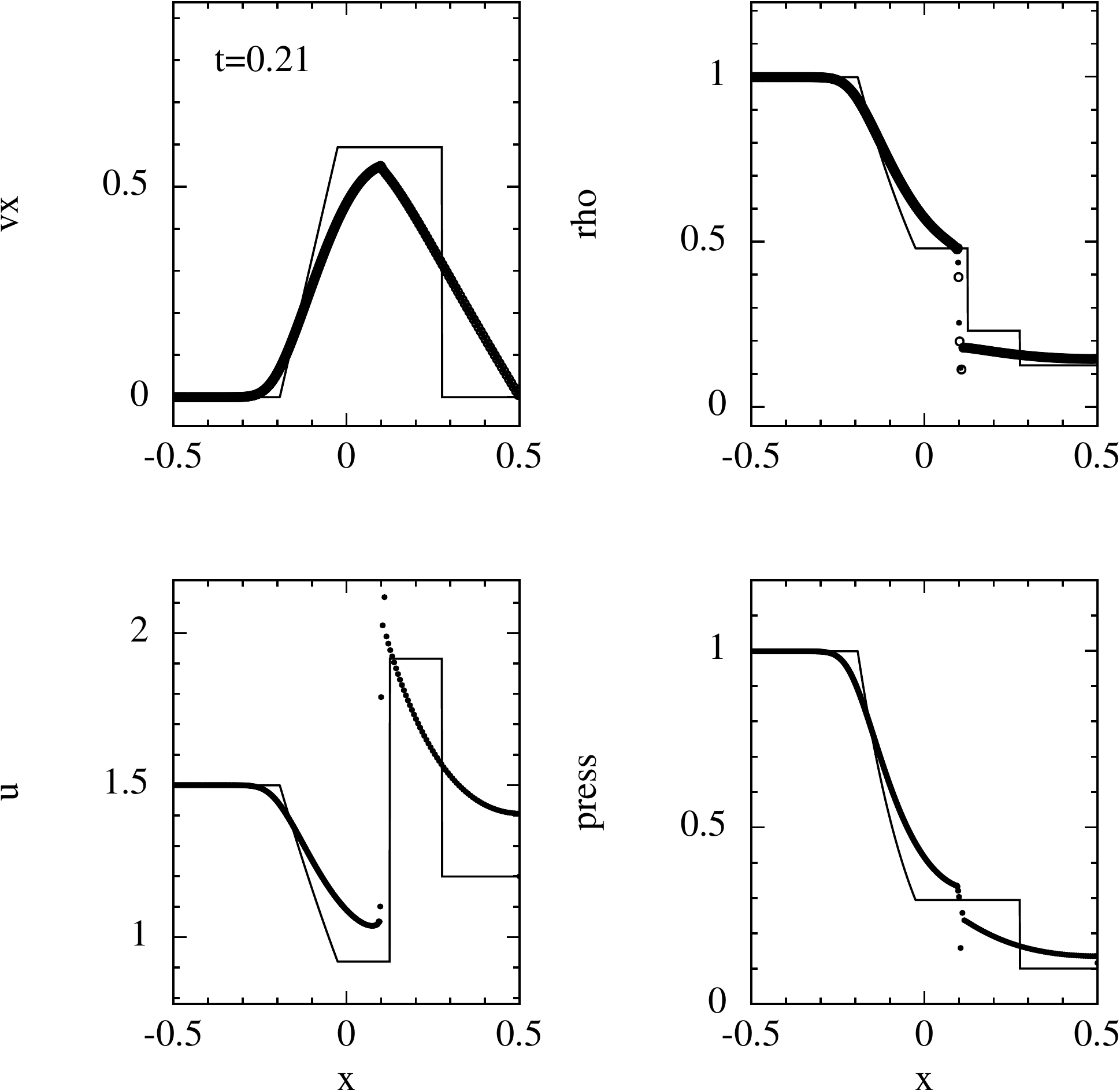}
\hspace{12mm} \includegraphics[width=80mm]{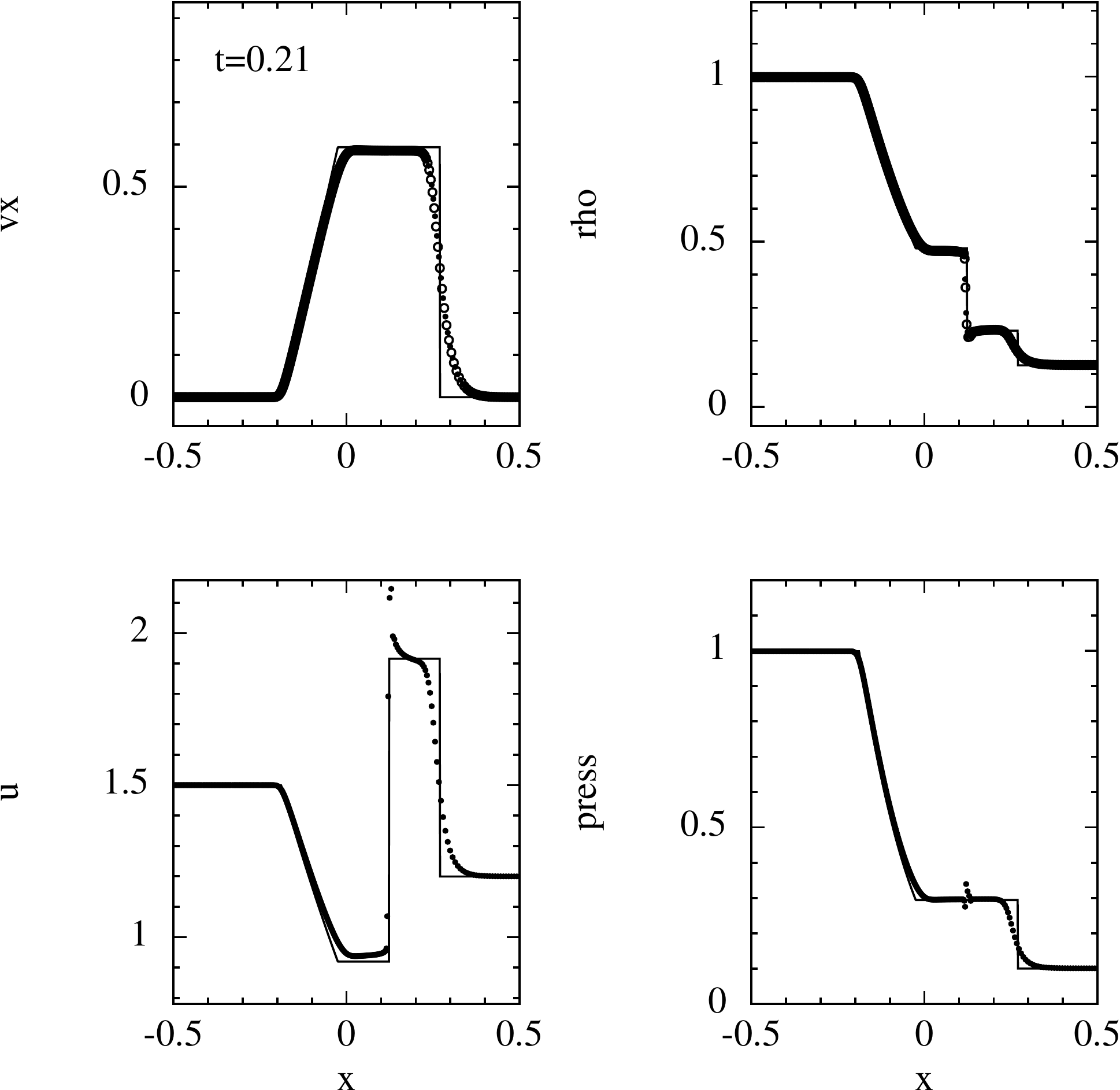}\\
\caption{Comparison of the $\rho_{\rm D}/\rho_{\rm G}=1$ shock tube test result with an 
explicit two-fluid approach (left plot) and our semi-implicit method (right plot) for a very high drag regime
with $K_{\rm s,const}/\hat{m}_{\rm D}=10^6$. The resolution is the same 
in both cases; a total of 569 particles per phase. The semi-implicit method 
is capable of generating a much better solution, even without satisfying the 
resolution criteria $h<c_{\rm s}t_{\rm s}$. The explicit method requires many 
more integration time steps to reach the same moment in time, due to the Courant
condition of the dust. Because of the error committed due to the lack of resolution
at every step, a very high deviation from the analytical solution is found.
The semi-implicit method, on the contrary, being only limited by the gas Courant
condition, largely avoids this problem.} \label{fig:test3c} \end{figure*}

In Fig.~\ref{fig:test2a}, four different snapshots of the time evolution
of the sound wave velocity in a $\rho_{\rm D}=1$, $\rho_{\rm G}=1$ case with
$\delta t/t_{\rm s} \approx 10^{-3}$ ($K_{\rm const}=1$) (left panel), and
$\delta t/t_{\rm s} \approx 10^{-1}$ ($K_{\rm const}=100$) (right panel),
are presented. In Fig.~\ref{fig:test2b}, four different snapshots of the 
time evolution of the sound wave velocity in a $\rho_{\rm D}=0.01$, $\rho_{\rm
G}=1$ case with $\delta t/t_{\rm s} \approx 10^{-2}$ ($K_{\rm const}=0.1$)
(left panel), and $\delta t/t_{\rm s} \approx 10$ ($K_{\rm const}=100$) 
(right panel), are presented. As can be seen, a good agreement with the
analytical solutions has been obtained in both cases. In order to quantify
the deviation from the analytical solution several error norms are calculated
(see figure captions)
\begin{equation}
L_1 = \frac{1}{Nf_{\rm max}}\sum_{\rm i}^{N}(f_{\rm i}-f_{\rm exact}),
\end{equation}
\begin{equation}
L_2 = \left[\frac{1}{N}\left(f^2_{\rm max}\sum_{\rm i}^{N}(f_{\rm i}-f_{\rm exact})^2\right)\right]^{1/2},
\end{equation}
\begin{equation}
L_{\infty} = \frac{1}{f_{\rm max}}{\rm max}_{\rm i}|f_{\rm i}-f_{\rm exact}|,
\end{equation}
where $f_{\rm max}$ is the maximum value of the exact solution in 
the plotted region, $f_{\rm exact}$ is the analytical solution for the $i$-th
point of the plot, and $N$ is the number of plotted points (http://users.monash.edu.au/dprice/$\sim$splash/userguide/). As previously 
mentioned, one of the most important characteristics of dust and gas 
mixtures is that the local sound speed modification as a function of 
the dust/gas fraction (equation \ref{cs}). Since the analytical solutions seen in 
Fig.~\ref{fig:test2a} and ~\ref{fig:test2b} take into account such a 
modification, the test confirms the capacity of the algorithm to 
reproduce this feature of dust/gas mixtures.

The results of Figs.~\ref{fig:test2a} and ~\ref{fig:test2b} also confirm that,
whenever the amount of momentum transferred between the phases is small
compared with the total momentum of the gas, an arbitrarily low number
of dust particles can be used. Both in the lower drag case of 
Fig.~\ref{fig:test2a}, and in Fig.~\ref{fig:test2b} only 8 dust particles
per wavelength are necessary to obtain reasonable results. Unfortunately,
as can be seen in  Fig.~\ref{fig:test2c},  a certain excess of energy
dissipation by drag in the high drag regime  becomes unavoidable. 
This is not a new phenomena and was already found by  \cite{LPa,LPb}
in their simulations. The SPH two fluid scheme needs a  minimum 
resolution ($h<c_{\rm s}t_{\rm s}$) in order to correctly resolve the
small position and velocity differences between the dust and gas phases,
otherwise overdissipation becomes unavoidable.
However, because our method treats only the gas as a fluid, and not
the dust, this resolution criterion must only be satisfied by the gas component,
not the dust.
In Fig.~\ref{fig:test2c},
the effect of particle resolution is investigated in the high drag
regime for two different dust-to-gas ratios. From top to bottom a total
of 256, 128, 64 and 32 gas particles have been used. In the left panel
of  Fig.~\ref{fig:test2c} a $\rho_{\rm D}/\rho_{\rm G}=1$ case with
$K_{\rm const}=100$ is presented. In this case, equal numbers of gas
and dust particles have been used. In the right panel of  
Fig.~\ref{fig:test2c} a $\rho_{\rm D}/\rho_{\rm G}=0.01$ case with 
$K_{\rm const}=100$ is presented. In this case only 8 dust particles
have been used. As can be seen, and in complete agreement with the 
minimum resolution condition, only the ones with a minimum number of
$256$ gas and dust particles is capable of matching the expected 
solution in the $\rho_{\rm D}/\rho_{\rm G}=1$ case. However, in the
$\rho_{\rm D}/\rho_{\rm G}=0.01$ case, overdissipation effects become
much less dramatic, even with a very low gas and dust particle 
resolution. This is important, since most astrophysical applications
have low dust-to-gas ratios.
%
% MISSING in the submitted version. Changes in the paragraph!!!
%
It is also important to note that the present method is
less dissipative than the previous ones, because we need to perform
many fewer integration time-steps, in order to evolve the simulation to
a given time. In the present method, the interpolation error is only committed once
per gas integration time step, in contrast with explicit or implicit 
methods where the error can be committed hundreds or thousands of times
per gas integration time step. In fact, in earlier versions of the present
method an iterative procedure was tried in order to achieve a higher
precision in the final relative velocities between the components, but instead it resulted
in a degree of overdissipation comparable to the one using a standard integration
method.
%
% MISSING in the submitted version. Changes in the paragraph!!!
%
\subsection{Shocks in a dust-gas mixture}

The next two tests are the shock tube test, and the Sedov blast test \citep{SD}.
They were both performed in order to test the behaviour of the scheme, in the
presence of strong density and pressure gradients. Since equation (16) will only
be valid as long as no big changes in the density or the pressure gradient occur
during the integration time-step, these experiments are critical to prove the
usefulness of the method. In these experiments, thermal energy plays an essential
role in the evolution of the system, so this time an adiabatic equation of state
with $P=(\gamma-1)u\rho_{\rm G}$ and $\gamma=5/3$ is used. Also, to correctly
model the shocks, \cite{MNa} artificial viscosity is used in both cases with
coefficients $\alpha=2$ and $\alpha_{\rm u}=1$ for the thermal conduction 
parameter. The signal velocities are $\textbf{v}_{\rm sig}=c_{\rm ij} -
\textbf{v}_{\rm ij}\cdot\hat{\textbf{r}}_{\rm ij}$ and $\textbf{v}_{\rm sig,u}=
|\textbf{v}_{\rm ij}\cdot\hat{\textbf{r}}_{\rm ij}|$ respectively. The time-step
$\delta t$ is calculated in both cases by finding the minimum value for all 
gas particles between
\begin{equation}
\delta t = \frac{h}{|\textbf{v}_{\rm sig}|},
\label{dt_v} \end{equation}
and
\begin{equation}
\delta t = 0.1\left(\frac{h}{|\textbf{a}|}\right)^{0.5},
\label {dt_a} \end{equation}
\begin{figure*} \centering \hspace{-1cm} \includegraphics[width=60mm]{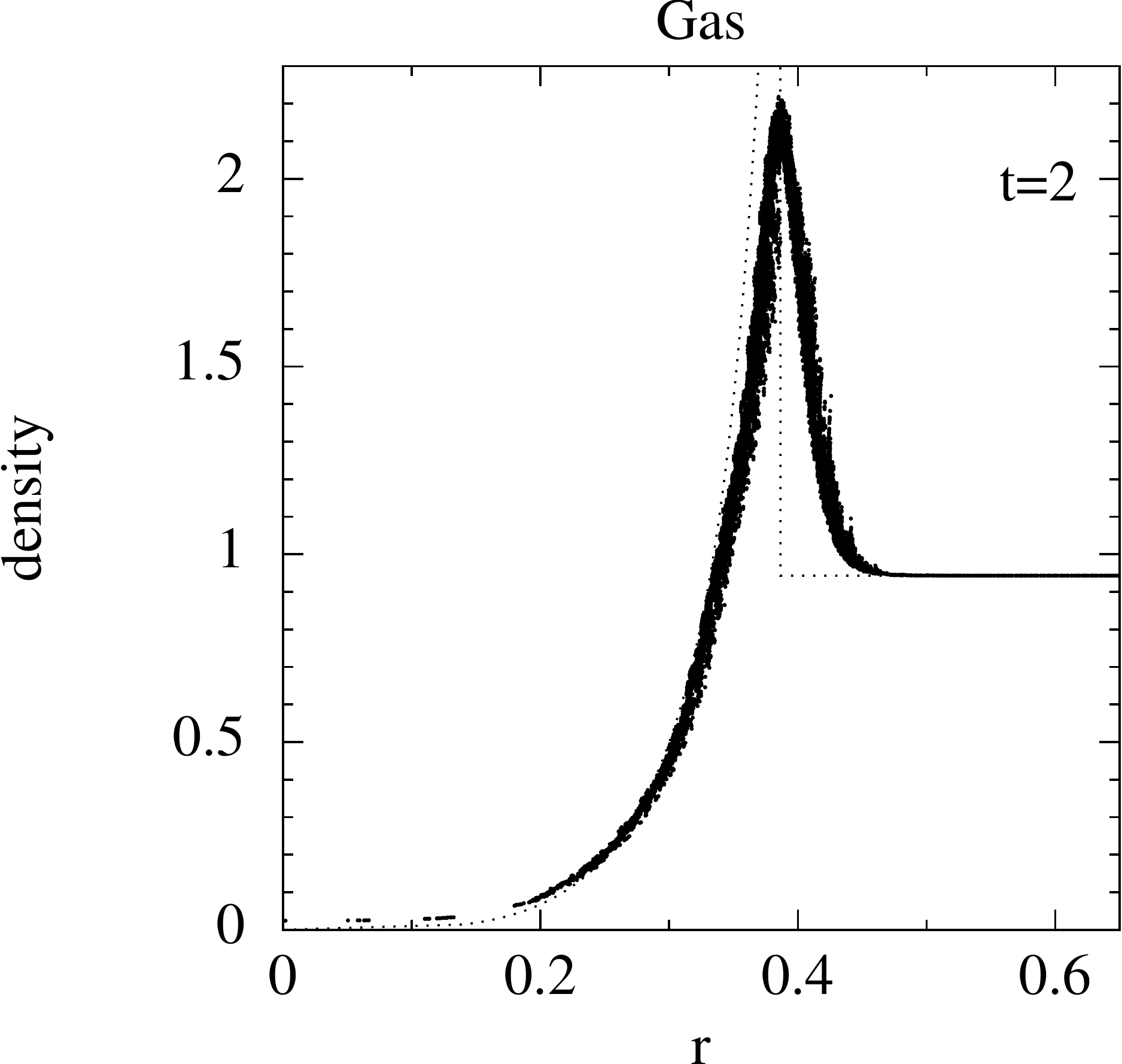}
\hspace{1cm} \includegraphics[width=60mm]{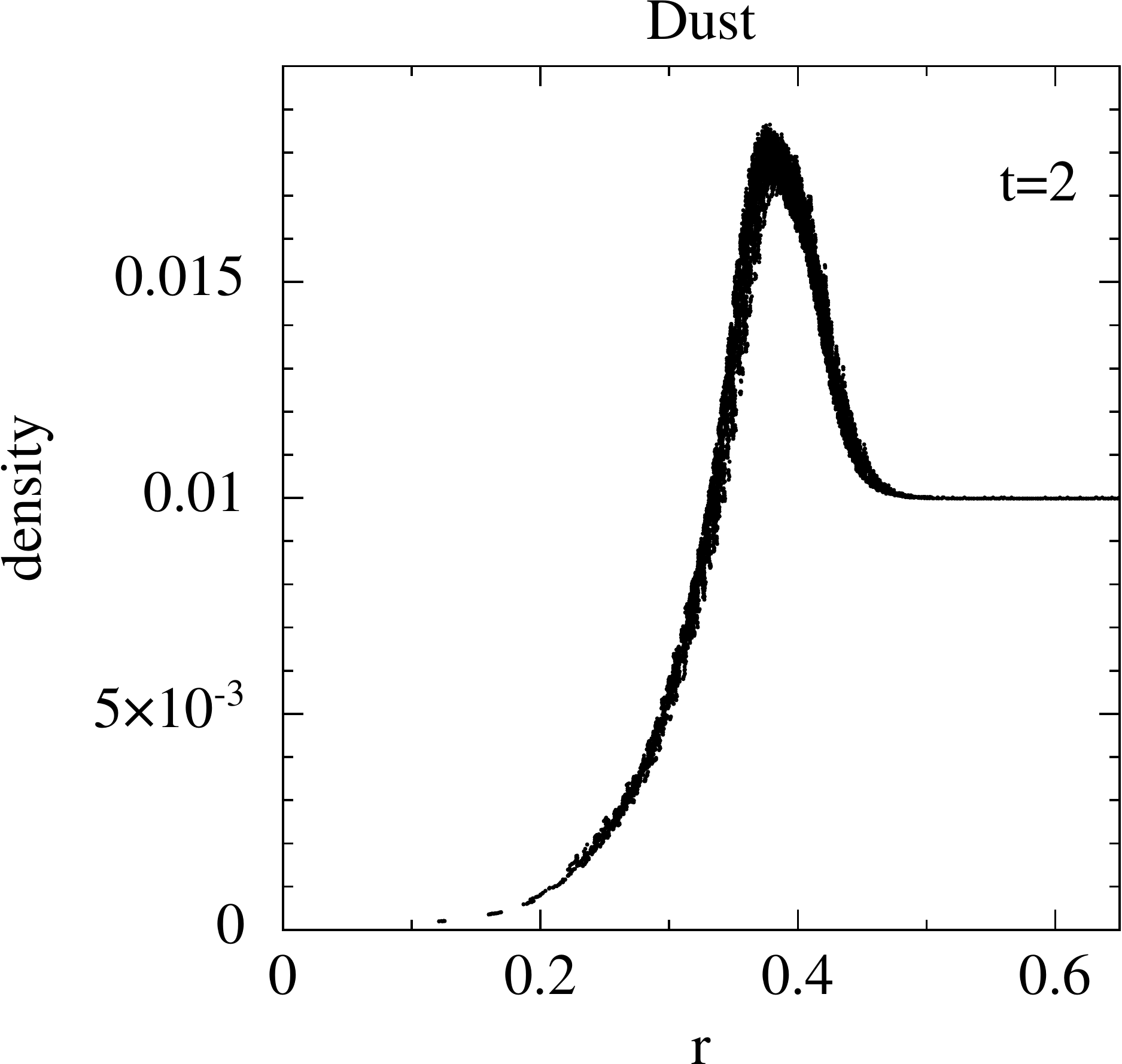}\\ \hspace{-1cm}
\includegraphics[width=60mm]{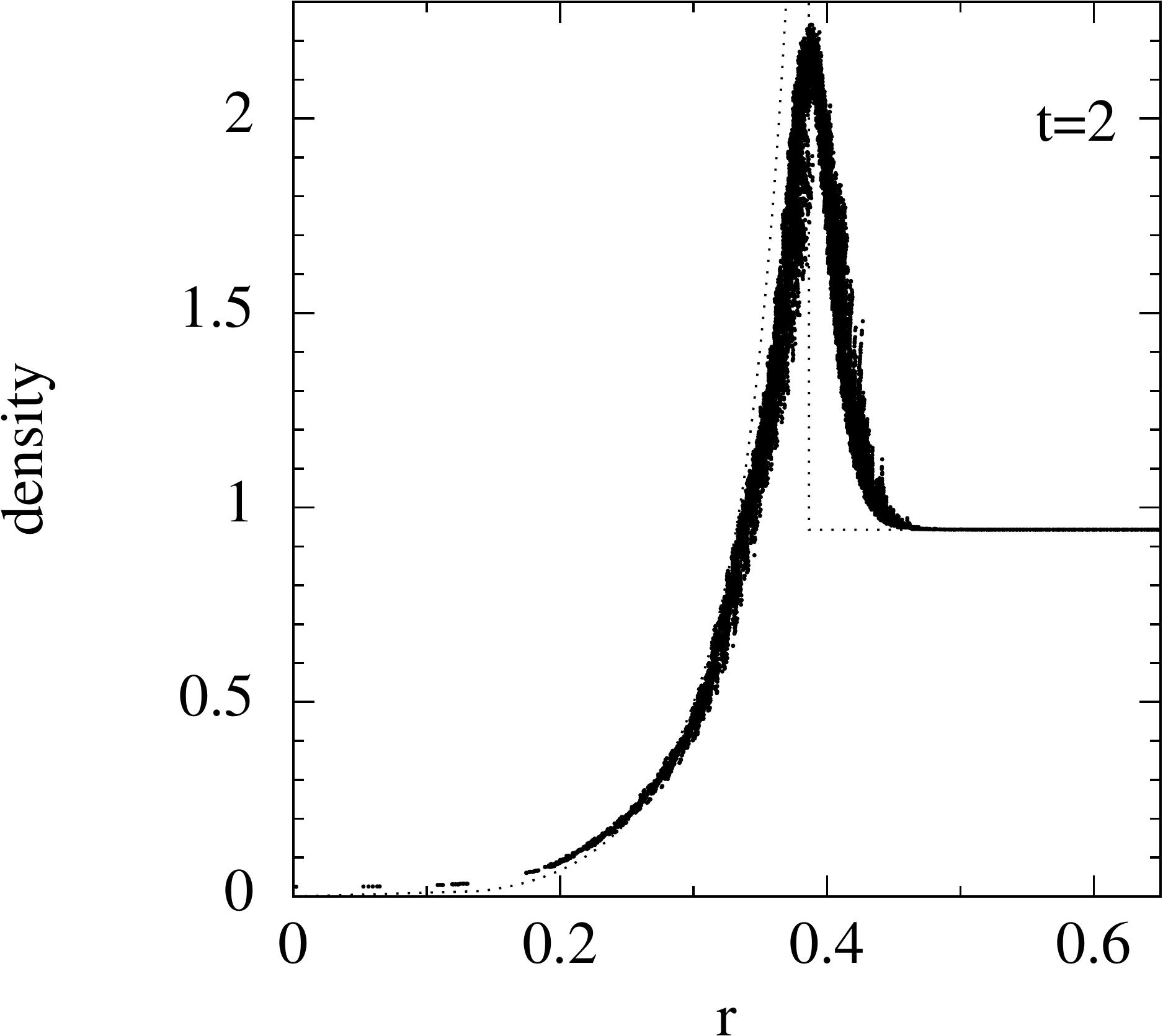} \hspace{1cm}
\includegraphics[width=60mm]{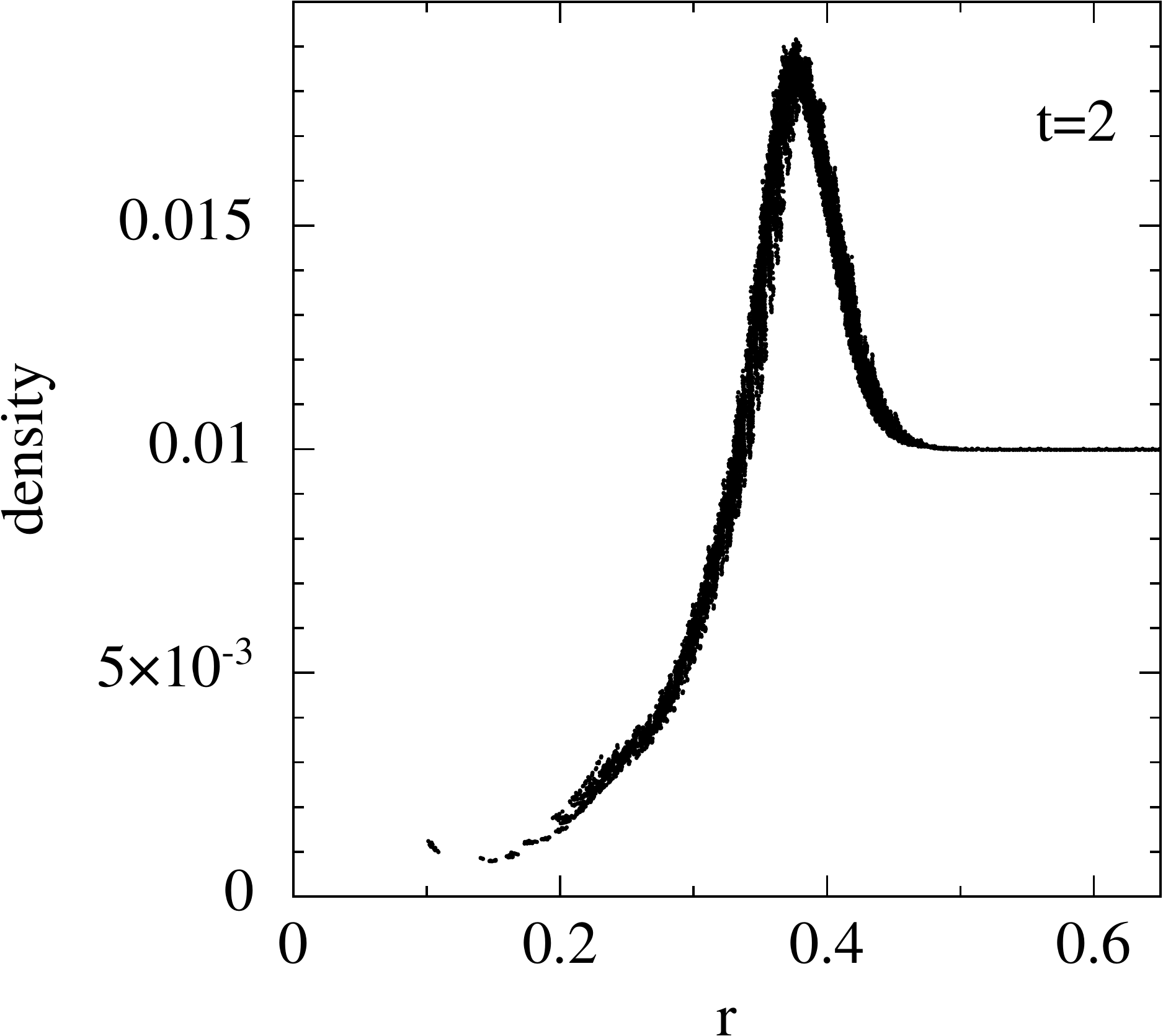} \caption{Particle density as a function
of radius in the Sedov blast test in the $\rho_{\rm G}/\rho_{\rm D}=0.01$,
and $\delta t/t_s \approx 1$ ($K_{\rm s,const}/\hat{m}_{\rm D}=100$) case. The
dotted line corresponds to the self-similar solution of the gas-only Sedov 
problem and has been added as an approximate guide. Uppermost panels correspond
to the solution obtained with the present algorithm and the lowermost panels
correspond to the result obtained using an explicit integrator. Left plots
correspond to the gas component, while right plots correspond to the dust
component.} \label{fig:test4a} \end{figure*}
where $\textbf{a}$ is the SPH gas particle acceleration. Note that if
additional forces affecting both phases (like radiation pressure for example)
were introduced in the simulation, condition \ref{dt_a} should also be 
taken into account for the dust particles. In these tests, the more restrictive
conditions will occur at the shock front. To set up the shock tube test,
an ensemble of particles with $\rho_{\rm L,G}=\rho_{\rm L,D}=1.0$, 
$\rho_{\rm R,G}= \rho_{\rm R,D}=0.125$, $P_{\rm L}=1.0$, $P_{\rm R}=0.1$,
are evenly distributed in a one-dimensional bounded domain $-0.5 \le x \le 0.5$. 
To model the density jump, a different number of particles is used at each
side of the discontinuity. In particular, since it is a one-dimensional case
\begin{equation} \frac{N_{\rm left}}{N_{\rm right}}= \left(\frac{\rho_{\rm
left}}{\rho_{\rm right}}\right)N, \end{equation}
where $N$ is the total number of particles. Particle masses are
calculated as in the previous sections.

Fig.~\ref{fig:test3}, presents the results for two weakly dragged cases in  
a constant drag regime. The left panel represents a case  with 
$K_{\rm s,const}/\hat{m}_{\rm D}=2$, while the right panel represents 
non linear drag regime (equations \ref{KSt} and \ref{CD3})
with $K^{\rm St}_{\rm s}/\hat{m}_{\rm D}=2$. Despite not having an
analytical solution for the transient phase, in both figures, 
the long-term analytical solution of the problem has been added
(dotted line) as a guideline. In both cases, the obtained solution compares very
favourably with the results previously obtained by \cite{LPa,LPb} through the
use of explicit/implicit methods. In the left panel of Fig.~\ref{fig:test3b}, a 
strongly dragged case with $K_{\rm s}^{\rm St}/\hat{m}_{\rm D}=100$ is
presented for two different resolutions in the non-linear regime. In this case,
the analytical solution is known (solid line), and as can be seen, it is well 
matched by the numerical results if enough resolution is used. In the right
panel of Fig.~\ref{fig:test3b} the same case is presented for a 
$\rho_{\rm D}/\rho_{\rm G}=0.01$ case with 569 gas and 50 dust particles.
As can be seen, despite the reduced dust resolution, the correct result is
obtained and there is no evidence of overdissipation.

\begin{figure} \centering \includegraphics[width=80mm]{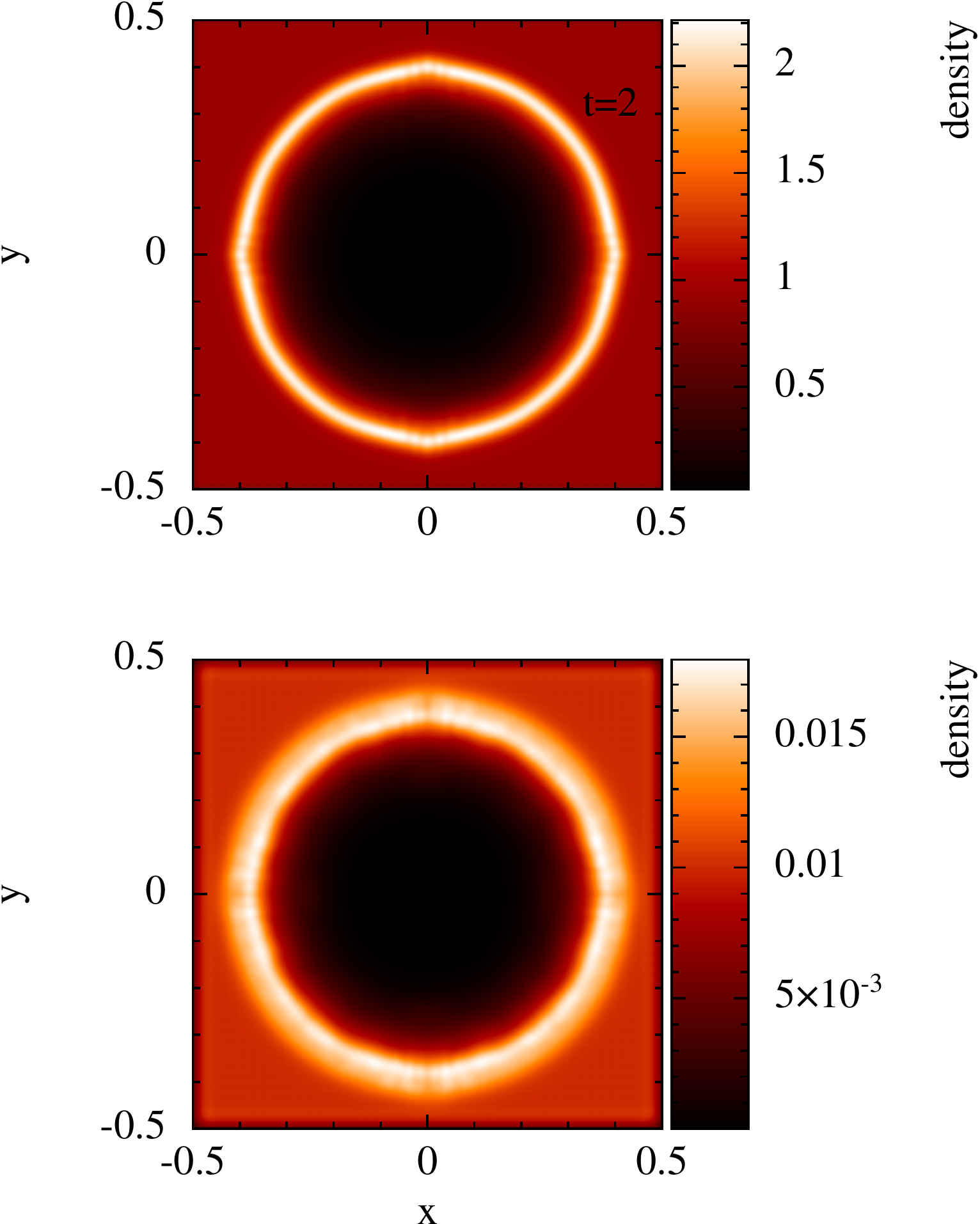} \caption{Cross 
sections of the mid-plane gas (top panel) and dust (lower panel) densities in the 
Sedov blast test for a extremely strong dragged case with 
$K_{\rm s,const}/\hat{m}_{\rm D}=10^{20}$.  Such a calculation would be 
impossible with an explicit time step.} \label{fig:test4b} \end{figure}
%
% MISSING in the submitted version. Should be COMMITTED and not COMMITED!!!
%
It is also very interesting to compare the results obtained using the semi-implicit
method, with those obtained with an explicit integration scheme, for a very high
drag regime with $K_{\rm s, const}/\hat{m}_{\rm D}=10^6$. As can 
be seen in Fig.~\ref{fig:test3c}, whereas excess dissipation in the explicit calculation
gives an incorrect solution, our semi-implicit method avoids the problem. 
As previously mentioned, the 
source of overdissipation in the semi-implicit method comes from the incapacity
of the algorithm to estimate the local barycentric velocity, due to the
lack of resolution. However, and in contrast with an ordinary explicit method,
if the semi-implicit method is used, the error in the barycentric velocity 
estimation is only committed once per gas integration time-step. On the contrary,
if an explicit method is used, due to the Courant condition of the drag interaction,
the error in the estimation of the drag acceleration is committed a lot more times
per gas integration time-step, leading to a very poor result. This effect will also
occur if a conventional implicit integration scheme is used \citep[e.g.][]{LPb}.
%
% MISSING in the submitted version. Should be COMMITTED and not COMMITED!!!
%
Finally, in the uppermost panels of Fig.~\ref{fig:test4a} the result of a Sedov
blast test with $\rho_{\rm D}=0.01$, $\rho_{\rm G}=1$, and $\delta t/t_s \approx
1$ ($K_{\rm s,const}/\hat{m}_{\rm D}=100$) is presented. In this case, the gas 
integration time-step is set by the Courant time condition at the shock front. 
In the test, a total of $2\times 50^3$ particles are evenly distributed in a
three-dimensional Cartesian grid with $-0.5\le x,y,z \le 0.5$. Particle masses
are calculated as in the previous sections. The dust grid is displaced with
respect to the gas one by half of the gas particle separation in each direction.
To model the explosion a total thermal energy of $10^{-3}$ code units is
distributed over the particles inside a certain radius ($r< 2h$). As a
comparison, the Sedov blast test performed with an ordinary explicit integration
scheme, is also presented in the lower panels of Fig.~\ref{fig:test4a}. As can
be seen, there are no significant differences.

\begin{table} \centering \caption{Computational time increase factors as a
function of the drag strength in the Sedov test. The computational time of each
simulation with the explicit method is divided by the computational time of
the semi-implicit method. In the semi-implicit method, since the time-step 
of the simulation is exclusively determined by the gas Courant time condition,
and the dust-to-gas ratio is small, no  noticeable extra computational effort
is needed if the drag strength is increased.}
\begin{tabular}{cc} \hline $K_{\rm s, const}/\hat{m}_{\rm D}$ & Explicit/semi-implicit computational time\\ 
\hline $10^2$ & 10 \\ $10^3$ & 200 \\ $10^4$ & 1500 \\ 
\hline \end{tabular} \end{table}

Again, in this case, no evidence of the resolution limitation has been found,
due to the lower density ratio between the gas and dust components. Since the
typical gas-to-dust ratios in the interstellar medium are very similar to the
ones used in the Sedov test, we expect the method to be useful in realistic
astrophysical simulations. Additionally, we have used this test to compare the
computational time of the method, with that of a traditional explicit
integration. In table 1, a comparison of the computational time for several
different drag strengths, is presented for both cases. In each case, the
computational time spent by each simulation is divided by the computational time
of the semi-implicit method in the $K_{\rm s,const}/\hat{m}_{\rm D}=100$ case.
As can be seen, as the drag strength is increased, the explicit integrator
computational time is increased, by several orders of magnitude, with respect 
to the computational time spent by semi-implicit method for 
$K_{\rm s,const}/\hat{m}_{\rm D}=100$. On the contrary, the computational 
time of the semi-implicit method remains stable, since the integration time-step
is exclusively determined by the gas Courant condition, and is independent 
of the drag strength. As can be seen in Fig.~\ref{fig:test4b}, arbitrarily
large values for the drag coefficient can be used. This value would be 
completely prohibitive in any two-fluid explicit integration method. 

\subsection{Dust settling in a gaseous disk in the Epstein regime.}

\begin{figure} \centering \includegraphics[width=80mm]{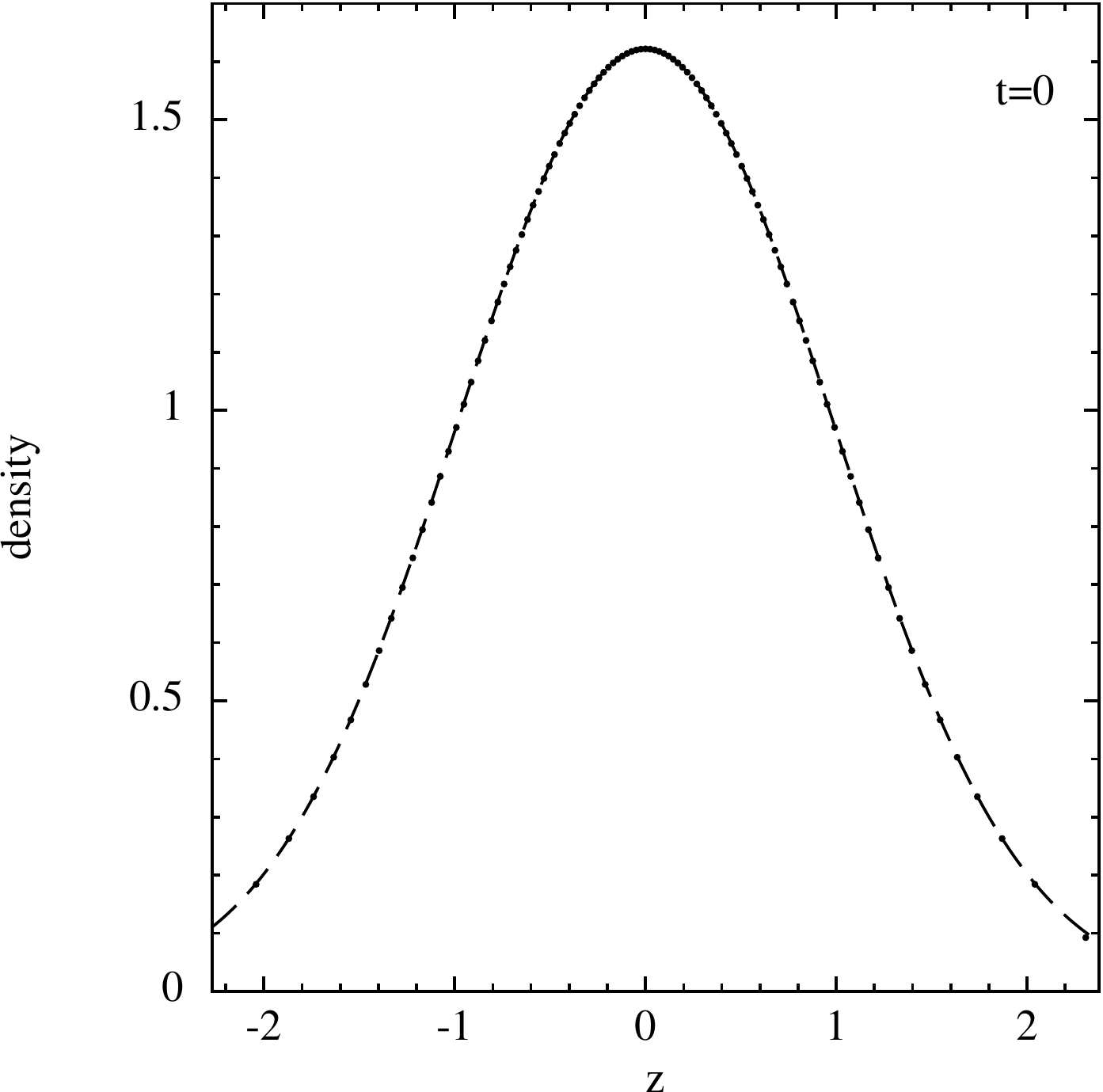}
\caption{Initial gas density profile of the relaxed disk as a function of z. A
total of 100 gas particles have been used to model the vertical disk profile.
Dots correspond to the gas particles whereas the dashed line corresponds to a
Gaussian profile, as predicted by equation \ref{Gauss}.} \label{fig:test5a}
\end{figure}

The final test performed was the mid-plane settling of dust particles in a
one-dimensional vertical section of an isothermal disk with 
$P=c^2_{\rm s}\rho_{\rm G}$ and $c_{\rm s}=1$. To set up the test, 
100 gas and 100 dust particles, with $\rho_{\rm D}=0.01$ and 
$\rho_{\rm G} = 1$, are evenly distributed over a one-dimensional domain
($-2<z<2$). Particle masses are assigned following the same procedure as
in the previous sections. 
%
% MISSING in the submitted version. Changes in the paragraph!!!
%
An external acceleration 
$a_{\rm ext,D} = a_{\rm ext,G} = -\Omega^2 z$ is used to simulate the vertical
component of the gravitational field from the star at the centre of the disk,
where $\Omega$ is the angular frequency.
%
% MISSING in the submitted version. Changes in the paragraph!!!
%
(see Appendices A and B for a detailed explanation about how to implement
external forces in the integration scheme). No boundaries have been used,
and since one does not expect shocks to be important, the use of artificial
viscosity is avoided. The evolution equations of the system  are given by
\begin{equation} \mathcal{D}_{\rm t,D}v_{\rm D} = - \frac{K^{\rm E}_{\rm s}}{\hat{m}_{\rm D}} \rho_{\rm G}\left(v_{\rm D}-v_{\rm G}\right) - \Omega^2 z, 
\label{Eu_disk1}\end{equation}
\begin{equation} \mathcal{D}_{\rm t,G}v_{\rm G} = \frac{K_{\rm s}^{\rm E}}{\hat{m}_{\rm
D}} \rho_{\rm D}\left(v_{\rm D}-v_{\rm G}\right) - \Omega^2 z - \frac{1}{\rho_{\rm G}}
\left(\frac{\partial P}{\partial z}\right). \label{Eu_disk2} \end{equation}
In order for the system to relax, the gas particles are evolved under
gravitational and pressure forces, until the hydrostatic equilibrium condition
is attained. Whenever hydrostatic equilibrium is reached, equations \ref{Eu_disk1} and
\ref{Eu_disk2} can be solved
\begin{equation} 0 = -K_{\rm s}^{\rm E}\left(\frac{\rho_{\rm G}}{\hat{m}_{\rm
D}}\right)\left(v_{\rm D}-v_{\rm G}\right) - \Omega^2z,
\label{Eq1}\end{equation}
\begin{equation}
0 = K_{\rm s}^{\rm E}\left(v_{\rm D}-v_{\rm G}\right) - \Omega^2z - \frac{c_{\rm
s}^2}{\rho_{\rm G}}\left(\frac{\partial \rho_{\rm G}}{\partial z}\right),
\label{Eq2} \end{equation}
giving the gas hydrostatic density profile:
\begin{equation} \begin{aligned} \rho_{\rm G} (z) &\approx \rho_{\rm
G}(0)e^{-\Omega^2 z^2/(2c^2_{\rm s})}, \end{aligned} \label{Gauss} \end{equation}
which is valid as long as $\rho_{\rm G} \gg \rho_{\rm D}$. In
Fig.~\ref{fig:test5a}, the initial gas density profile of the isothermal 
disk is presented. As can be seen, the gas perfectly reproduces a Gaussian
density profile with $\rho_{\rm G}(0) = 1.622$, $c_{\rm s} = 0.98$ and 
$\Omega=1$. After gas relaxation, drag forces are switched on, and evolution is started
again. If the drag coefficient $K_{\rm s}^{\rm E}/\hat{m}_{\rm D}$ is high
enough, dust particles reach a limiting velocity, given by the solution
of equations \ref{Eq1} and \ref{Eq2}.
\begin{equation} v_{\rm D} (z) - v_{\rm G} (z) = -\left(\frac{\Omega^2\hat{m}_{\rm D}}
{K_{\rm s}^{\rm E}\rho_{\rm G}(0)}\right)ze^{\Omega^2z^2/(2c_{\rm s}^2)}. 
\label{v_lim} \end{equation}

\begin{figure*} \centering \includegraphics[width=80mm]{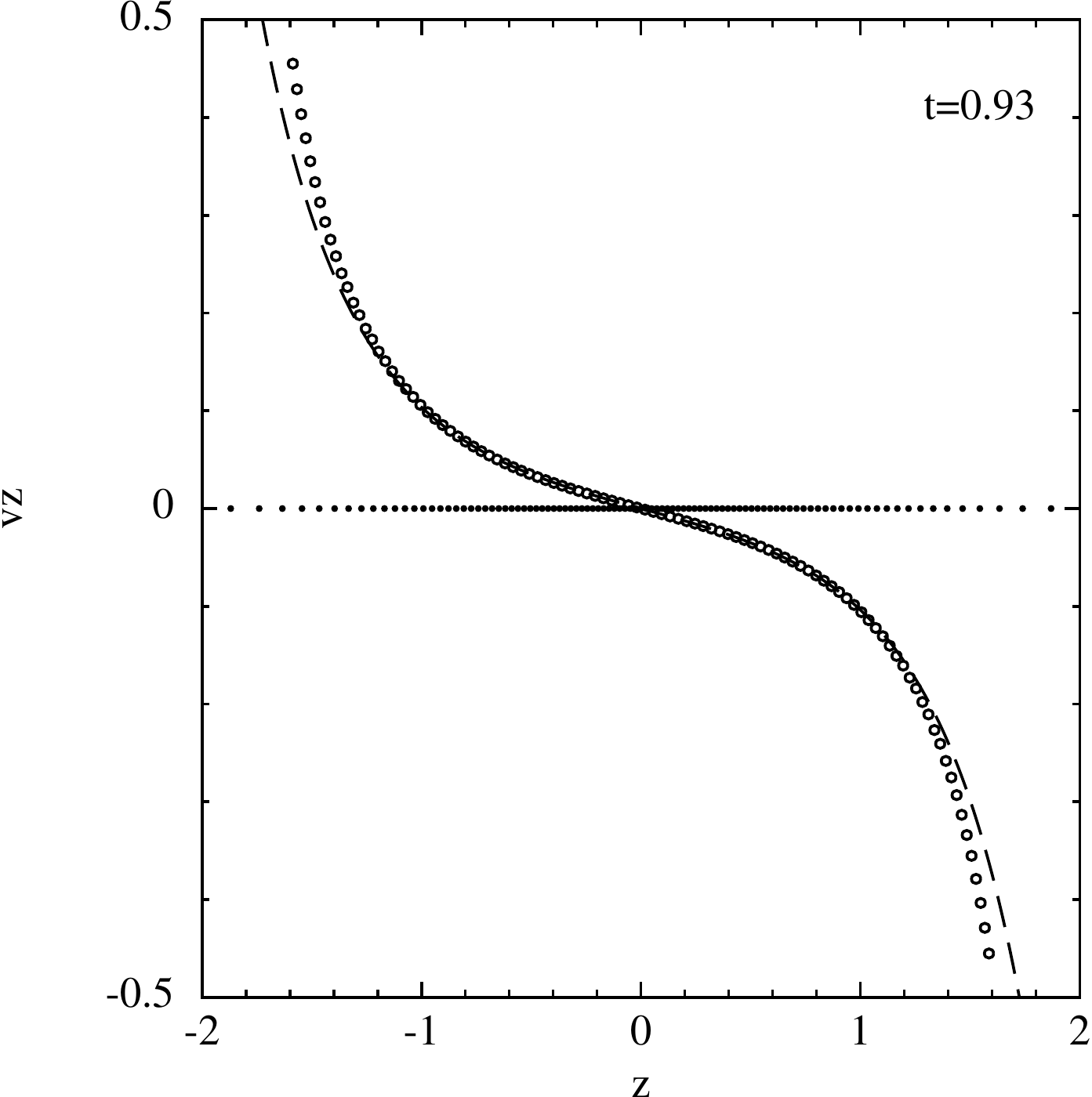}
\includegraphics[width=80mm]{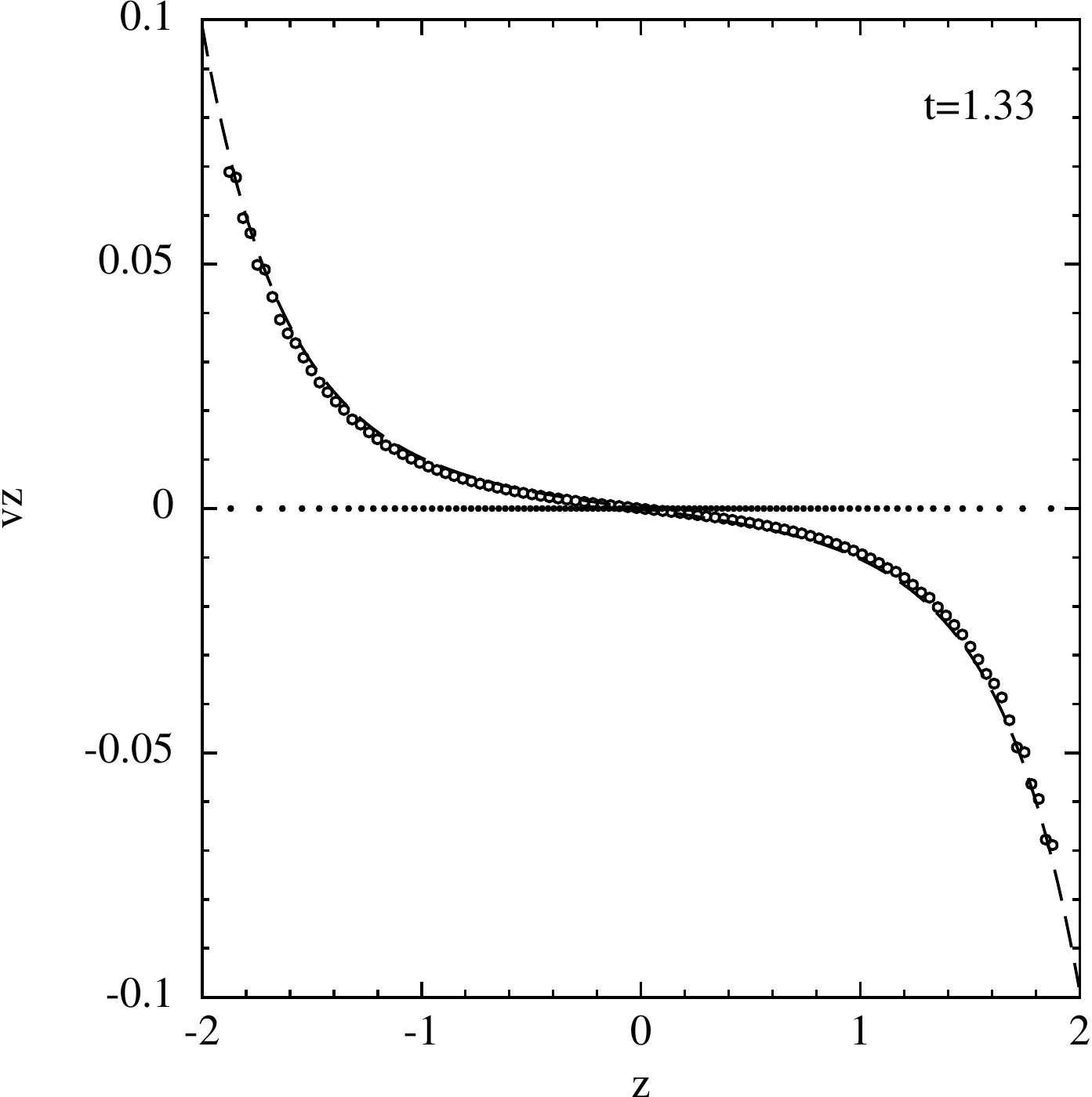} \caption{Velocity of the dust component
as a function of $z$ in the dust settling test. Dots correspond to the gas particles, while open circles
correspond to the dust particles. The left plot corresponds to a $K^{\rm E}_{\rm s}=0.1$ case,
whereas the right plot corresponds to a $K^{\rm E}_{\rm s}=1.0$ case. The dashed line
corresponds to the stationary solution of the problem in each case, as shown
in equation \ref{v_lim}. In the $K^{\rm E}_{\rm s}=0.1$ case, due to the weakness 
of the drag force, the outermost dust particles ($|z| > 1.2$) are still
in the transient state.} \label{fig:test5b} \end{figure*}

\begin{figure*} \centering \includegraphics[width=80mm]{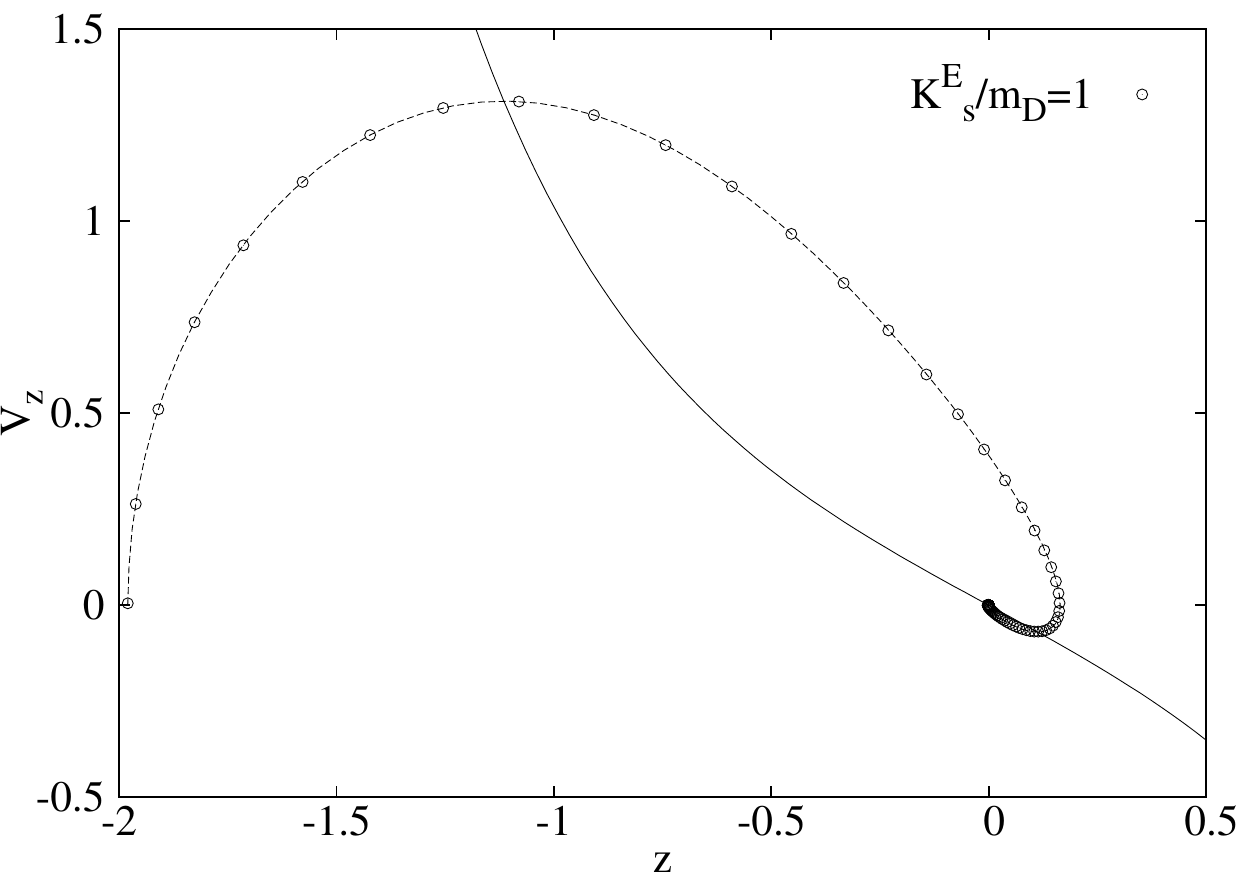}
\includegraphics[width=80mm]{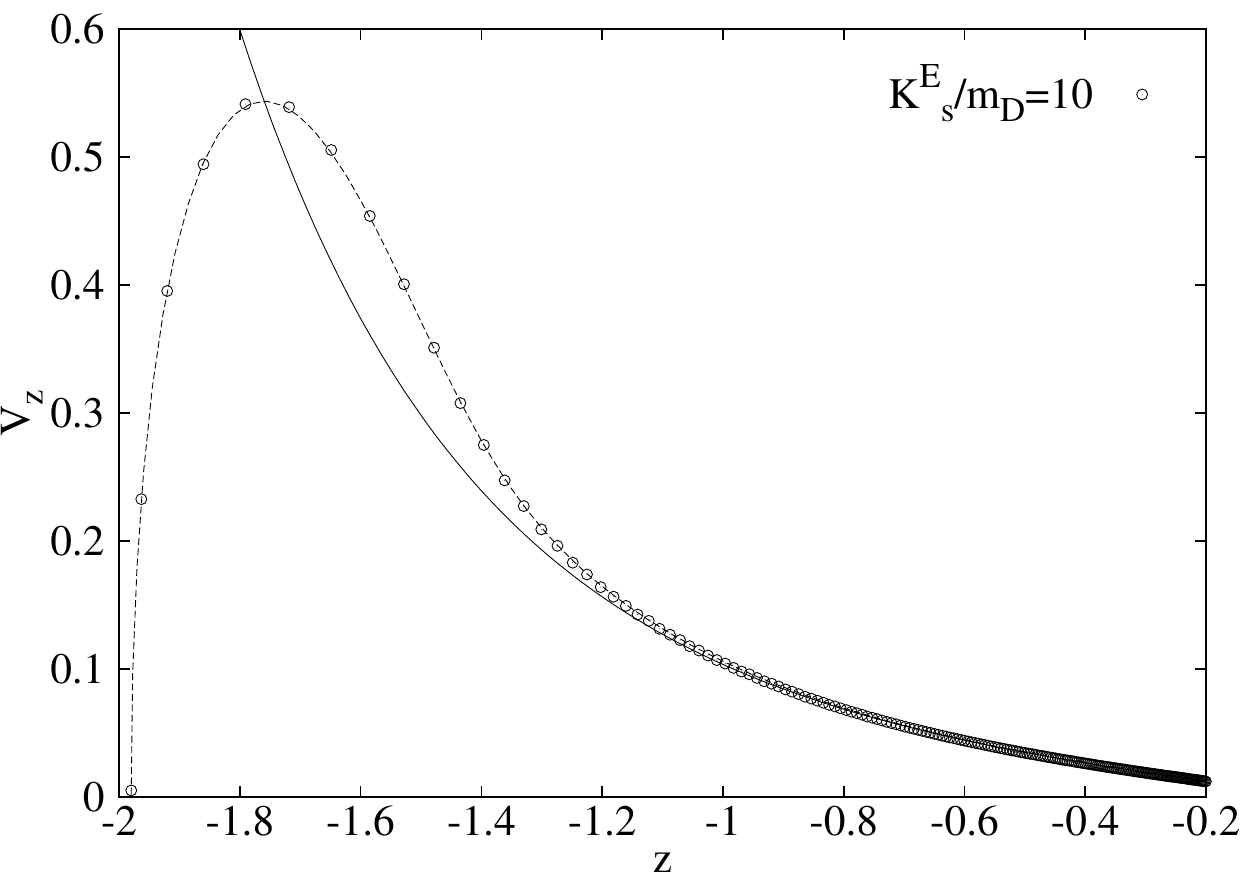}\\
\includegraphics[width=80mm]{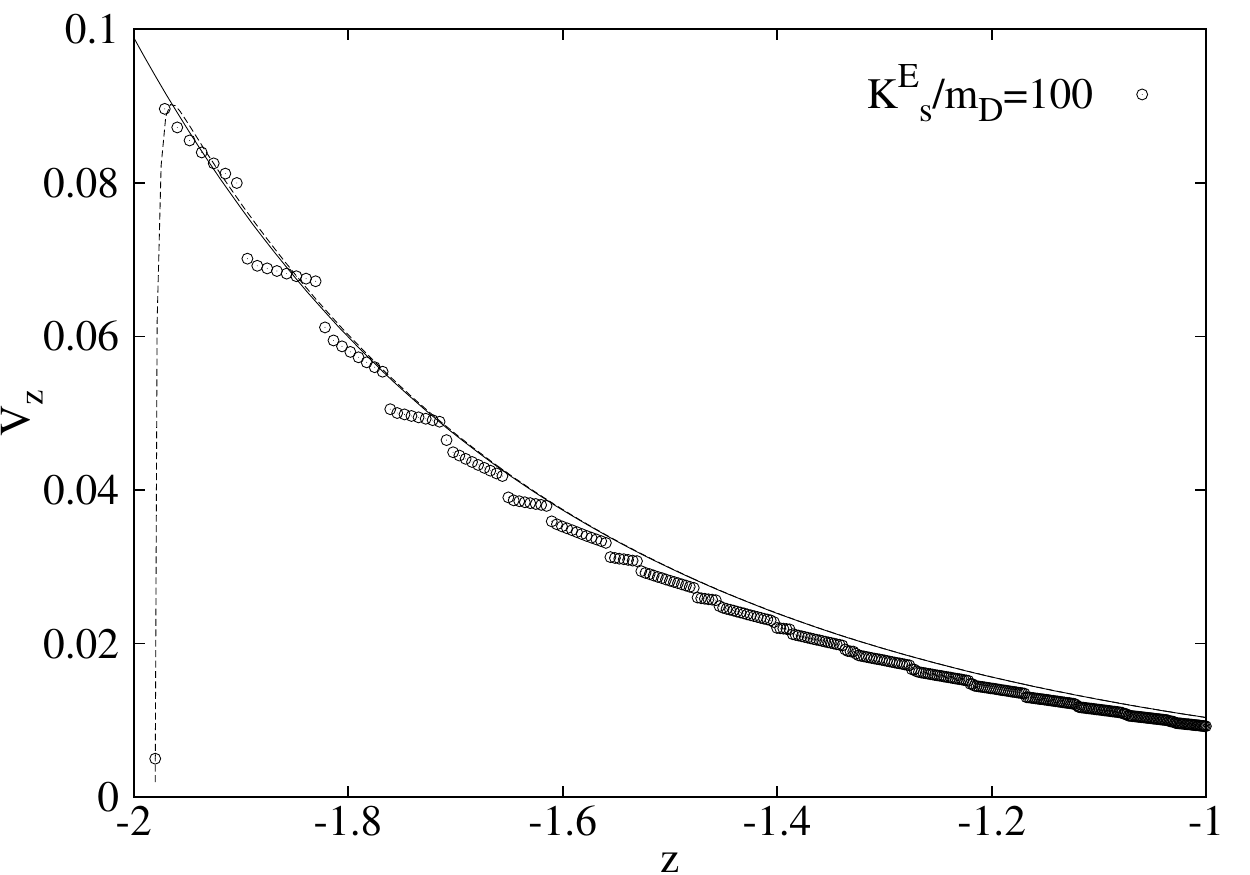}
\includegraphics[width=80mm]{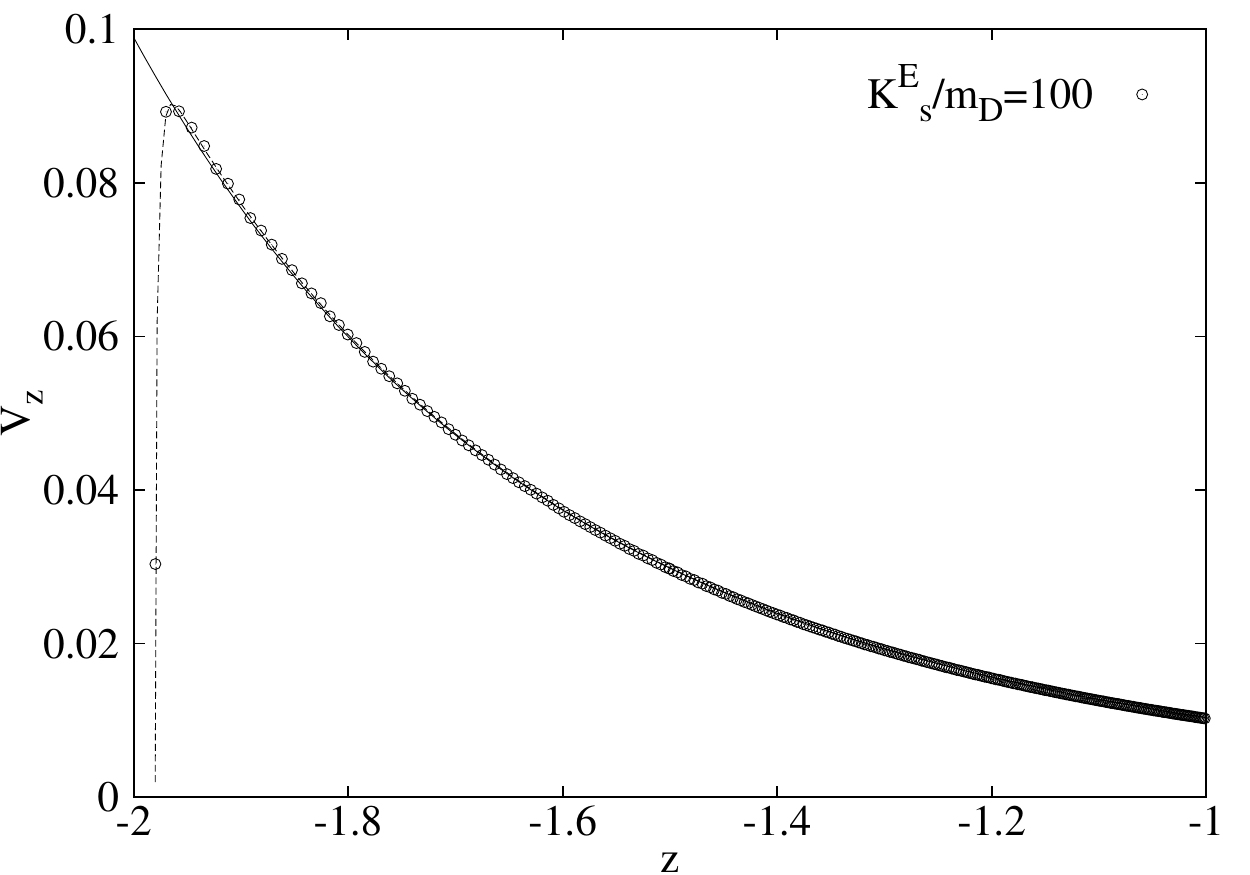} \caption{Velocity of a single
dust particle as a function of $z$ for different $K^{\rm E}_{\rm s}$ values
in the dust settling test.
Circles correspond to the particle velocity at different time steps. Dashed
lines correspond to the numerical solution of equations \ref{Eu_disk1} and 
\ref{Eu_disk2}, while solid lines represent the limiting velocity for each
case as given by equation \ref{v_lim}. As can be seen, the higher 
$K^{\rm E}_{\rm s}$ is, the sooner the limiting velocity is reached, as expected.
In the $K^{\rm E}_{\rm s}=1$ case, some oscillations of the dust particle velocity
are found in the outermost part of the disk, due to the low number of gas
particles. If a second simulation with 1000 gas and 1000 dust particles is
performed (low right plot), the trajectory of the dust particle becomes 
free from oscillations, and closely matches the analytical solution.} 
\label{fig:test5c} \end{figure*}

In Fig.~\ref{fig:test5b}, the dust component velocity as a function of $z$ is
presented for two cases ($K_{\rm s}^{\rm E}/\hat{m}_{\rm D}=10$ and 
$K_{\rm s}^{\rm E}/\hat{m}_{\rm D}=100$). As can be seen, the correct 
limiting velocity of the dust component is reached in both cases. Because
$\rho_{\rm D}/\rho_{G}=0.01$, the momentum transferred between the dust
and gas phases is rather small, and the gas component remains very close
to the hydrostatic equilibrium. As can be seen in the 
$K_{\rm s}^{\rm E}/\hat{m}_{\rm D}=100$ case (right plot), dust particles
almost instantaneously reach its limiting velocity. On the contrary, 
if $K_{\rm s}^{\rm E}/\hat{m}_{\rm D}=10$ (left plot), particles need more
time to reach the limiting velocity and the transitory state can be seen
for $\mid z \mid > 1.2$. In order to check whether the algorithm is capable
of correctly reproducing such a transitory regime, the velocity as a function
of $z$ for a single SPH dust particle can be compared with the numerical
solution of equations \ref{Eu_disk1} and \ref{Eu_disk2}. In 
Fig.~\ref{fig:test5c} the evolution of a single SPH dust particle is plotted 
for three different $K_{\rm s}^{\rm E}$ values. Circles represent the 
velocity of the particle, for different time steps, as it falls down
towards the disk mid-plane. Dashed lines represent the numerical solution 
of equations \ref{Eu_disk1} and \ref{Eu_disk2} for each case, while solid
lines represent the limiting velocity for each case as given by equation 
\ref{v_lim}. If $K_{\rm s}^{\rm E}/\hat{m}_{\rm D}=0.01$, the particle does
not have time to reach the limiting velocity, and simply suffers velocity
damping while it oscillates around the disk mid-plane. As can be seen, 
a perfect agreement is achieved with the theoretical behaviour. If 
$K_{\rm s}^{\rm E}/\hat{m}_{\rm D}=10$, the dust particle reaches the
limiting velocity at $z\simeq -1.2$, in perfect agreement with the numerical
solution of equations \ref{Eu_disk1} and \ref{Eu_disk2}, and explaining the 
global velocity profile of the dust component (as seen in Fig.~\ref{fig:test5b}).
For $K_{\rm s}^{\rm E}/\hat{m}_{\rm D}=100$, although the theoretical solution
is approximately obtained, some oscillations of the particle velocity can 
be observed. Such oscillations occur due to the low number of gas particles
present in the outermost parts of the disk. If a higher resolution simulation
is performed (1000 gas particles), the oscillations disappear, and the 
velocity of the dust particle closely  matches the analytical solution.

\section{Conclusions}

A new method has been proposed to avoid explicit integration of the time
evolution equations of small dust grains in the two fluid SPH approach. Through
the use of semi-analytic solutions for the decay of the gas and dust
relative velocity, the present method has been able to reproduce all the
features of the previous two fluid SPH approach of \cite{LPa,LPb}, with the
advantage of a considerable gain in computational time in strong drag
regimes. Due to its strictly dissipative nature, the velocity changes
induced by the drag force can be estimated without the need for explicit
acceleration recalculations or iterative procedures, even when the
stopping time becomes much shorter than the gas
evolutionary time-scale. The method is numerically stable, and always provides
convergence towards the analytical solutions as the resolution is increased.

The method has also been capable of reproducing the correct behaviour of
the drag force for all regimes. In the weak drag regime, the method is
theoretically equivalent to a standard explicit integration, both in accuracy
and computational efficiency, as long as strong gradients are not
present in the immediate neighbourhood of dust particles. In the high drag
regime, the method is capable of reproducing all the expected features of 
dust/gas mixtures. The results obtained in the test
cases are completely analogous to those found by \cite{LPa,LPb} through the use
of standard explicit and implicit methods.

In agreement with previous studies \citep{LPa,LPb}, a resolution limit has been
found for the method in the \textsc{dustywave} experiment. For high drag regimes with 
dust-to-gas ratios of order unity, the resolution should exceed 
$h<c_{\rm s}t_{\rm s}$ in order to avoid overdissipation. However, 
in the shock tube experiment, our method avoids the effects of 
overdissipation, which until now has been considered to be one of the
main limitations of the two-fluid SPH approach. Furthermore, it has also
been demonstrated that for low drag regimes, and even for high drag regimes
with low dust-to-gas ratios, the number of dust particles present in the
simulation becomes irrelevant, and the accuracy of the solution is only 
dependent on having sufficient gas resolution. Since
in the vast majority of astrophysical applications the dust-to-gas ratio is
expected to be rather low, only a good gas resolution will be necessary to avoid
overdissipation. However, special attention must be payed to this limitation, 
since it will be very difficult to completely avoid overdissipation in
complex global simulations, especially if one expects abrupt changes in the
dust-to-gas ratios.

\section*{Acknowledgments}

We thank the anonymous referee, whose thoughtful and thorough report not only resulted
in substantial improvements to the paper, but as a by-product also meant that we improved 
the numerical algorithm itself.  
We also thank Joe Monaghan and Daniel Price for very useful discussions.
Figures 2 to 9 have been created
using SPLASH \citep{DPb}, a SPH visualization tool publicly available at
http://users.monash.edu.au/$\sim$dprice/splash. The calculations for
this paper were performed on the DiRAC Complexity machine, jointly funded by
STFC and the Large Facilities Capital Fund of BIS, and the University of Exeter
Supercomputer, a DiRAC Facility jointly funded by STFC, the Large Facilities
Capital Fund of BIS and the University of Exeter. This work was also supported
by the STFC consolidated grant ST/J001627/1.

\appendix

\section{Runge-Kutta-Fehlberg integrator}

In order to  capture both the gas and dust evolution described by equations 
\ref{Eu1}, \ref{Eu2}, \ref{Eu3}, \ref{Cont1}, and \ref{Cont2}, the dust-gas
drag equations must be coupled with an explicit hydrodynamical integrator.
Combination with an integrator incorporating the gas pressure gradients is
needed. One chosen integrator is a second order Runge-Kutta-Fehlberg
scheme and the combined scheme can be summarized as follows
\begin{equation} \begin{aligned} \tilde{\textbf{v}}^{t+1/2}_{\rm D} &=
\textbf{v}^t_{\rm D} +\textbf{a}_{\rm ext,D}^{\rm t}\delta t/2,\\ 
\tilde{\textbf{v}}^{t+1/2}_{\rm G} &= \textbf{v}^t_{\rm G} - 
\left(\frac{\mathbf{\nabla} P}{\rho_{\rm G}}\right)_t \delta t /2+
\textbf{a}_{\rm ext,G}^{\rm t}\delta t/2,\\
\textbf{v}^{t+1/2}_{\rm D} &= \tilde{\textbf{v}}^{t+1/2}_{\rm D}
-\xi(\tilde{\textbf{v}}^{t+1/2}_{\rm D}-\tilde{\textbf{v}}^{t+1/2}_{\rm G}),\\
\textbf{v}^{t+1/2}_{\rm G} &= \tilde{\textbf{v}}^{t+1/2}_{\rm G} +
\frac{\rho_{\rm D}}{\rho_{\rm
G}}\xi(\tilde{\textbf{v}}^{t+1/2}_{\rm D}-
\tilde{\textbf{v}}^{t+1/2}_{\rm G}),\\ \textbf{r}^{t+1/2}_{\rm D} &=
\textbf{r}^t_{\rm D} + \textbf{v}^{t}_{\rm D}\delta t/2,\\
\textbf{r}^{t+1/2}_{\rm G} &= \textbf{r}^t_{\rm G} + \textbf{v}^{t}_{\rm
G}\delta t/2, \end{aligned} \end{equation}
for the first half of the time-step, and
\begin{equation} \begin{aligned} \tilde{\textbf{v}}^{t+1}_{\rm D} &=
\textbf{v}^{t}_{\rm D} + \frac{1}{256} \textbf{a}_{\rm ext,D}^{\rm t}
\delta t + \frac{255}{256}\textbf{a}_{\rm ext,D}^{t+1/2}\delta t,\\ 
\tilde{\textbf{v}}^{t+1}_{\rm G} &= \textbf{v}^{t}_{\rm G} - \frac{1}{256} \left(\frac{\mathbf{\nabla} P}{\rho_{\rm G}}\right)_t\delta t -
\frac{255}{256}\left(\frac{\mathbf{\nabla} P}{\rho_{\rm
G}}\right)_{t+1/2}\delta t \\
& + \frac{1}{256} \textbf{a}_{\rm ext,G}^{\rm t}
\delta t + \frac{255}{256} \textbf{a}_{\rm ext,G}^{t+1/2}\delta t,\\
\textbf{v}^{t+1}_{\rm D} &= \tilde{\textbf{v}}^{t+1}_{\rm D} -
\xi(\tilde{\textbf{v}}^{t+1}_{\rm D}-\tilde{\textbf{v}}^{t+1}_{\rm G}),\\
\textbf{v}^{t+1}_{\rm G} &= \tilde{\textbf{v}}^{t+1}_{\rm G} +
\frac{\rho_{\rm D}}{\rho_{\rm G}}\xi(\tilde{\textbf{v}}^{t+1}_{\rm D}-
\tilde{\textbf{v}}^{t+1}_{\rm G}),\\
\textbf{r}^{t+1}_{\rm D} &= \textbf{r}^{t}_{\rm D}+
\frac{1}{256}\textbf{v}^{t}_{\rm D}\delta t +
\frac{255}{256}\textbf{v}^{t+1}_{\rm D}\delta t,\\
\textbf{r}^{t+1}_{\rm G} &= \textbf{r}^{t}_{\rm G}+
\frac{1}{256}\textbf{v}^{t}_{\rm G}\delta t +
\frac{255}{256}\textbf{v}^{t+1}_{\rm G}\delta t, \end{aligned}
\end{equation}
for the full time-step. We have introduced the dust and gas components external accelerations $\textbf{a}_{\rm ext,D}$ and $\textbf{a}_{\rm ext,G}$, in order to account for
forces like gravity, physical viscosity or radiation pressure. This method relies 
on the possibility of
considering pressure and drag forces as separable interactions. As the
performed tests have shown, it seems to be a good assumption.\\

Another particularly useful property of the present method is its capacity to
predict the correct modified sound speed of the dust/gas mixture, as a function
of the dust/gas ratio. By substituting the pre-dragged quantities
$\tilde{\textbf{v}}^{\rm G}$ and $\tilde{\textbf{v}}^{\rm D}$ and the expression
for the $\xi$ parameter into the $\textbf{v}^{\rm G}$ and $\textbf{v}^{\rm D}$
equations, one can convert the two-step method into an equivalent one-step
method given by the set of equations for the first half of the time-step
\begin{equation} \begin{aligned} \textbf{v}^{t+1/2}_{\rm D} &= \textbf{v}^t_{\rm
D} +\textbf{a}^{\rm t}_{\rm ext,D}\delta t/2 - \frac{\rho_{\rm G}}{\rho^{*}}
\left(\textbf{v}^t_{\rm DG} + \textbf{a}^t_{\rm ext,DG}\delta t/2\right) \\
&- \left(\frac{\mathbf{\nabla} P}{\rho^{*}}\right)_t\delta t/2,\\
\textbf{v}^{t+1/2}_{\rm G} &= \textbf{v}^t_{\rm G} +
\textbf{a}^{\rm t}_{\rm ext,G}\delta t/2 + 
\frac{\rho_{\rm D}}{\rho^{*}}\left(\textbf{v}^t_{\rm DG} +
\textbf{a}^t_{\rm ext,DG}\delta t/2\right) \\
&- \left(\frac{\mathbf{\nabla}P}{\rho^{**}}\right)_t\delta t/2,\\
\textbf{r}^{t+1/2}_{\rm D} &= \textbf{r}^t_{\rm D} + \textbf{v}^{t}_{\rm D}\delta t/2,\\
\textbf{r}^{t+1/2}_{\rm G} &= \textbf{r}^t_{\rm G} + \textbf{v}^{t}_{\rm
G}\delta t/2, \end{aligned} \end{equation}
where $\textbf{v}^t_{\rm DG} \equiv \textbf{v}^t_{\rm
D}-\textbf{v}^t_{\rm G}$, $\textbf{a}^t_{\rm DG} \equiv \textbf{a}^t_{\rm
D}-\textbf{a}^t_{\rm G}$ and
\begin{equation} \begin{aligned} \textbf{v}^{t+1}_{\rm D} &= \textbf{v}^{t}_{\rm
D} +\frac{1}{256}\textbf{a}^{\rm t}_{\rm ext,D}\delta t + 
\frac{255}{256}\textbf{a}^{\rm t+1/2}_{\rm ext,D}\delta t \\
&-\frac{1}{256}\left(\frac{\mathbf{\nabla} P}{\rho^*}\right)_t\delta
t -\frac{255}{256}\left(\frac{\mathbf{\nabla}
P}{\rho^*}\right)_{t+1/2}\delta t \\
&- \frac{\rho_{\rm G}}{\rho^{*}}\left(\textbf{v}^{t}_{\rm DG}+
\frac{1}{256}\textbf{a}^{\rm t}_{\rm ext,DG}\delta t + 
\frac{255}{256}\textbf{a}^{\rm t+1/2}_{\rm ext,DG}\delta t\right) ,\\ 
\textbf{v}^{t+1}_{\rm G} &= \textbf{v}^{t}_{\rm
G} +\frac{1}{256}\textbf{a}^{\rm t}_{\rm ext,G}\delta t + 
\frac{255}{256}\textbf{a}^{\rm t+1/2}_{\rm ext,G}\delta t \\
&-\frac{1}{256}\left(\frac{\mathbf{\nabla} P}{\rho^{**}}\right)_t\delta
t -\frac{255}{256}\left(\frac{\mathbf{\nabla}
P}{\rho^{**}}\right)_{t+1/2}\delta t \\
&+ \frac{\rho_{\rm D}}{\rho^{*}}\left(\textbf{v}^{t}_{\rm DG}+
\frac{1}{256}\textbf{a}^{\rm t}_{\rm ext,DG}\delta t + 
\frac{255}{256}\textbf{a}^{\rm t+1/2}_{\rm ext,DG}\delta t\right),
\end{aligned} \end{equation}
for the full time-step, where we have defined
\begin{equation} \begin{aligned} \rho^* &\equiv \frac{\rho_{\rm D}+\rho_{\rm
G}}{1-e^{-\delta t/t_{\rm s}}},\\ \rho^{**} &\equiv \frac{\rho_{\rm D}+\rho_{\rm
G}}{1+\left(\frac{\rho_{\rm D}}{\rho_{\rm G}}\right)e^{-\delta t/t_{\rm s}}}.
\end{aligned} \end{equation}
As we can see, in this set of equations, dust can be no longer considered 
pressureless. It suffers an acceleration due to pressure gradient, and
possesses an effective density $\rho^*$. This result can be understood if one
realizes that a purely dissipative force does not always lead to a velocity
decrease. Because drag is a purely dissipative force, it will always lead to a
decrease in the relative velocity between dust and gas components. But
sometimes, the only way to decrease such a relative velocity is to accelerate
the dust component. Also, the effective densities $\rho^*$ and $\rho^{**}$ can
be understood as the effective inertial response of the dust and gas components
to the effective pressure terms. The weaker the drag force is, the higher the
pressure gradient must be to accelerate the dust component. If $\delta t/t_{\rm
s} \ll 1$, $\rho^* \rightarrow \infty$, $\rho^{**} \rightarrow \rho_{\rm G}$,
and the equations for the change in velocity of the dust and gas components
become
\begin{equation} \begin{aligned} \textbf{v}^{t+1/2}_{\rm D} &= \textbf{v}^t_{\rm
D}+\textbf{a}^{\rm t}_{\rm ext,D}\delta t/2,\\ \textbf{v}^{t+1/2}_{\rm G} &= 
\textbf{v}^t_{\rm G} - \left(\frac{\mathbf{\nabla} P}{\rho_{\rm G}}\right)_t\delta t/2 + 
\textbf{a}^{\rm t}_{\rm ext,G}\delta t/2, \end{aligned}
\end{equation}
for the first half time-step and
\begin{equation} \begin{aligned} \textbf{v}^{t+1}_{\rm D} &=
\textbf{v}^{t+1/2}_{\rm D}+\frac{1}{256}\textbf{a}^{\rm t}_{\rm ext,D}\delta t 
+\frac{255}{256}\textbf{a}^{\rm t+1/2}_{\rm ext,D}\delta t,\\
\textbf{v}^{t+1}_{\rm G} &= \textbf{v}^{t+1/2}_{\rm G}
-\frac{1}{256}\left(\frac{\mathbf{\nabla} P}{\rho_{\rm
G}}\right)_t\delta t -\frac{255}{256}\left(\frac{\mathbf{\nabla} P}{\rho_{\rm
G}}\right)_{t+1/2}\delta t\\
&+\frac{1}{256}\textbf{a}^{\rm t}_{\rm ext,G}\delta t 
+\frac{255}{256}\textbf{a}^{\rm t+1/2}_{\rm ext,G}\delta t,\end{aligned} \end{equation}
for the full time-step. The effective dust density term $\rho^*$ has
become infinitely big, so the dust does not respond at all to the pressure
gradient terms. That is, gas and dust decouple, and gas evolves as a single
component fluid with sound speed $c_{\rm s}$. If, on the contrary, $\delta
t/t_{\rm s} \gg 1$, $\rho^* \rightarrow \rho_{\rm D} + \rho_{\rm G}$, $\rho^{**}
\rightarrow \rho_{\rm D} +\rho_{\rm G}$, and the equations for the change in
velocity of the dust and gas components become this time
\begin{equation} \begin{aligned} \textbf{v}^{t+1/2}_{\rm D} &= \frac{\rho_{\rm
D}\textbf{v}_{\rm D}^t + \rho_{\rm G}\textbf{v}_G^t} {\rho_{\rm D}+ \rho_{\rm
G}} - \left(\frac{\mathbf{\nabla} P}{\rho_{\rm D}+\rho_{\rm G}}\right)_t\delta
t/2\\
&+\frac{\rho_{\rm D}\textbf{a}_{\rm ext,D}^t + \rho_{\rm G}
\textbf{a}_{\rm ext,G}^t} {\rho_{\rm D}+ \rho_{\rm G}}\delta t/2,\\
\textbf{v}^{t+1/2}_{\rm G} &= \frac{\rho_{\rm D}\textbf{v}_{\rm D}^t +
\rho_{\rm G}\textbf{v}_G^t} {\rho_{\rm D}+ \rho_{\rm G}} -
\left(\frac{\mathbf{\nabla} P}{\rho_{\rm D} +\rho_{\rm G}}\right)_t\delta t/2\\
&+\frac{\rho_{\rm D}\textbf{a}_{\rm ext,D}^t + \rho_{\rm G}
\textbf{a}_{\rm ext,G}^t} {\rho_{\rm D}+ \rho_{\rm G}}\delta t/2,
\end{aligned} \end{equation}
for the first half time-step and
\begin{equation} \begin{aligned} \textbf{v}^{t+1}_{\rm D} &=\frac{\rho_{\rm
D}\textbf{v}_{\rm D}^t + \rho_{\rm G}\textbf{v}_G^t} {\rho_{\rm D}+ \rho_{\rm
G}} - \left(\frac{1}{256}\right)\left(\frac{\mathbf{\nabla} P}{\rho_{\rm D}
+\rho_{\rm G}}\right)_t\delta t\\
&-\frac{255}{256}\left(\frac{\mathbf{\nabla} P}{\rho_{\rm
D}+\rho_{\rm G}}\right)_{t+1/2}\delta t\\
&+\frac{1}{255}
\frac{\rho_{\rm D}\textbf{a}_{\rm ext,D}^t + \rho_{\rm G}
\textbf{a}_{\rm ext,G}^t}{\rho_{\rm
D}+\rho_{\rm G}}\delta t \\
&+\frac{255}{256}
\frac{\rho_{\rm D}\textbf{a}_{\rm ext,D}^{t+1/2} + \rho_{\rm G}
\textbf{a}_{\rm ext,G}^{t+1/2}}{\rho_{\rm D}+\rho_{\rm G}}\delta t,\\
 \textbf{v}^{t+1}_{\rm G} &=
\frac{\rho_{\rm D}\textbf{v}_{\rm D}^t + \rho_{\rm G}\textbf{v}_G^t} {\rho_{\rm
D}+ \rho_{\rm G}} -\frac{1}{256}\left(\frac{\mathbf{\nabla}
P}{\rho_{\rm D}+\rho_{\rm G}}\right)_t\delta t -\\
&-\frac{255}{256}\left(\frac{\mathbf{\nabla} P}{\rho_{\rm
D}+\rho_{\rm G}}\right)_{t+1/2}\delta t\\
&+\frac{1}{255}
\frac{\rho_{\rm D}\textbf{a}_{\rm ext,D}^t + \rho_{\rm G}
\textbf{a}_{\rm ext,G}^t}{\rho_{\rm
D}+\rho_{\rm G}}\delta t \\
&+\frac{255}{256}
\frac{\rho_{\rm D}\textbf{a}_{\rm ext,D}^{t+1/2} + \rho_{\rm G}
\textbf{a}_{\rm ext,G}^{t+1/2}}{\rho_{\rm D}+\rho_{\rm G}}\delta t,
\end{aligned} \end{equation}
for the full time-step. Both effective density terms $\rho^*$ and
$\rho^{**}$ become equal, so both dust and gas components evolve as a single
component fluid, with the total mass of the mixture being advected. However, and
since only the gas component can produce real pressure, they travel with a
modified sound speed $\hat{c}_{\rm s} = c_{\rm s}/\sqrt{1+\rho_{\rm
D}/\rho_{\rm G}}$, exactly as predicted by theory (see for example
\cite{MRB}). As can be seen in equations A9, both phases adopt in this regime
the barycentric velocity in just one time-step, as it corresponds to a case
where $\delta t/t_{\rm s} \gg 0$.

Despite being particularly useful to visualize the behaviour of dust and gas
mixtures, and to show that the effective sound speed of the mixture is the
expected one in the strong drag regime, we still recommend using the two-step
method given by equations A1 and A2. It is clearly
technically easier to implement into a pre-existing SPH code.

\section{Predictor-corrector integrator}

The second chosen integrator was a modification of the predictor-corrector
scheme of \cite{Sea} and it can be summarized as follows
\begin{equation} \begin{aligned} \tilde{\textbf{v}}^{t+1/2}_{\rm D} &=
\textbf{v}^t_{\rm D} + \textbf{a}^{\rm t}_{\rm ext,D}\delta t,\\ 
\tilde{\textbf{v}}^{t+1/2}_{\rm G} &= \textbf{v}^t_{\rm
G} - \left(\frac{\mathbf{\nabla} P}{\rho_{\rm G}}\right)_t \delta t + 
\textbf{a}^{\rm t}_{\rm ext,G}\delta t, \\
\textbf{v}^{t+1/2}_{\rm D} &= \tilde{\textbf{v}}^{t+1/2}_{\rm D}
-\xi(\tilde{\textbf{v}}^{t+1/2}_{\rm D}-\tilde{\textbf{v}}^{t+1/2}_{\rm G}),\\
\textbf{v}^{t+1/2}_{\rm G} &= \tilde{\textbf{v}}^{t+1/2}_{\rm G} +
\frac{\rho_{\rm D}}{\rho_{\rm
G}}\xi(\tilde{\textbf{v}}^{t+1/2}_{\rm D}-
\tilde{\textbf{v}}^{t+1/2}_{\rm G}),\\ \textbf{r}^{t+1/2}_{\rm D} &=
\textbf{r}^t_{\rm D} + (\textbf{v}^{t+1/2}_{\rm D}+\textbf{v}^{t}_{\rm D})\delta
t/2,\\ \textbf{r}^{t+1/2}_{\rm G} &= \textbf{r}^t_{\rm G} +
(\textbf{v}^{t+1/2}_{\rm G}+\textbf{v}^{t}_{\rm G})\delta t/2,\end{aligned}
\end{equation}
for the predictor phase and
\begin{equation} \begin{aligned} \tilde{\textbf{v}}^{t+1}_{\rm D} &=
\textbf{v}^{t+1/2}_{\rm D}+\left[\textbf{a}_{\rm ext,D}^{t+1/2} - 
\textbf{a}_{\rm ext,D}^{t}\right]\delta t/2,\\ \tilde{\textbf{v}}^{t+1}_{\rm G} &=
\textbf{v}^{t+1/2}_{\rm G} - \left[\left(\frac{\mathbf{\nabla} P}{\rho_{\rm
G}}\right)_{t+1/2} - \left(\frac{\mathbf{\nabla} P}{\rho_{\rm
G}}\right)_{t}\right]\delta t/2 + \left[\textbf{a}_{\rm ext,G}^{t+1/2} - 
\textbf{a}_{\rm ext,G}^{t}\right]\delta t/2, \\ \textbf{v}^{t+1}_{\rm D} &=
\tilde{\textbf{v}}^{t+1}_{\rm D} -\xi\left[(\tilde{\textbf{v}}^{t+1}_{\rm
D}-\tilde{\textbf{v}}^{t+1}_{\rm G})-(\textbf{v}^{t+1/2}_{\rm
D}-\textbf{v}^{t+1/2}_{\rm G})\right],\\ \textbf{v}^{t+1}_{\rm G} &=
\tilde{\textbf{v}}^{t+1}_{\rm G} + \frac{\rho_{\rm D}}{\rho_{\rm
G}}\xi\left[(\tilde{\textbf{v}}^{t+1}_{\rm
D}-\tilde{\textbf{v}}^{t+1}_{\rm G})-(\textbf{v}^{t+1/2}_{\rm
D}-\textbf{v}^{t+1/2}_{\rm G})\right],\\ \textbf{r}^{t+1}_{\rm D} &=
\textbf{r}^{t+1/2}_{\rm D} + (\textbf{v}^{t+1}_{\rm D}-\textbf{v}^{t+1/2}_{\rm
D})\delta t/3,\\ \textbf{r}^{t+1}_{\rm G} &= \textbf{r}^{t+1/2}_{\rm G} +
(\textbf{v}^{t+1}_{\rm G}-\textbf{v}^{t+1/2}_{\rm G})\delta t/3, \end{aligned}
\end{equation}
for the corrector phase.

\bsp
\label{lastpage}

\begin{thebibliography}{23}

\bibitem[\protect\citeauthoryear{Armitage}{2010}]{Ar} Armitage P.~J.,2010,
Cambridge University Press

\bibitem[\protect\citeauthoryear{Ayliffe et. al }{2012}]{Aea} Ayliffe B.~A., 
Laibe G., Price D.~J., Bate M.~R, 2011, MNRAS, 423, 1450

\bibitem[\protect\citeauthoryear{Bate}{1995}]{MB} Bate M., 1995, PhD thesis,
Univ.~Cambridge

\bibitem[\protect\citeauthoryear{Fehlberg}{1968}]{RK} Fehlberg E., Low-order
classical Runge-Kutta formulas with step size control and their application to
some heat transfer problems, NASA Technical Report 315

\bibitem[\protect\citeauthoryear{Fulk \& Quinn}{1996}]{FQ} Fulk D.A., Quinn
D.W., 1996, J. Comput. Phys., 17, 19

\bibitem[\protect\citeauthoryear{Laibe \& Price}{2011}]{LPc} Laibe G., Price
D.~J., 2011, MNRAS, 418, 1491

\bibitem[\protect\citeauthoryear{Laibe \& Price}{2012a}]{LPa} Laibe G., Price
D.~J., 2012a, MNRAS, 420, 2345

\bibitem[\protect\citeauthoryear{Laibe \& Price}{2012b}]{LPb} Laibe G., Price
D.~J., 2012b, MNRAS, 420, 2365

\bibitem[\protect\citeauthoryear{Laibe \& Price}{2014a}]{LaiPri2014a} Laibe G., Price D.~J., 2014, MNRAS, 440, 2136

\bibitem[\protect\citeauthoryear{Laibe \& Price}{2014b}]{LaiPri2014b} Laibe G., Price D.~J., 2014, MNRAS, 440, 2147

\bibitem[\protect\citeauthoryear{Marble}{1970}]{MRB} Marble, F., 1970, Annual
Review of Fluid Mechanics, 2, 397

\bibitem[\protect\citeauthoryear{Monaghan}{1992}]{MN} Monaghan J. J., 1992,
Annual review of astronomy and astrophysics, 30, 543

\bibitem[\protect\citeauthoryear{Monaghan}{1997}]{MNa} Monaghan J. J., 1997,
Journal of Computational Physics, 136, 298

\bibitem[\protect\citeauthoryear{Monaghan}{2002}]{MNb} Monaghan J. J., 2002,
MNRAS, 335, 843

\bibitem[\protect\citeauthoryear{Monaghan}{1997}]{MNc} Monaghan J. J., 1997,
Journal of Computational Physics, 138, 801

\bibitem[\protect\citeauthoryear{Monaghan \& Kocharyan}{1995}]{MK} Monaghan J.
J., Kocharyan A., 1995, Computer Physics Communications, 87, 225

\bibitem[\protect\citeauthoryear{Price}{2007}]{DPb} Price D.~J., 2007,
Publications of the Astronomical Society of Australia, 24, 159

\bibitem[\protect\citeauthoryear{Price}{2008}]{DP} Price D.~J., 2008, Journal of
Computational Physics, 227, 10040

\bibitem[\protect\citeauthoryear{Price \& Monaghan}{2004}]{PriMon2004} Price D.J., Monaghan J. J., 2004,
MNRAS, 348, 139

\bibitem[\protect\citeauthoryear{Randles \& Libersky}{1996}]{RL} Randles P.W.,
and Libersky L.D., 1996, Smoothed Particle Hydrodynamics some recent
improvements and applications, Computer Methods in Applied Mechanics and
Engineering, 138, 375

\bibitem[\protect\citeauthoryear{Saffman}{1962}]{S} Saffman P.G.,On the
stability of laminar flow of a dusty gas, J. Fluid Mechanics, 13, 120

\bibitem[\protect\citeauthoryear{Serna et al.}{1995}]{Sea} Serna A., Alimi
J.-M., Chieze J.-P., 1995, Astrophysical Journal, 461, 884

\bibitem[\protect\citeauthoryear{Sedov}{1959}]{SD} Sedov L., 1959, Similarity
and Dimensional Methods in Mechanics, New York: Academic

\bibitem[\protect\citeauthoryear{Springel \& Hernquist}{2002}]{SH} Springel V.,
Hernquist L., 2002, MNRAS, 333, 649

\bibitem[\protect\citeauthoryear{Weidenschilling}{1977}]{Wei} Weidenschilling,
S.J., 1977, MNRAS, 180, 57

\bibitem[\protect\citeauthoryear{Whipple}{1972}]{Whi} Whipple, F.L., 1972, From
plasma to planet, ed A. Elvius, Wiley, London, p.211 

\bibitem[\protect\citeauthoryear{Wetzstein}{2009}]{Wetz} Wtezstein M., Nelson A.~F.,
Naab T., Burkert A., 2009, ApJS, 184, 298

\end{thebibliography}
\end{document}